\documentclass[fleqn,usenatbib]{mnras}
\usepackage{chngcntr}
\counterwithout{table}{section}
\counterwithout{figure}{section}
\usepackage{graphicx}
\usepackage{amsmath}
\DeclareMathOperator{\sech}{sech}
\title[Solar System in Context]{Placing the Solar System in its Astrophysical Context}
\author[G. Gonzalez]{Guillermo Gonzalez$^{1}$\thanks{E-mail:
ggonzo@mac.com}\\
Tellus1 Scientific, LLC, 8401 Whitesburg Dr SE, Unit 4662, Huntsville, AL 35802 USA}
\hypersetup{draft}
\begin{document}

\date{Accepted ??. Received ??; in original form ??}

\pagerange{\pageref{firstpage}--\pageref{lastpage}} \pubyear{??}

\maketitle

\label{firstpage}

\begin{abstract}
We examine recent astronomical data to assess whether the sun and Solar System possess anomalous properties compared to other stars and exoplanetary systems, providing context for astrobiology research. Utilising data primarily from large surveys like {\it Gaia}, {\it Kepler}, {\it TESS}, and ground-based spectroscopy (e.g., GALAH, LAMOST, HARPS), we construct comparison samples (e.g., nearby stars, solar analogues and twins within 20-200 pc) and employ statistical methods, including regression analysis, to account for parameter dependencies. We find that the sun is modestly metal-rich compared to nearby solar-age stars. More anomalous solar properties include its mass (top $\sim$8 percent locally), low photometric variability on short timescales ($\sim$0.2 percent), specific light and heavy element abundance patterns (high beryllium, low lithium, low carbon/oxygen and nitrogen/oxygen ratios, and low heavy neutron capture and refractory elements), slow rotation, and low superflare rate. The sun has average $\alpha$/iron, phosphorus/iron, and Ytterbium/iron abundance ratios. It also has average chromospheric activity as measured by R$^{\rm '}_{\rm HK}$(T$_{\rm eff}$), R$^{\rm +}_{\rm HK}$, and H$\alpha$ indices. The Solar System is unusual in its lack of super-Earths despite hosting a cold Jupiter ($\sim$3 percent), the low eccentricities of its planets (especially considering detectability, $<2$ percent), its large size scale for a multi-planet system ($\sim$6 percent), and potentially the sun's obliquity. The sun's galactic orbit is less eccentric and has lower vertical excursions than $\sim$95 percent of nearby solar analogues. Its current position is near perigalacticon and minimum distance from the Galactic plane, resulting in a higher local star density than 98.8 percent of randomly chosen times from $-0.5$ to $+0.5$ Gyrs.
\end{abstract}


\section{Introduction}
\label{sec:intro}
Is the sun an average or ordinary star, as it is often described? Over the past few decades the question of the sun's or Solar System's ``specialness'' has been addressed directly in a number of studies \citep{gus98,gon99a,gon99b,rob08,mel09,livio19}. The conclusions have varied, but there has been evidence that the sun and planets are anomalous in at least some respects.

This topic is often discussed within the context of the Anthropic Principle, as an observer selection bias \citep{carter74,carter83}. If the sun or Solar System differs significantly in some parameter relative to a carefully prepared comparison sample, this can be taken as evidence that our existence requires it to have that particular value. Given this, a search for anomalous solar or Solar System parameter values can serve the broad goals of astrobiology research.

Both the quality and quantity of data on stars and exoplanets have improved in recent years due to ongoing and recently completed ground- and space-based surveys. Of particular importance for the present study are: the stellar observations from {\it Gaia} \citep{vall23}, the stellar and planetary transit observations made with {\it Kepler} \citep{bor10}, and various stellar and planetary RV observations made with ground-based spectroscopic surveys.\footnote{Example ground-based spectroscopic survey results cited in the present study include: GALAH DR3 \citep{bud21}, LAMOST DR7 \citep{zhang22}, and \citep{bed18}.} They will serve as the foundation for most of the comparisons we make with the sun and the planets in the Solar System.

In the present work, when we state that the sun or Solar System is anomalous in some property, we mean the following. First, we construct a comparison sample consisting of stars that are similar to the sun in all the relevant aspects, except the property that is being compared. In most cases this will entail drawing from surveys of nearby stars in a volume centered on the sun; the radius of the sample volume ranges from 10 pc to 200 pc, depending on the survey. Next, the value of the sun's property is compared in a statistical way to the comparison sample, which could be as simple as the difference from the sample mean and the sample standard deviation. If the value of the solar property differs from the comparison sample by at least 1-$\sigma$, then we can say that the sun is anomalous in that property in relation to the local sample considered. We do have to be careful whether we use the standard error (se) or standard error of the mean (sem) in the comparison; we justify our choice in each case we examine below.

This approach doesn't capture a category of anomalies best illustrated by an example. There is evidence that the photometric variability of sun-like stars in the blue is at a minimum at or very near the solar metallicity. If we were to compare the sun's blue variability to that of other stars and correct for differences in fundamental stellar parameters, we wouldn't find that it deviates from the mean. In this case, if we correct for all differences except metallicity, we would find that the sun occupies a special place at the minimum of the variability versus metallicity plot.

The purpose of the present work is to review the recent literature and employ the best astronomical data relevant to answering the question, ``Is the sun and the Solar System typical or anomalous?''. The comparisons between the sun/Solar System and other stars/exoplanets is divided up into three sections: comparing the sun to nearby dwarfs (Section~\ref{sec:nearby}), comparing the Solar System planets to exoplanets (Section~\ref{sec:exoplanets}), and the galactic context (Section~\ref{sec:galactic}). The primary product of the present work is Table \ref{tab:summary}, where we list, for each property of the sun or Solar System examined, whether it is anomalous and an estimate of the probability that the sun was drawn from a suitable comparison sample at random. In Section~\ref{sec:disc} we offer some possible explanations for our findings.

\section{Comparing the sun to other nearby dwarfs}
\label{sec:nearby}

The observable properties of the sun and other stars include luminosity, size, effective temperature, surface gravity, surface composition, rotation period, activity (irradiance variations, Ca II H\&K emission, flares), binarity, presence of planets and galactic orbit. Derived properties include age and mass. Arguably the most fundamental stellar parameter is mass, as it determines a star's full lifecycle; secondary in importance is the composition \citep{eggen21}. Although a star is specifically defined as fusing hydrogen in its core while on the main sequence, in the following analysis brown dwarfs are also included in this category unless stated otherwise.

Comparing the sun's properties to other stars is not always straightforward, as the properties are not always independent. One of the more difficult cases is comparison of metallicity. It depends both on the location and kinematics in the galaxy as well as age \citep{wu21,gond23}. Older stars in hotter components of the galaxy are generally more metal-poor. Observational biases can result when these correlations are not taken into account. The following analyses will sometimes include several ways of comparing the sun's properties to other stars in order to reduce the biases.

Our approach in this section involves construction of comparison samples, each tailored to the specific solar property being examined. For mass, we examine volume-limited samples of the nearest stellar and sub-stellar objects. For metallicity and age, we construct multiple samples with varying parameter cuts and assumptions to try to capture the diversity of processes involved, such as the galactic radial metallicity and local age gradients, and initial versus present surface metallicity. For specific element abundances and also for the stellar activity metrics, we follow a modified approach. In these cases, we mostly rely on the sample produced by the work in question and apply usually modest parameter cuts. Then, we fit a simple model with least-squares to capture the significant dependencies of a particular quantity on the fundamental stellar parameters. With this fit, we predict its value for the solar parameters. In this way, we remove the trends from the comparison sample in order to compare to the observed solar value fairly.

The ages of stars in the analyses we cite below are obtained from isochrone fitting. The results are somewhat sensitive to the particular models used (a diversity of models are used by the various studies). Modern studies also employ Bayesian methods to derive ages, and these depend on the assumed priors. Ages from isochrone-fitting have large uncertainties for lower main sequence stars given the convergence of stellar evolutionary tracks there. However, for the types of stars we examine in the present work (generally with  T$_{eff} > 5600$ K) and the typical uncertainties in the stellar parameters for bright stars, ages can be estimated with low uncertainties (as low as about 10 percent).

Table \ref{tab:sun_params} lists the basic properties of the sun, many of which we will be comparing to other stars in this work. The most precise value was selected for each parameter, along with its associated uncertainty and source(s). In some cases an average (or mid-point in the case of $R_{\rm 0}$) was calculated based on multiple estimates from the literature. In the case of \( z_{0} \), the sun's distance above the mid-plane of the galaxy, two values are listed, which represent the low and high estimates.

\begin{table}
\centering
\caption{Properties of the sun.}
\label{xmm}
\begin{tabular}{lcc}
\hline
Parameter & value & reference \\
\hline
Spectroscopic classification & G2V & --\\
\( T_{\text{eff}} \) & 5772.0(8) K & 1\\
\( \text{log~}{g} \) & 4.438 dex & 1\\
\( G_{\odot} \) & -26.90(1) mag & 2\\
\( M_{\text{G},\odot} \) & 4.67(1) mag & 2\\
\( (G_{\text{BP}}-G)_\odot \) & 0.33(1) mag & 2\\
\( (G-G_{\text{RP}})_\odot \) & 0.49(1) mag & 2\\
\( \text{log~}{L_{X}/L_{bol}} \) & -6.78 to -5.68 & 3\\
Age & 4.60(4) Gyr & 4 \\
\text{[M/H], [Fe/H]} & $\equiv$ 0.00 dex & --\\
A(Li) & $1.07^{+0.03}_{-0.02}$ dex & 5 \\
\( P_{\text{rot,eq}}\text{(sid)} \) & 25.033(6) d & 6\\
\( P_{\text{rot,Carr}}\text{(sid)} \) & 25.38 d & --\\
\( V_{\text{eq}} \) & 2.016(13) km s$^{-1}$ & 7\\
Galactic orbit: & & \\
~~~\( R_{0} \) & 8.23(5) kpc & 8, 9, 10\\
~~~\( z_{0} \) & 16(4) pc & 11, 12\\
~~~\( z_{0} \) & 5(2) pc & 13\\
~~~\( U_{\odot}, V_{\odot}, W_{\odot} \) & 10(2), 11(2), 7.3(3) km s$^{-1}$ & 14, 15 \\
\hline
\label{tab:sun_params}
\end{tabular}
\medskip
\parbox{80mm}{\footnotesize Notes: The standard error is given in the second column within the parentheses and refers to the last one or two digits of the quoted quantity. The references are as follows: 1 - \citet{prsa16}, 2 - \citet{cas18}, 3 - \citet{linsky20}, 4 - \citet{hg11}, 5 - \citet{mart23}, 6 - \citet{jha21}, 7 - \citet{sch80}, 8 - \citet{reid19}, 9 - \citet{grav22}, 10 - \citet{leu23}, 11 - Table 2 of \citet{griv21}, 12 - \citet{jm23}, 13 - \citet{wid19}, 14 - Table 1 of \citet{wang21b}, 15 - \citet{rob22}. In some of our analyses we will use \text{[M/H]} as a metallicity index, while in others we will use \text{[M/H]}.}
\end{table}

\subsection{Mass}
\label{subsec:mass}
{\it Gaia} has made possible the compilation of highly reliable and nearly complete samples of the nearest objects to the sun. \citet{reyle21} compiled a catalog of all the stellar and non-stellar objects within 10 pc of the sun, making use of {\it Gaia} EDR3. \citet{gol23} compiled the Fifth Catalog of Nearby Stars (CNS5), which employs {\it Gaia} EDR3 and ground-based observations of stars within 25 pc of the sun. For the purpose of comparing the sun's mass to nearby stars, these catalogs have their advantages and disadvantages. The 10 pc sample is more complete than the 25 pc sample, but the statistical uncertainties of the frequencies are greater.

The \citeauthor{reyle21} catalog contains 463 stellar objects and includes 20 white dwarfs, 4 subgiants and 86 brown dwarfs. Of these, 18 main sequence stars (3.9 percent) have spectral types equal to or earlier than the sun's. This roughly translates to the same percentage for mass; if the 4 subgiants are more massive than than the sun, then the fraction rises to 4.7 percent. Including also the 20 white dwarfs and assuming they all had initially larger masses than the sun, it can be concluded that 9.5 percent of stars were more massive than the sun at birth.

CNS5 contains 5931 stellar objects and includes 4946 main sequence stars, 264 white dwarfs, 20 red giants, and 701 brown dwarfs. If we go by luminosity\footnote{For this calculation we adopt the same value of the sun's absolute G magnitude as employed by \citet{gol23}, M$_{G_{\odot}} = 4.67$.} as a stand-in for mass for the main sequence stars, then 294, or 5.9 percent, of the main sequence stars are more massive than the sun. In comparison to the 10 pc sample, about one-third of the brown dwarfs are missing from CNS5. Correcting for this incompleteness and adding the white dwarfs and red giants, implies that 9.4 percent of the CNS5 stars had an initial mass greater than the sun's.

Given the sun's location near the mid-plane of the Milky Way (see Section~\ref{sec:galactic}), stars that reach large distances from the mid-plane are underrepresented in the 10 and 25 pc samples. According to \citet{wiel96} this bias can be corrected by applying a weight to each star (or binned group of stars) equal to $|$W$+$W$_{\odot} |$, where W is the velocity of a star toward the north galactic pole relative to the sun's, and W$_{\odot}$ is about 8 km~s$^{\rm -1}$. Applying this weight to mass bins for those stars in the CNS5 sample having the required kinematic data yields an occurrence frequency correction ratio (low-mass/high-mass) of 1.19. This results in a true incidence of stars born more massive than the sun in the solar neighborhood\footnote{The phrase ``solar neighborhood'' as employed here is not specific with respect to volume of space included.} of 8.0 percent. 

We can compare this result to estimates based on the initial mass function (IMF). The widely cited \citet{kroupa01} and \citet{chab03} IMFs imply that about 10 and 11 percent of stars are more massive than the sun, respectively. This assumes an upper limit of 100 M$_{\odot}$ and a lower limit of 0.08 M$_{\odot}$; lowering the limit to 0.05 M$_{\odot}$, in order to include brown dwarfs, reduces the fractions to 8 and 9 percent, respectively.

\subsection{Composition and age}
\label{subsec:comp}

There are many ways to compare the sun's composition and age to nearby sun-like stars. For the solar neighborhood metallicity distribution, the most basic approach is to prepare a volume-limited sample, similar to the one described above for the mass comparisons. In the past this was done in part with spectroscopic data supplemented with photometry \citep[e.g.,][]{rpm96}. At the time of this writing, the highest quality and most complete samples include some combination of {\it Gaia} EDR3 or DR3 and ground-based observations.

On the other hand, when comparing the abundances of individual elements or their ratios among nearby stars, it is not necessary to construct a strictly volume-limited or magnitude-limited sample. In the analyses summarized below, the comparisons are always performed relative to solar twins or analogues with mean parameters close to solar. In some cases, non-LTE corrections might be important, or the abundance might be based on a line in a very crowded region of the spectrum, where the linelist is incomplete. 

There are two approaches we have taken to mitigate these potential sources of systematic error. First, multiple regression fits of a given elemental abundance to the fundamental stellar parameters (e.g., T$_{\rm eff}$, [Fe/H], age, etc) allow us to estimate its value for an effective solar doppelganger. This has two advantages: it doesn't require that we have stars in the comparison sample that are extremely close to the sun in their properties, and it permits any dependences of the abundance on the stellar parameters to be empirically modelled (and removed). Second, for a given study, the value the authors obtain for the solar abundance is adopted when conducting the comparison (rather than a ``standard'' abundance based on some other source). This has the advantage that systematic errors (such as neglecting non-LTE and missing lines in the linelist) should cancel when comparing the solar values to the comparison stars, and any residuals that become significant for stars that differ most from the sun will be automatically incorporated into the regression fits.

\citet{kordo23} produced a catalogue containing over 5 million stars with spectroscopic parameters from {\it Gaia} DR3 and photometric data from {\it Gaia} EDR3 and 2MASS. They also derive isochrone-based ``projected'' ages and initial masses\footnote{See Section 2.2 of \citet{kordo23} for a description of the calculation of the stellar parameters with the projection method. From comparison with ages of field and cluster stars, they find that reliable ages are obtained for stars younger than 9-10 Gyrs for relative age uncertainties less than 50 percent; they underestimate ages for older stars.} as well as galactic orbits. We make use of these results below, as well as their projected stellar parameters: $T_{\rm eff}$, $\log g$, and \text{[M/H]}.

We performed the following cuts on the columns of their catalogue to produce the baseline volume-limited sample ({\it K-sample 1}):

\begin{itemize}
\item[(i)]   \texttt{\( \text{r\_med\_geo} \) $<$ 200}. This criterion restricts stars to be within 200 pc in order to minimize the effects of extinction and incompleteness;
\item[(ii)] \texttt{\( \text{M\_ini} \) $>$ 0.90}. This cut includes stars with initial isochrone-derived masses greater than 0.90 solar masses (M$_\odot$);
\item[(iii)] \texttt{\( \text{logg} \) $>$ 4.2}. This cut on the logarithm of the surface gravity includes only main sequence and some turnoff stars.
\end{itemize}

This sample contains 74,803 stars and includes stars up to 2.52 M$_{\odot}$. We also created a second sample ({\it K-sample 2}), only differing with the addition of an upper initial mass limit of 1.1 M$_{\odot}$; it has 64,371 stars. Finally, we made a third sample ({\it K-sample 3}), which additionally constrains the ages to be between 3.6 and 5.6 Gyr; it contains 12,496 stars. The mean uncertainty in the age for the stars in {\it K-sample 3} is 0.8 Gyr. We show the resulting simple and W-corrected means in Table~\ref{tab:star_means}.

\begin{table*}
\centering
\begin{minipage}{160mm}
\caption{Mean values of stellar samples drawn from the \citet{kordo23} catalogue.}
\label{xmm}
\label{tab:star_means}
\begin{tabular}{lcccccccc}
\hline
K-sample & \( \langle \text{age} \rangle \) & \( \langle \text{age} \rangle_{\text{corr}} \) & \( \langle M_{\text{ini}} \rangle \) & \( \langle M_{\text{ini}} \rangle_{\text{corr}} \) & \( \langle T_{\text{eff}} \rangle \) & \( \langle T_{\text{eff}} \rangle_{\text{corr}} \) & \( \langle \text{[M/H]} \rangle \) & \( \langle \text{[M/H]} \rangle_{\text{corr}} \)\\
 & (Gyr) & (Gyr) & (M$_\odot$) & (M$_\odot$) & (K) & (K) & (dex) & (dex)\\
\hline
1 & 4.09 & 4.43 & 1.009 & 1.003 & 5863 & 5852 & 0.009 & 0.003\\
2 & 4.37 & 4.70 & 0.983 & 0.979 & 5784 & 5786 & 0.013 & 0.005\\
3 & 4.62 & 4.62 & 1.006 & 1.001 & 5960 & 5945 & -0.066 & -0.069\\
\hline
\end{tabular}
\end{minipage}
\end{table*}

{\it K-sample 1} is the least restrictive of the three samples. It shows that the W-weighted means of the non-restricted parameters (age and metallicity) are close to the solar values and that the differences between the simple and W-weighted means are minor. {\it K-sample 2} values are closer to the solar values for the ages and temperatures. It is not surprising that the masses and temperatures are close to the solar values, given the construction of the samples. More interestingly, the W-corrected ages and [M/H] values are very close to solar. From these two samples we can conclude that the mean metallicity of solar-type stars in the solar neighborhood is solar.

{\it K-sample 3} shows that the mean W-weighted [M/H] value is about 0.07 dex below the solar value. About 65 percent of stars in this sample have lower [M/H] than the sun. Therefore, when we restrict the comparison sample to match the age of the sun, we find that the sun is modestly metal-rich.

We can also compare the sun's metallicity to ``zero-age'' objects in the solar neighborhood, such as B stars, young open clusters, Type I Cepheids and H II regions. \citet{este22} reviewed the results of recent spectroscopic analyses of these kinds of objects and concluded that the chemical composition of the local interstellar medium is highly homogeneous and near the solar value. The mean metallicity for the young stellar objects is about 0.04 dex with a scatter of about 0.04 dex. \citep{nit05}, who employed measurements of isotope ratios in presolar grains, argued that supernova ejecta do not produce more than one percent inhomogeneity in the interstellar medium.

We need to correct the observed surface metallicity of a star for the accumulated effects of microscopic element diffusion in its outer convection zone. \citet{yild19} calculated that the sun's initial metallicity was 0.1 higher than its present measured photospheric value. In order to better compare the sun's initial metallicity to the current metallicity of the local ISM, we combined several catalogues from the GALAH DR3 survey \citep{bud21}\footnote{\url{https://www.galah-survey.org/dr3/the_catalogues/}}: \( \text{GALAH\_DR3\_main\_allstar\_v2} \), \( \text{GALAH\_DR3\_VAC\_GaiaEDR3\_v2} \), \( \text{GALAH\_DR3\_VAC\_ages\_v2} \). These catalogues include stars observed with the HERMES spectrograph on the Anglo-Australian spectrograph. The main catalogue includes 588,571 stars. The value-added catalogues (VAC) include {\it Gaia} EDR3 and the Bayesian Stellar Parameters Estimation Code (BSTEP) to estimate fundamental parameters (e.g., age, mass, initial metallicity) using stellar isochrones \citep{sharma18}. The following cuts were applied to create a sample of solar twin stars:

\begin{itemize}
\item[(i)]   2.6 $<$ \texttt{\( \text{age\_bstep} \) $<$ 6.6}, restricts stars to be close to the sun's age (4.6 Gyr);
\item[(ii)] 0.95 $<$ \texttt{\( \text{m\_act\_bstep} \) $<$ 1.05}, includes stars with actual/observed masses between 0.95 and 1.05 M$_\odot$;
\item[(iii)] \texttt{\( \text{logg} \) $>$ 4.3}, restricts the logarithm of the surface gravity to include only main sequence stars;
\item[(iv)] 5670 $<$ \texttt{\( \text{teff\_bstep} \) $<$ 5870 K}, includes stars with similar T$_{\rm eff}$ to the sun;
\item[(v)] -0.1 $<$ \texttt{\( \text{fe\_h} \) $<$ 0.1}, includes stars with similar measured [Fe/H] to the sun;
\item[(vi)] -0.1 $<$ \texttt{\( \text{alpha\_fe} \) $<$ 0.1}, includes stars with similar [$\alpha$/Fe] to the sun;
\item[(vii)] \texttt{\( \text{distance\_bstep} \) $<$ 0.8}, includes only stars within 0.8 kpc distance,
\item[(viii)] \texttt{\( \text{snr\_c1\_iraf, snr\_c2\_iraf, snr\_c3\_iraf, snr\_c4\_iraf} \) $>$ 10}, includes only stars with signal-to-noise ratio (SNR) $>$ 10 in all four HERMES CCDs.
\end{itemize}

The ranges of the cuts are based, in part, on the typical uncertainties of the parameters; for example, the mean uncertainties in T$_{\rm eff}$ and age are 90 K and 2.4 Gyr, respectively. The resulting sample, {\it GALAH Solar Sample 1}, contains 3165 stars. The mean initial [M/H] (from the parameter \texttt{\( \text{meh\_ini\_bstep} \)}) for this sample is 0.055 dex, while the actual/current [M/H] (from the parameter \texttt{\( \text{meh\_act\_bstep} \)}) is 0.011 dex. Applying this difference to the sun, then, we find that its original [M/H] was 0.044 dex, indistinguishable from the current metallicity of the local zero-age objects.

Over the last few Gyrs the metallicity of the ISM has been increasing at about 0.02 dex/Gyr \citep{neto21}. This places the sun's metallicity almost 0.10 dex above the metallicity of the local ISM at the time of its formation. Furthermore, the sun is currently very close to its perigalacticon in its orbit, and its mean galactocentric distance is about 0.5 kpc farther than the current position (see Section~\ref{sec:galactic}). Given the galactic radial metallicity gradient of about $-0.06$ dex/kpc \citep{neto21}, this implies that the metallicity of the ISM was lower still by another 0.03 dex relative to the sun at its mean galactocentric distance. In summary, the metallicity of the local ISM was about 0.13 dex smaller at the time of the sun's birth at its average galactocentric distance.

We can also compare the distribution of [$\alpha$/Fe] in the {\it GALAH Solar Sample 1} by removing the restriction on this quantity. Doing so, we find that the mean of [$\alpha$/Fe] is 0.01 dex. The sun appears to be typical in this regard among solar twins.

We also compared the sun's age to the {\it GALAH Solar Sample 1} by removing the age, [Fe/H], and [$\alpha$/Fe] restrictions. This resulted in 8340 stars. The mean age of this sample is 5.8 Gyr, and the sun is younger than 75 percent of the stars. This differs from the mean age for the much larger {\it K-sample 2}. It is not obvious why the two estimates differ. Therefore, we cannot conclude that the sun's age is anomalous relative to sun-like stars.

\subsubsection{Other element abundances}
\label{subsec:other_elem}

\citet{wang24} report non-LTE lithium (Li) abundances for the {\it GALAH} DR3 stars. For the {\it GALAH} Solar Sample 1 stars, 2266 have Li detections; the remainder are only upper limits. Only 17 stars have Li abundances below 1.1 dex (their assumed solar abundance). This illustrates the difficulty of comparing Li abundances of other stars to the sun's Li abundance. The sun's Li line at 6707 \( \text{\AA} \) is weak and blended \citep[e.g., see Figure 1 of ][]{car19}. Spectra with high resolving power and SNR are required to perform a proper comparison when the Li abundance is comparable to the sun's value. This also applies to other elements with weak lines in sun-like stars. For the purpose of searching for subtle solar abundance anomalies, then, it is best to employ the highest quality spectra of solar twins. Fortunately, a number of high quality spectroscopic studies of FGK dwarfs are available.

For the remainder of this section and the next, we will build our comparison samples based on a loose definition of solar twin. For some of the analyses we adopt the definition used by \citet{bot20}; to be a solar twin a star must be within $\pm100$ K in T$_{\rm eff}$ and $\pm0.1$ dex in log g and [Fe/H] of the solar values. For others we will relax these constraints somewhat, for example, so the sample size is not too small (at least several tens of stars to have enough degrees of freedom for the fitting). Since we will be correcting for differences in stellar properties relative to the sun, the precise definition is not important for our use cases.

Li is a particularly informative element in the atmosphere of a star given its relatively rapid destruction and its sensitivity to several stellar parameters \citep{rath23,mart23}. Currently, the highest quality (in terms of resolving power and SNR) spectroscopic dataset for Li abundances in solar twins and analogues is the one assembled by \citeauthor{mart23}\footnote{We selected the sample stars from \citeauthor{mart23} for analysis instead of the larger sample from \citeauthor{rath23}, since the latter study mostly adds younger and also cooler main sequence stars than the sun. Instead, our analysis focuses on stars close to the sun's mass, and \citeauthor{mart23} already have a sufficiently large range in age to establish the correlation of this parameter with Li abundance.} They derived non-LTE Li abundances as well as ages, masses and other fundamental stellar properties for 118 stars, which range in temperature from 5513 to 5916 K. The ages and masses, as with other studies cited below, were estimated with Bayesian analyses using stellar isochrones in combination with the measured stellar properties. \citeauthor{mart23} also applied the same analysis to the sun using spectra of reflected light from Vesta or the Moon obtained with the same instrument.

\begin{figure}
\includegraphics[width=3.2in]{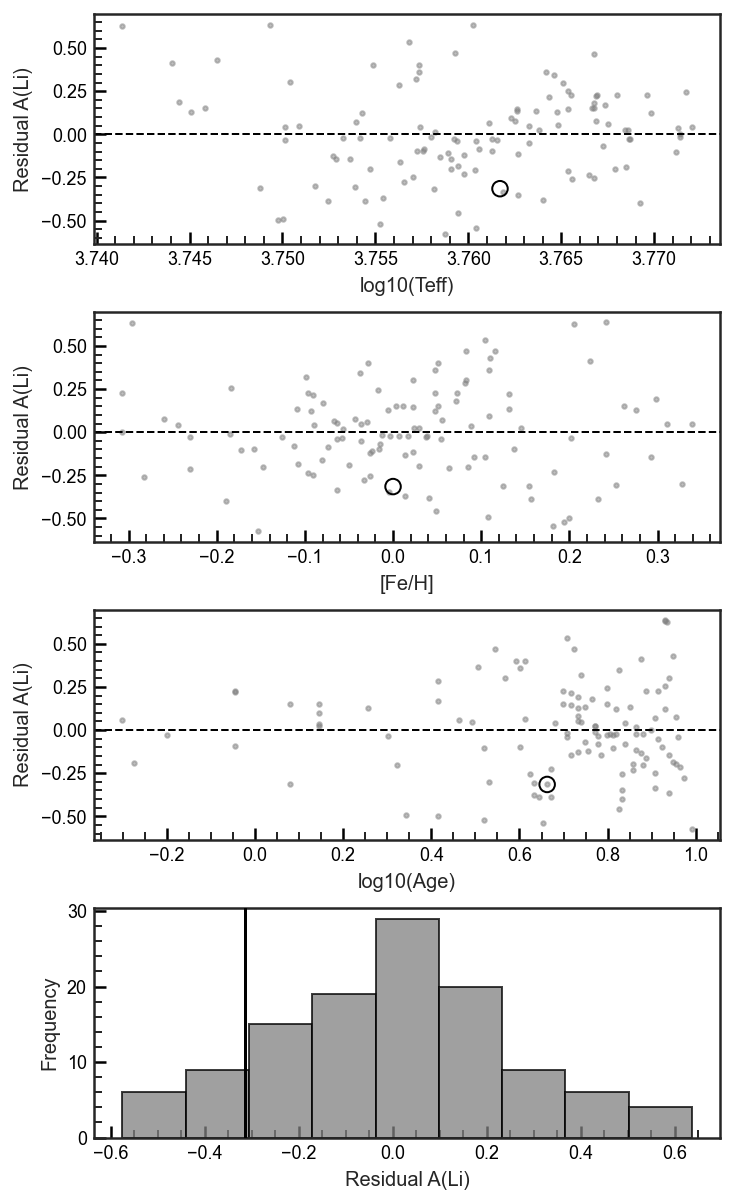}
\caption{Residual A(Li) values from regression fit using Equation~\eqref{eq:eq_Li_all} to the full dataset from \citet{mart23}. The sun is shown as an open circle in the top three panel and as a vertical solid line in the histogram panel.}
\label{fig:Li_all_fig}
\end{figure}

In order to compare the Li abundances of the stars in their sample to the sun's, it is necessary to account for differences in stellar properties with multiple regression least-squares analysis.\footnote{This regression fit (as well as the others in the present work) was performed using the ordinary least-squares function within the \texttt{statsmodels} Python module. See \url{https://www.statsmodels.org/stable/index.html}} After trying fits with various combinations of parameters (guided by the trends in Figures 3 and 5 of \citeauthor{mart23}), we settled on the following using their full dataset (with the non-LTE Li abundances):
\begin{equation}
\begin{split}
\text{A(Li)} = -(139.51 \pm 20.89) + (37.61 \pm 5.55) \times \log_{10}(\text{T$_{\rm eff}$}) \\
- (1.42 \pm 0.27) \times [\text{Fe/H}] + (1.75 \pm 0.93) \times [\text{Fe/H}]^{2} \\
- (0.88 \pm 0.30) \times (\log_{10}(\text{Age}))
+ (0.79 \pm 0.90) \times (\log_{10}(\text{Age}))^{2} \\
- (1.22 \pm 0.71) \times (\log_{10}(\text{Age}))^{3}
\end{split}
\label{eq:eq_Li_all}
\end{equation}
The adjusted R-squared of the fit is 0.80, and the standard deviation of the residuals is 0.26 dex.\footnote{Note, one star with a residual from the first regression fit round near 1 dex was removed, HIP 54287, leaving 117 stars in the sample.} The predicted solar A(Li) from this model is $1.39 \pm 0.27$(se) $\pm~0.04$(sem) dex. Given that the typical uncertainty in A(Li) is only about 0.05 dex among the sample stars, it is clear that there are unmodeled errors much larger than the measurement errors. As is evident from Fig.~\ref{fig:Li_all_fig}, the residuals appear to be random relative to the independent variables employed in the regression. We don't know their source; for lack of a better label, we will attribute them to some physical ``cosmic variance''. \citet{mart23} determined the solar A(Li) to be $1.07^{+0.03}_{-0.02}$. In this case, it we must adopt the standard error, which makes the sun's A(Li) 1.1-$\sigma$ smaller than the prediction; 11 percent of the stars have smaller A(Li) than the sun. There are a couple ways to improve upon this analysis.

\begin{figure}
\includegraphics[width=3.2in]{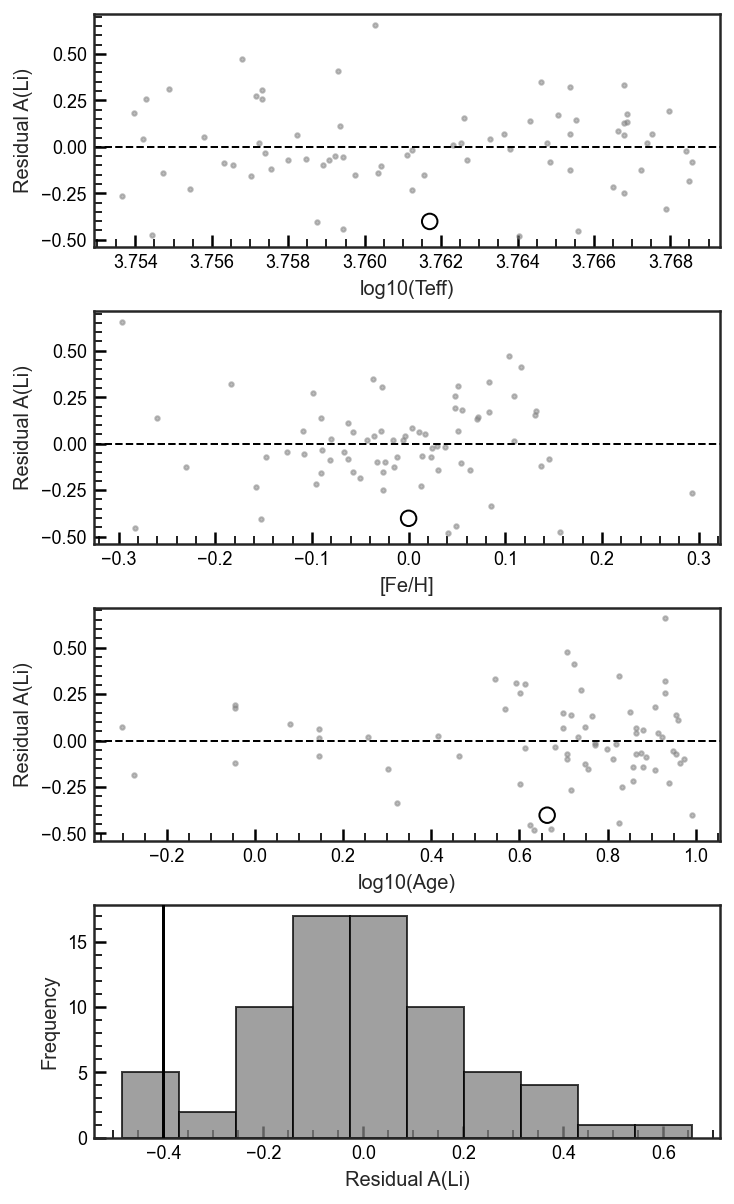}
\caption{Residual A(Li) values from regression fit using Equation~\eqref{eq:eq_Li_small} to the reduced size dataset from \citet{mart23}. All else as in Fig.~\ref{fig:Li_all_fig}.}
\label{fig:Li_small_fig}
\end{figure}

First, \citet{gonz10a} and \citet{israel09} noted that stars hosting planets tend to have lower Li abundances than otherwise similar stars lacking detected planets. \cite{rath23} and \citet{mart23} confirmed this trend, finding a difference of 0.23 dex between these two groups of stars. Second, given the strong dependence of Li abundance on temperature and the small uncertainties that \citet{mart23} quote for T$_{\rm eff}$, the analysis can benefit from further restriction on this quantity. Restricting the analysis to the stars without detected planets\footnote{\url{https://exoplanet.eu/catalog/all_fields/}} and T$_{\rm eff}$ between 5670 and 5870 K yields a sample size of 72 stars (after removing the outlier); the age ranges from 0.5 to 9.8 Gyr, and [Fe/H] ranges from -0.3 to 0.3 dex. The model fit for this new sample is:
\begin{equation}
\begin{split}
\text{A(Li)} = -(148.52 \pm 28.04) + (40.02 \pm 7.44) \times \log_{10}(\text{T$_{\rm eff}$}) \\
- (1.44 \pm 0.32) \times [\text{Fe/H}] + (2.7 \pm 1.5) \times [\text{Fe/H}]^{2} \\
- (0.59 \pm 0.29) \times (\log_{10}(\text{Age}))
+ (0.89 \pm 0.88) \times (\log_{10}(\text{Age}))^{2} \\
- (1.85 \pm 0.70) \times (\log_{10}(\text{Age}))^{3}
\end{split}
\label{eq:eq_Li_small}
\end{equation}
The adjusted R-squared of the fit is 0.84, and the standard deviation of the residuals is 0.22 dex. We show the residuals in Fig.~\ref{fig:Li_small_fig}. The predicted solar A(Li) from this model is $1.47 \pm 0.24$(se) $\pm~0.05$(sem) dex, which is 1.7 $\sigma$ greater than the solar A(Li). This corresponds to a one-sided probability of 5 percent; 7 percent of the stars have smaller A(Li) than the sun.

Applying this model to the 15 stars with planets in the same temperature range in their sample yields a mean A(Li) difference of $-0.21 \pm 0.05$ dex. There is a weak trend in the A(Li) differences that increases with T$_{\rm eff}$, similar to that reported in \citet{gonz15}. Thus, corrected for differences in their properties, on average, solar twins and analogues are not as Li-poor as the sun, regardless of the presence of detected planets. This is consistent with the preliminary conclusions of \citet{car19} and \citet{gonz15}.

Beryllium (Be) is another element that has been depleted in the sun's envelope over its history, though not as much as lithium. \citet{boes22} is the largest study of Be abundances in solar twins and analogues. They included 52 stars within two percent of the sun's mass. Their LTE spectroscopic analysis found that the Be abundance increases very gradually with increasing age, and they observe additional weak trends with T$_{\rm eff}$ and [Fe/H] (see their Figures 6 and 7). They quote the typical uncertainty in their Be abundances to be $\pm 0.12$ dex. We will analyze their results in a similar way to our analysis of Li.

\begin{figure}
\includegraphics[width=3.2in]{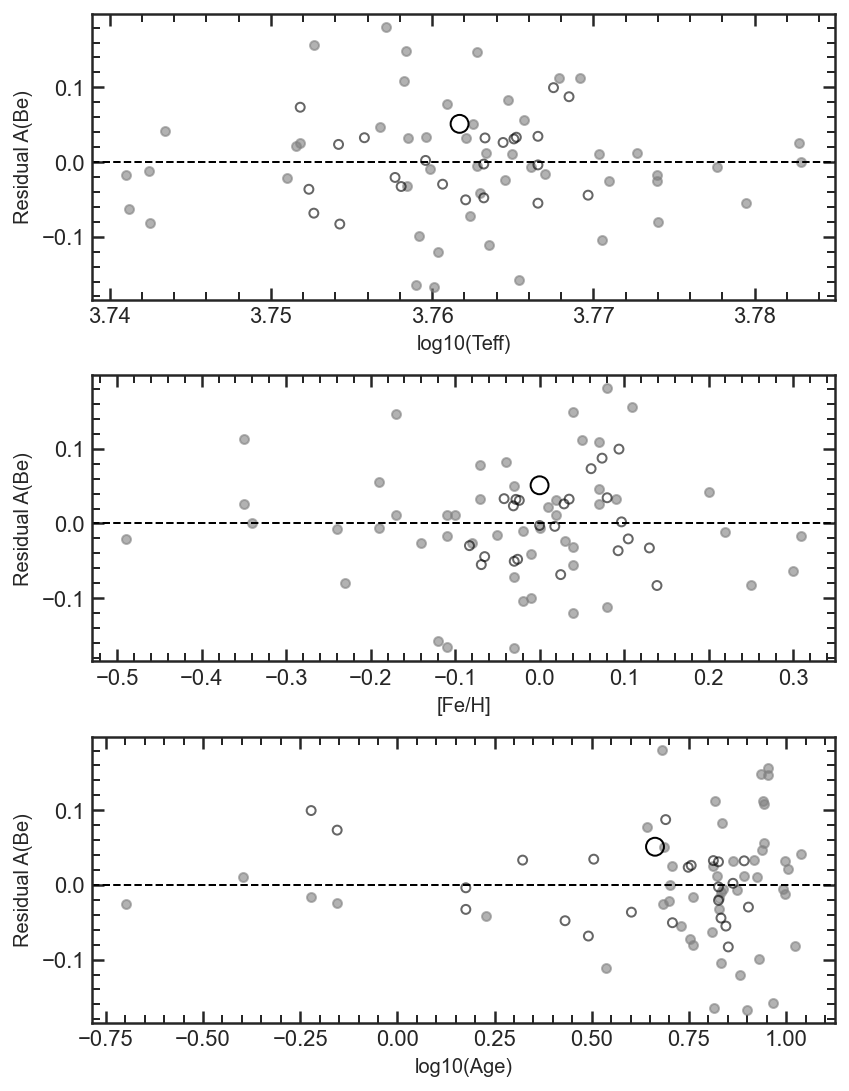}
\caption{Residual A(Be) values from regression fit using Equation~\eqref{eq:eq_Be_abund2} to the data from \citet{boes22} (dots) and \citet{regg25} (open circles).}
\label{fig:Be_abund_fig}
\end{figure}

First, we corrected their LTE Be abundances for NLTE (non-local thermodynamic equilibrium) using the lookup table of corrections published by \citet{koro22}; the solar Be abundance of \citeauthor{boes22}, corrected for NLTE, is 1.16 dex. We also excluded two stars with only upper limits on the Be abundance. We decided not to make any additional cuts to their dataset, given the small range of A(Be) and to get as much leverage as possible on the weak trends. The least-squares model fit to our 50-star subsample with NLTE A(Be) values is:
\begin{equation}
\begin{split}
\text{A(Be)} = -(8.76 \pm 7.80) + (2.60 \pm 2.07) \times \log_{10}(\text{T$_{\rm eff}$}) \\
+ (0.45 \pm 0.14) \times [\text{Fe/H}] - (0.48 \pm 0.35) \times [\text{Fe/H}]^{2} \\
+ (0.120 \pm 0.036) \times \log_{10}(\text{Age}) \\
\end{split}
\label{eq:eq_Be_abund}
\end{equation}
The adjusted R-squared of the fit is 0.46, and the standard deviation of the residuals is 0.08 dex. Given this, \citeauthor{boes22} slightly overestimated the typical uncertainty of their A(Be) values. We show the residuals in Fig.~\ref{fig:Be_abund_fig}. We can also conclude that the scatter in the residuals is fully captured by the measurement error. The predicted solar A(Be) from this model is $1.110 \pm 0.086$(se) $\pm~0.014$(sem) dex. In this case, the predicted solar A(Be) is 3.6 $\sigma$ smaller than the solar A(Be).

An independent analysis of Be abundances in solar twins was conducted by \citet{regg25}, who performed a differential analysis of 23 stars relative to the sun. Their dataset is about half the size of \citeauthor{boes22} and covers smaller ranges in T$_{\rm eff}$, [Fe/H] and ages, but the stellar parameters have smaller uncertainties. Their A(Be) uncertainties are similar. There are only 5 stars in common between the two datasets. In order to make an improved analysis of Be abundances, we have combined the two datasets and arrived at a new fit. The only difference is the addition of a parameter which gives the mean offset in A(Be) between them:
\begin{equation}
\begin{split}
\text{A(Be)} = -(11.21 \pm 5.86) + (3.26 \pm 1.56) \times \log_{10}(\text{T$_{\rm eff}$}) \\
+ (0.50 \pm 0.10) \times [\text{Fe/H}] - (0.39 \pm 0.29) \times [\text{Fe/H}]^{2} \\
+ (0.10 \pm 0.03) \times \log_{10}(\text{Age})\\
+ (0.023 \pm 0.020) \times \text{Offset}\\
\end{split}
\label{eq:eq_Be_abund2}
\end{equation}
where ``Offset'' equals 1 for the \citeauthor{regg25} stars 0 for the \citeauthor{boes22} stars. As expected, the coefficient uncertainties are smaller. The standard deviation of the residuals is 0.075 dex. The predicted solar A(Be) (on the scale of \citeauthor{boes22}) is $1.111 \pm 0.076$(se) $\pm~0.013$(sem) dex. This is 0.051 dex, or 3.9 $\sigma$, smaller than the solar non-LTE value. This confirms, though now with greater confidence, the comment made by \citeauthor{boes22} that compared to the stars most similar to it, the sun has a higher value of A(Be) by 0.05 dex.

The light elements carbon (C), nitrogen (N), and oxygen (O) are important for life and are also abundant in the universe. However, theirs are among some of the more difficult elemental abundances to determine with high precision and accuracy.\footnote{This statement is based on the author's experience in deriving CNO abundances from high-resolution high-SNR spectra of sun-like stars. Their atomic lines tend to be weak and often blended. When the CH and CN molecular lines are employed, spectrum synthesis must be used in the analysis, often in blended regions requiring well-calibrated line lists.} The highest quality (in terms of resolving power and SNR) spectroscopic datasets for the abundances of these elements in solar twins are those reported in \citet{bed18} and \citet{bot20}; both studies were done by the same research group, using the same HARPS instrument spectra. Although the authors already performed a detailed analysis to arrive at specific conclusions concerning galactic abundance trends and how the sun compares to them (see Figures 6 to 9 of \citeauthor{bot20}), we will employ their data to perform a slightly modified analysis.

First, we fit multiple linear regression models to the [C/H], [N/H], [C/O], and [C/N] values with [O/H], $\log_{10}(\text{age(Gyr)})$, and [$\alpha$/Fe] as the independent variables.\footnote{The value of $\alpha$ was calculated from the mean of [Mg/H], [Si/H], [Ca/H], and [Ti/H].} By using [$\alpha$/Fe] as one of the independent variables, we can fit the full sample of 63 stars from \citet{bot20}, rather than restricting the analysis to only the 55 thin disk solar twins. The [$\alpha$/Fe] values are taken from \citet{bed18}. Our motivation for selecting [O/H] as an independent variable, rather than [Fe/H], is the presence of trends with condensation temperature (see below), which affects the abundances of volatile elements (e.g., CNO) relative to refractories (e.g., Fe). As the CNO elements have comparable condensation temperatures, their relative abundances should not be affected by such differences.
\begin{equation}
\begin{split}
\text{[C/H]} = -(0.005 \pm 0.006) - (0.029 \pm 0.010) \times \log_{10}(\text{Age})\\
+ (0.910 \pm 0.040) \times [\text{O/H}] - (0.015 \pm 0.113) \times [\alpha/\text{Fe}]\\
\\
\text{[N/H]} = -(0.092 \pm 0.013) + (0.058 \pm 0.021) \times \log_{10}(\text{Age})\\
+ (1.253 \pm 0.080) \times [\text{O/H}] - (0.76 \pm 0.23) \times [\alpha/\text{Fe}]\\
\\
\text{[C/O]} = -(0.005 \pm 0.006) - (0.029 \pm 0.010) \times \log_{10}(\text{Age})\\
- (0.090 \pm 0.040) \times [\text{O/H}] - (0.015 \pm 0.113) \times [\alpha/\text{Fe}]\\
\\
\text{[C/N]} = (0.088 \pm 0.012) - (0.087 \pm 0.019) \times \log_{10}(\text{Age})\\
- (0.343 \pm 0.073) \times [\text{O/H}]  + (0.74 \pm 0.20) \times [\alpha/\text{Fe}]\\
\\
\text{[N/O]} = -(0.092 \pm 0.013) + (0.058 \pm 0.021) \times \log_{10}(\text{Age})\\
+ (0.253 \pm 0.080) \times [\text{O/H}] - (0.76 \pm 0.23) \times [\alpha/\text{Fe}]\\
\\
^{12}\text{C}/^{13}\text{C} = (88.6 \pm 1.4) - (3.6 \pm 2.2) \times \log_{10}(\text{Age})\\
+ (65.2 \pm 8.6) \times [\text{O/H}] + (55.7 \pm 24.0) \times [\alpha/\text{Fe}]\\
\end{split}
\label{eq:eq_CNO_abund}
\end{equation}
\renewcommand{\arraystretch}{1.5}

\begin{table}
\centering
\caption{Predicted solar CNO abundances and ratios from fits to solar twins.}
\label{xmm}
\begin{tabular}{lcccc}
\hline
Quantity & solar & solar & cosmic var & diff \\
 & measured(err) & predicted(err) & & \\
\hline
\( {\text{[C/H]}} \) & $+0.002(1)$ & $-0.026(4)$ & 0.022 & 1.3\\
\( {\text{[N/H]}} \) & $+0.002(18)$ & $-0.056(8)$ & 0.034 & 1.7\\
\( {\text{[C/O]}} \) & $+0.002(1)$ & $-0.024(4)$ & 0.018 & 1.4\\
\( {\text{[C/N]}} \) & $+0.000(18)$ & $+0.030(7)$ & 0.027 & 1.1\\
\( {\text{[N/O]}} \) & $+0.002(1)$ & $-0.054(8)$ & 0.032 & 1.8\\
$^{12}\text{C}/^{13}\text{C}$ & $88.7(5)$ & $86.2(9)$ & 2.3 & 1.1\\
\hline
\end{tabular}
\label{tab:CNO}
\medskip
\parbox{80mm}{\footnotesize Notes: The errors in the predicted solar values are sem values. The 'diff' column is the absolute value of the difference between the observed and predicted solar values based on the linear model regression fits given in Equations~\eqref{eq:eq_CNO_abund}, expressed in units of the cosmic variance (which is actually a standard deviation); the error in the solar value is neglected, except in the cases of [N/H] and $(^{12}\text{C}/^{13}\text{C})$. The solar measured values are from the first line of Table 8 of \citet{bot20}.}

\end{table}

\begin{figure}
\includegraphics[width=3.2in]{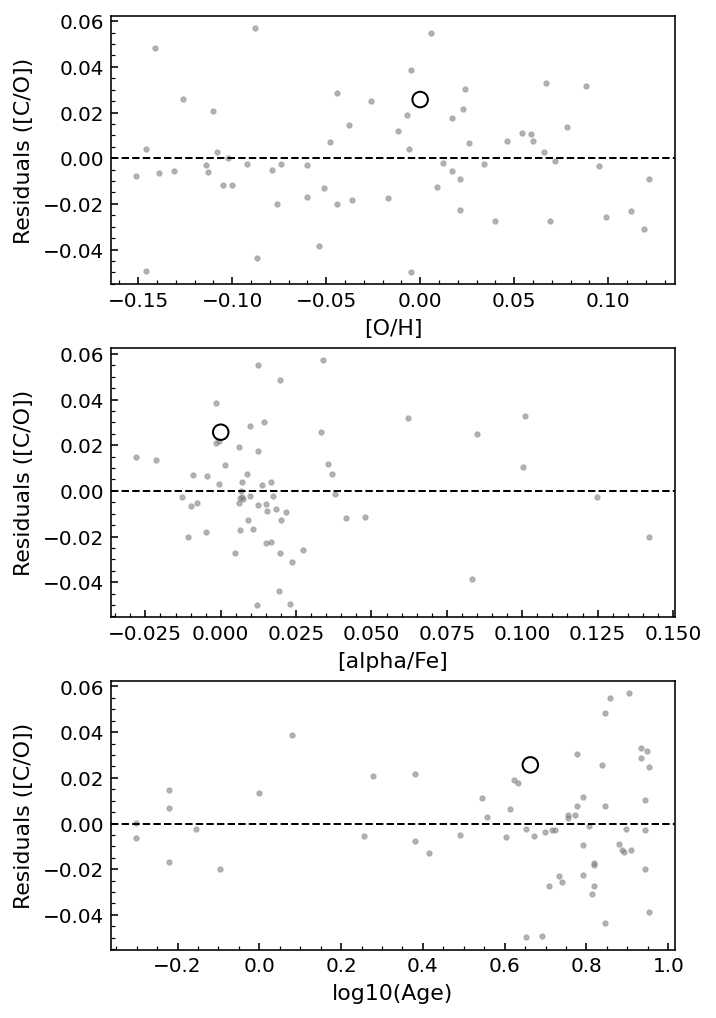}
\caption{Residual [C/O] values from regression fit using Equations~\eqref{eq:eq_CNO_abund} to the data from \citet{bot20}.}
\label{fig:C_O_abund_fig}
\end{figure}

\begin{figure}
\includegraphics[width=3.2in]{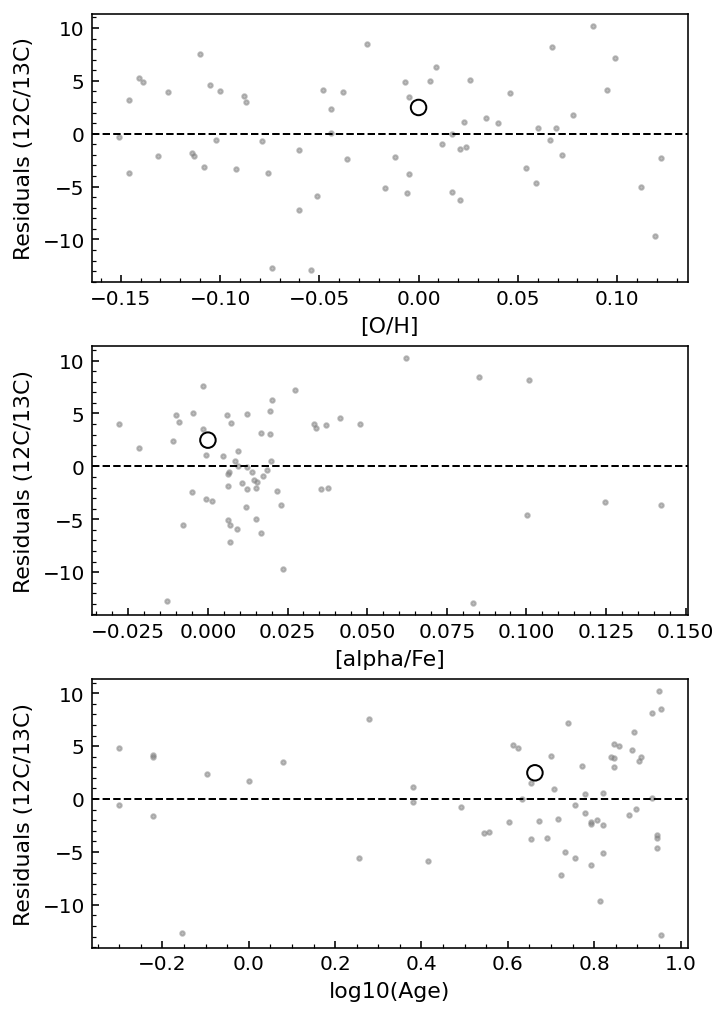}
\caption{Same as Fig.~\ref{fig:C_O_abund_fig} but for the $(^{12}\text{C}/^{13}\text{C})$ isotope ratios.}
\label{fig:12C_13C_fig}
\end{figure}

\renewcommand{\arraystretch}{1}
The adjusted R-squared of the regressions range from 0.23 to 0.89. We show sample residuals plots for [C/O] and $^{12}\text{C}/^{13}\text{C}$ in Fig.~\ref{fig:C_O_abund_fig} and Fig.~\ref{fig:12C_13C_fig}, respectively. Our analysis confirms the findings of \citeauthor{bot20} that the sun's absolute and relative CNO abundances and $(^{12}\text{C}/^{13}\text{C})$ isotope ratio differ from nearby solar twins when corrected for differences in age, metallicity, and $\alpha$-element abundance. 

To determine the significance of these differences, we need to compare the variance of the residuals to our regression fits to the uncertainties estimated by \citet{bot20}. For each case in Equation~\eqref{eq:eq_CNO_abund} the variance of the residuals is greater than the mean estimated variance. We estimated the cosmic variance (which is actually a standard deviation), as we did for A(Li) above, by subtracting the mean estimated variance in quadrature from the residuals variance and taking the square root. We list the resulting cosmic variance values in Tab.~\ref{tab:CNO}. Relative to the cosmic variance, then, the sun deviates in CNO abundances and ratios by an average factor near 1.5. We assign a probability of 5 percent, based on these results.

Following \citet{mel09}, \citet{bed18} calculated elemental abundance trends with condensation temperature (T$_{\rm c}$), and they also corrected for trends with galactic chemical evolution (GCE), which they parameterized with age. \citeauthor{mel09}, using a sample of 21 solar twins and analogues, had found that the sun is deficient in refractory elements relative to nearby solar twins. \citet{gonz10b} confirmed their findings. \citeauthor{bed18} not only confirmed these earlier studies, but they showed that the sun's abundance-T$_{\rm c}$ slope is even more anomalous than previously estimated. Only 5 of the 68 stars in their analysis have a corrected T$_{\rm c}$ slope smaller than the sun's. Statistically, then, the sun's T$_{\rm c}$ slope is smaller than 93 percent of their sample. 

This result was largely confirmed by the recent work of \citet{mart25}, who applied a different analysis method to 99 previously studied solar twins and analogs. They determined that of the stars lacking detected planets, the sun is more depleted in refractories than 89 percent of them with 9.5$\sigma$; for the stars with detected planets (sample of 10 stars), the significance is less at 4.3$\sigma$, but the sun is still more depleted.

\citeauthor{bed18} also noted that the solar abundances of the heavy neutron-capture elements, Cerium (Ce), Praseodymium (Pr), Neodymium (Nd), and Europium (Eu), are significantly deficient relative to the solar twins even after accounting for the abundance-T$_{\rm c}$ trend. The only explanation they offered is that their linear GCE corrections might be too simple. The contribution from the {\it r}-process to these elements is almost 100 percent for Eu and in the range 30-50 percent for each of the others. Certainly, the residual scatter in Figure 5 of \citeauthor{bed18} would be reduced if the sun's $r$-process fraction were increased. 

Thorium (Th) is rarely measured in the spectra of sun-like stars. \citet{bot19} compared the sun's thorium abundance to those of 59 nearby solar twins; their sample substantially overlaps with that of \citeauthor{bed18} They found that [Th/H] and [Th/Fe] decline with age and that the sun has smaller values of both these quantities than other stars of its age. They also calculated [Th/H]$_{\rm ZAMS}$ and [Th/Fe]$_{\rm ZAMS}$ values, which are the respective abundances at the ZAMS (zero age main sequence), accounting for the decay of thorium since arrival on the main sequence; on this comparison the sun also comes up short, by $0.081 \pm 0.003$ dex. This is close to the sun's [Eu/Fe] deficiency relative to solar twins \citep{bed18}. Thorium is an $r$-process element with a larger value of T$_{\rm c}$ than europium \citep{lod03}; its excess depletion in the sun follows the same pattern as the other heavy neutron capture elements examined by \citeauthor{bed18}.

\begin{figure}
\includegraphics[width=3.2in]{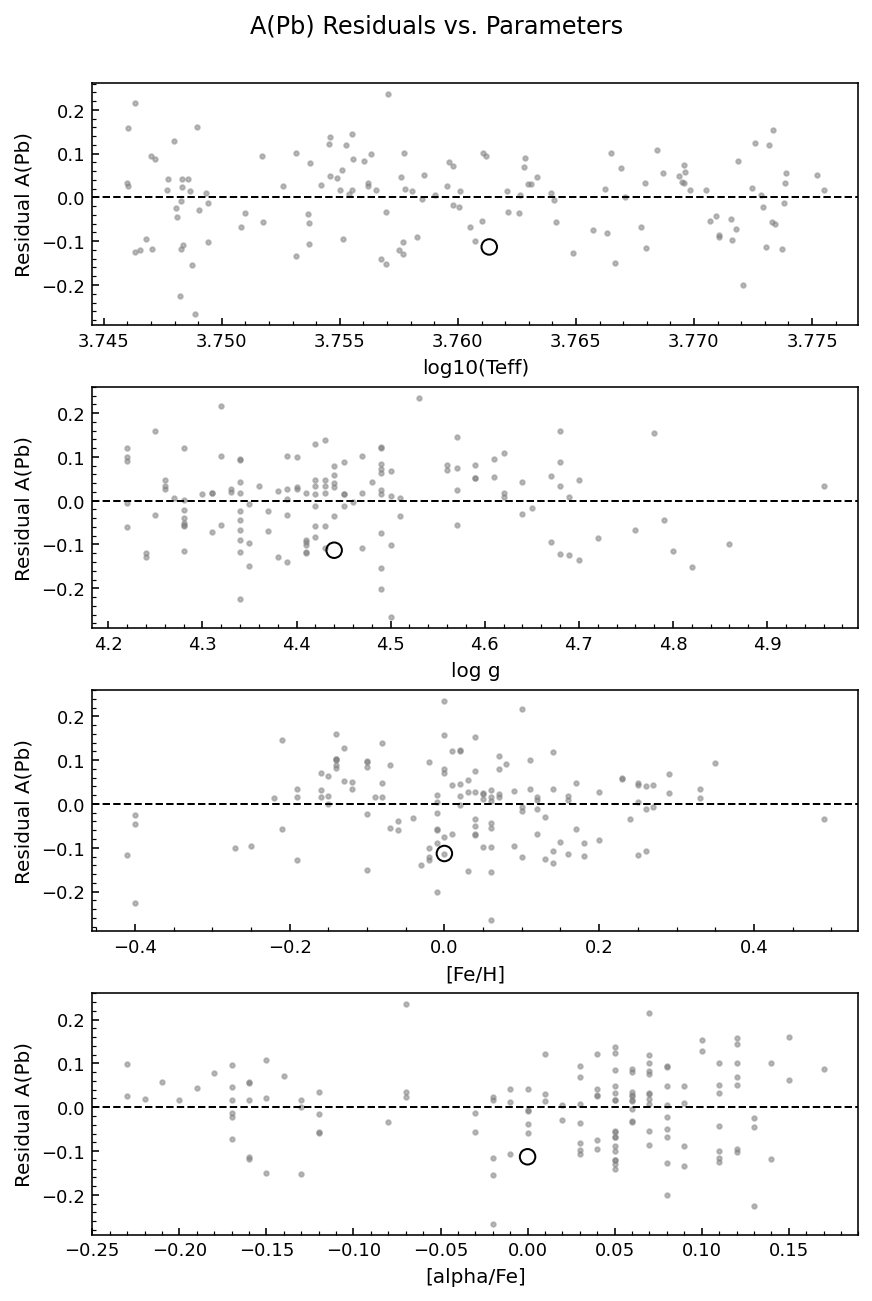}
\caption{Residual A(Pb) values from regression fit using Equation~\eqref{eq:eq_Pb_abund} to the data from \citet{cont24}.}
\label{fig:Pb_abund_fig}
\end{figure}

Another heavy $s$-process element is lead (Pb). \citet{cont24} derived high quality Pb abundances in 653 stars. They quote a mean A(Pb) uncertainty of 0.10 dex. We applied the following cuts to their catalog: $5570 <$ T$_{\rm eff} < 5970$ K, log g $ > 4.2$ dex, and [Fe/H] $> -0.50$. Using this subsample of 141 stars, we fit their LTE Pb abundances to the stellar parameters they reported, obtaining:
\begin{equation}
\begin{split}
\text{A(Pb)} = -(9.114 \pm 3.30) + (2.85 \pm 0.88) \times \log_{10}(\text{T$_{\rm eff}$}) \\
+ (0.07 \pm 0.05) \times [\text{log g}] + (0.70 \pm 0.05) \times [\text{Fe/H}]\\
- (0.12 \pm 0.08) \times [\text{$\alpha$/Fe}] \\
\end{split}
\label{eq:eq_Pb_abund}
\end{equation}
The adjusted R-squared of the fit is 0.63, and the standard deviation of the residuals is 0.088 dex; the residuals are plotted in Fig.~\ref{fig:Pb_abund_fig}. This implies that the scatter in the residuals is fully accounted for by measurement errors. The predicted solar Pb abundance is $1.913 \pm 0.008$(sem) dex. \citeauthor{cont24} give the solar LTE Pb abundance as 1.80 dex, which is over 10-$\sigma$ smaller. Pb is considered a moderately volatile element, with a T$_{\rm c}$ value similar to Zn. In Figure 5 of \citeauthor{bed18}, Zn is slightly enhanced in the sun relative to solar twins. They did not include Pb in their analysis, but if it follows the same trend with T$_{\rm c}$ as the bulk of the elements (including Zn), then it would not be depleted in the sun. Instead, our results reveal that Pb exhibits the same pattern of excess depletion as the other heavy neutron capture elements (Eu, Nd, Pr, Th). We can conclude with a high degree of confidence that the sun has anomalously low heavy neutron capture element abundances.

The abundance of the light $s$-process element Yttrium (Y) has been shown to correlate with age. \citet{she24} derived [Y/Fe], ages, and other basic stellar parameters for 233 solar-type stars. In their equation 6 they give a multivariate linear fit for [Y/Mg] in relation to [Fe/H] and age. From their equation the predicted solar value is $0.00 \pm 0.03$ dex. This is consistent with the finding of \citeauthor{bed18} that the sun's Mg and Y abundances are close to the average for solar twins when corrected for trends with age, despite these elements having significantly different T$_{\rm c}$ values (see their Figure 5).

Phosphorus (P) is another element rarely included in detailed spectroscopic abundance analyses of FGK stars. \citet{maas22} and \citet{sad22} are the highest quality such studies, which include 163 and 45 stars, respectively. Restricting their samples to 5570 $<$ T$_{\rm eff} <$ 5970 K, [Fe/H] $> -0.3$, and log g $> 4.0$, yields 56 stars. From this subsample, we fit the following linear model with ordinary least-squares:
\begin{equation}
\begin{split}
\text{[P/Fe]} = (17.046 \pm 5.75) - (4.37 \pm 1.53) \times \log_{10}(\text{T$_{\rm eff}$})\\
- (0.160 \pm 0.070) \times [\text{Fe/H}] - (0.14 \pm 0.10) \times (\text{log g}) \\
\end{split}
\end{equation}
The adjusted R-squared of the fit is 0.18. The predicted value of [P/Fe] for the solar parameters is $0.00 \pm 0.02$ dex. Although this linear model only explains a small fraction of the variance in [P/Fe], what is important for our application is that the model captures the most significant sources of systematic variation. We assume any variance not captured by the model is random.

\citeauthor{lod03} list the value of T$_{\rm c}$ for P as 1229 K, about halfway between Mn and Cr. From Figure 5 of \citeauthor{bed18} the difference between the solar [X/Fe] and the mean [X/Fe] for solar twins near a T$_{\rm c}$ value of 1229 K is about 0.01 dex. Given the uncertainties, then, the solar [P/Fe] value does not deviate significantly from the T$_{\rm c}$ trendline of the solar twins.

\subsection{Photometric variability, chromospheric activity, and coronal emission}
\label{subsec:phot_var}
Nearly continuous space-based high-precision total solar irradiance (TSI) observations of the sun began in 1978. According to the measurements \citep{will14}, the amplitude of TSI variations over the course of an 11-year activity cycle is about 0.1 percent (one millimagnitude, mmag). TSI variations are 0.25 mmag on annual timescales \citep{rad18}, and on timescales of 6.5 hours they are about 0.015 mmag \citep{gill11}. Comparison of the sun's TSI to stellar observations made in particular bandpasses requires correction factors \citep{rad18}. 

Differential photometric observations of bright stars from the ground have achieved a precision just good enough to detect sun-like variability \citep{lock13}, while the {\it Kepler} primary mission was able to reach 0.02 mmag precision for a 12th magnitude sun-like star over 6.5 hours \citep{gill11}. Ground based observations have the advantage of long duration ($>$4 years), which also permits the study of activity cycles (similar to the sun's 11-year cycle).

The sun's full disk chromospheric Ca II H+K emission has been monitored since the 1970s, while the chromospheric emission of about 100 sun-like stars has been monitored for just over 50 years \citep{rad18}. The chromospheric activity is most often reported as the log of the emission ratio index, R$^{\rm '}_{\rm HK}$. It has traditionally been calibrated in terms of the $B-V$ color index, but recently it has also been calibrated with spectroscopic T$_{\rm eff}$. The sun's mean $\log$ R$^{\rm '}_{\rm HK}(B-V)$ index value has been determined to be $-4.91$ by \citeauthor{rad18}; the T$_{\rm eff}$ version was determined to be $-5.02$ by \citet{lor18}. In contrast with photometric variations, chromospheric Ca II H+K emission variability is relatively easy to measure from the ground and is used to estimate stellar rotation and activity periods \citep{bal95}.

\subsubsection{Nearby field stars}
\label{subsubsec:phot_var_nearby}
The longest running ground-based photometric monitoring program of bright nearby sun-like stars was conducted at the Fairborn Observatory. \citeauthor{rad18} reported the results of Str\"omgren {\it b, y} photometry of 72 stars ranging in spectral type from late-F to mid-K. Transformed to the same bandpass (({\it b} + {\it y})/2) and timescale as their stellar observations, the sun's photometric variability averages near 0.40 mmag. They found that the sun's variability generally follows the trends seen among nearby sun-like stars. \citeauthor{rad18} admit, however, that the relationship between TSI and Str\"omgren variability is uncertain.

\citeauthor{rad18} also concluded that the sun appears to be anomalous relative to the stars in their sample in two respects. First, the sun has a relatively smoother and more regular activity cycle. Second, it displays smaller photometric variability for its chromospheric activity level (see their figures 12a and 13b). 

\citet{wit18} explored possible theoretical explanations for the second observation by testing for the dependence of the ({\it b} + {\it y})/2 variability on a star's metallicity, temperature and inclination of the rotation axis. They found a high sensitivity to stellar parameters from modeling stellar atmospheres with magnetic features and that ``the photometric brightness variations in the Str\"omgren {\it b}, and Str\"omgren ({\it b} + {\it y})/2 filter for the solar case is close to a local minimum for the parameter space of the metallicity and effective temperature.'' In particular, a small increase in metallicity as well as small decreases or increases in T$_{\rm eff}$ result in larger variability with the {\it b} filter. They also explored the effect of changing the inclination angle of the sun, confirming previous studies \citep[e.g., ][]{shap14} that our privileged observing position on Earth results in reduced variability. They found that the effect of inclination on variability is highly metallicity dependent, with a minimum occurring for a metallicity of -0.1 dex; at this metallicity the variability is just under 0.1 mmag. In summary, the sun's particular combination of metallicity and temperature results in a minimum photometric variability in the blue.

In a follow up study, \citet{wit20} confirmed the metallicity dependence of variability with additional modeling of solar twins with the {\it Kepler} bandpass. They confirm that the variability due to rotational modulation in the 4-year {\it Kepler} data is a minimum for metallicity near the solar value.

The sun occupies an unusual position in the plot of activity cycle period versus rotation period. In particular, it is located between two sequences: one extends from short to long activity cycles and occupies short rotation periods, and the other extends over a large range in rotation periods, but occupies short activity cycles \citep[see Figure 5 of][]{jeff23}. They speculate that the sun is in a special transitional dynamo state. However, \citet{boro18} question the presence of the upper (active) sequence. In addition, \citet{bona22} suggest that metallicity explains the differences in the two sequences; their sample consists of 67 stars. Thus, the sun may not be anomalous in this respect, but additional comparison stars are required.

\begin{figure}
\includegraphics[width=3.2in]{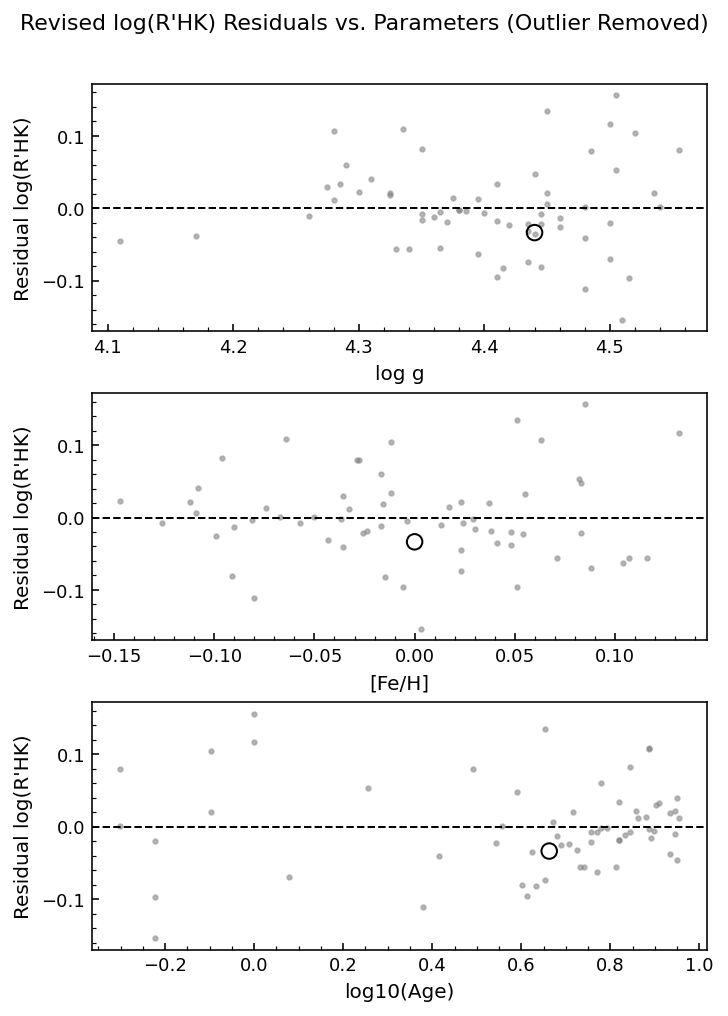}
\caption{Residual log R$^{\rm '}_{\rm HK}$(T$_{\rm eff}$) values from regression fit using Equation~\eqref{eq:eq_R_HK_TEFF} to the data from \citet{cont24}.}
\label{fig:logR_HK_TEFF_fig}
\end{figure}

\citet{lor18} provide log R$^{\rm '}_{\rm HK}$(T$_{\rm eff}$) values along with small-uncertainty stellar parameters for 63 nearby single stars, based on high-resolution and high-cadence time-series HARPS spectra; they also determined the solar value, based on HARPS solar spectra for cycles 23 to 24. The mean uncertainty of their measurements is 0.024 dex. We fit the following equation to their data:
\begin{equation}
\begin{split}
\text{log R$^{\rm '}_{\rm HK}$(T$_{\rm eff}$)} = -(5.441 \pm 0.67) + (0.17 \pm 0.15) \times (\text{log g}) \\
- (0.057 \pm 0.13) \times [\text{Fe/H}] - (0.482 \pm 0.037) \times \log_{10}(\text{Age}) \\
\end{split}
\label{eq:eq_R_HK_TEFF}
\end{equation}
After the first round of fitting, we excluded one outlier (HIP 114615). The adjusted R-squared of the final regression fit is 0.90, and the standard deviation of the residuals is 0.061 dex. We show the residuals in Fig.~\ref{fig:logR_HK_TEFF_fig}. This is larger than the mean measurement error. The predicted solar log R$^{\rm '}_{\rm HK}$(T$_{\rm eff}$) is $-4.986 \pm 0.061$(se) $\pm 0.011$(sem) dex. \citet{lor18} give the solar value as $-5.020 \pm 0.016$ dex, which is about 0.6-$\sigma$ smaller than the prediction. We conclude that the sun is not anomalous in this parameter.

\citet{dos24} measured H$\alpha$ chromospheric fluxes for 511 solar-type stars. From examination of the subsample of their targets with multiple observations (spanning up to three decades), they conclude that stellar activity cycles have very little effect on the H$\alpha$ flux. From their adopted stellar atmospheric parameters (Table 3), derived stellar parameters (Table 4), and total H$\alpha$ flux (F$_{\rm Total}$, Table 5) we prepared a subsample of 23 stars satisfying the following constraints: $5570 <$ T$_{\rm eff} < 5970$ K, log g $ > 4.1$ dex, and age $ > 1.0$ Gyr. Their ages are based on a Bayesian analysis using PARSEC \citep{bress12} stellar isochrones with their stellar parameters. We excluded the youngest stars from the subsample given the very steep dependence of F$_{\rm Total}$ on age (see their Figure 10), which would otherwise make it difficult to describe the age-dependence with a low-order function. 

From examination of their Figure 6, we assume that the dependence of F$_{\rm Total}$ on T$_{\rm eff}$ can be captured adequately with the log of T$_{\rm eff}$ over the temperatures covered by our subsample. Given these considerations, we fit the following equation to the subsample (in units of 10$^6$ erg cm$^{-2}$ s$^{-1}$):
\begin{equation}
\begin{split}
\text{F$_{\rm Total}$} = -(95.57 \pm 21.58) + (27.14 \pm 5.58) \times \log_{10}(\text{T$_{\rm eff}$})\\
- (0.21 \pm 0.31) \times (\text{log g}) + (0.34 \pm 0.23) \times [\text{Fe/H}]\\
- (0.136 \pm 0.067) \times \text{Age} + (0.0077 \pm 0.0050) \times \text{Age}^2 \\
\end{split}
\end{equation}
The adjusted R-squared of the fit is 0.66, and the standard deviation of the residuals is 0.087. The predicted solar F$_{\rm Total}$ is $5.121 \pm 0.072$(sem) $\times$ 10$^6$ erg cm$^{-2}$ s$^{-1}$. \citeauthor{dos24} give the solar value as $5.120 \pm 0.082$\footnote{Note, there is a typo in the uncertainty for the solar value of F$_{\rm Total}$ listed in Table 6 of \citeauthor{dos24} It should be decreased by a factor of 10.} $\times$ 10$^6$ erg cm$^{-2}$ s$^{-1}$ (their Table 6), which is nearly identical with the prediction and well within the 1-$\sigma$ error.

Space X-ray telescope observations of nearby stars allows us to study their coronal emission. \citet{zhu25} present X-ray luminosities for 107 G dwarfs within 20 pc of the Sun compiled from archival observations from {\it Chandra}, {\it XMM-Newton}, eROSITA, and ROSAT. Cross-referencing their table with stellar parameters from the {\it Gaia} FGK benchmark stars \citep{soub24} and the {\it Gaia} DR3 astrophysical parameters supplement catalog,\footnote{\url{https://gaia.aip.de/metadata/gaiadr3/astrophysical_parameters/}} yields 29 stars in common; we estimated the ages of the FGK benchmark stars from the isoclassify code\footnote{\url{https://github.com/danxhuber/isoclassify}}, averaging results from the MESA and PARSEC isochrones. The mean uncertainty in $\log_{10}$ (L$_{X}$/L$_{bol}$) is 0.08 dex. 

One of the stars in this subsample, HIP33277, has an upper limit on L$_{X}$. To properly incorporate this star in the analysis, we performed our regression using an Accelerated Failure Time (AFT) model. Unlike the Ordinary Least Squares (OLS) method, which minimizes the sum of squared residuals and would treat the upper limit as a biased detection, the AFT model is a survival analysis technique that uses maximum likelihood estimation. This statistical approach correctly uses the information from both the detected sources and the censored data point to derive an unbiased set of model parameters. We implemented this AFT model using the \texttt{WeibullAFTFitter} class from the Python package \texttt{lifelines}.\footnote{\url{https://doi.org/10.5281/zenodo.10281631}} We fit the following equation to the ratio of X-ray to bolometric luminosities:
\begin{equation}
\begin{split}
\text{$\log_{10}$ (L$_{X}$/L$_{bol}$)} = (45.12 \pm 40.70) \\
- (15.4 \pm 10.3) \times \log_{10}(\text{T$_{\rm eff}$}) + (2.11 \pm 0.74) \times (\text{log g}) \\
- (1.54 \pm 0.54) \times [\text{Fe/H}] - (0.31 \pm 0.06) \times \text{Age} \\
\end{split}
\label{eq:eq_X-ray}
\end{equation}

The goodness-of-fit for the AFT model was assessed using the concordance index, which was 0.78, indicating good predictive discrimination. The standard deviation of the residuals is 0.57 dex among the stars with detections. Cosmic variance completely dominates measurement error. Although the ratio does display a high sensitivity to age, the addition of a quadratic term was not justified. The predicted solar $\log_{10}$ (L$_{X}$/L$_{bol}$) from this fit is $-5.00 \pm 0.49$(se) $\pm 0.09$(sem). The log of the observed average ratio for the sun is $-5.95$ \citep{judge03}, which places it almost 2-$\sigma$ below the predicted value. However, given the small sample size, the large intrinsic variability and the strong dependence on age, it would be helpful to study a larger sample before making definitive conclusions on the sun's L$_{X}$/L$_{bol}$ ratio relative to nearby sun-like stars. For now, we can tentatively conclude that the sun has a moderately lower X-ray to bolometric luminosity ratio relative to other nearby stars like it.

\begin{figure}
\includegraphics[width=3.2in]{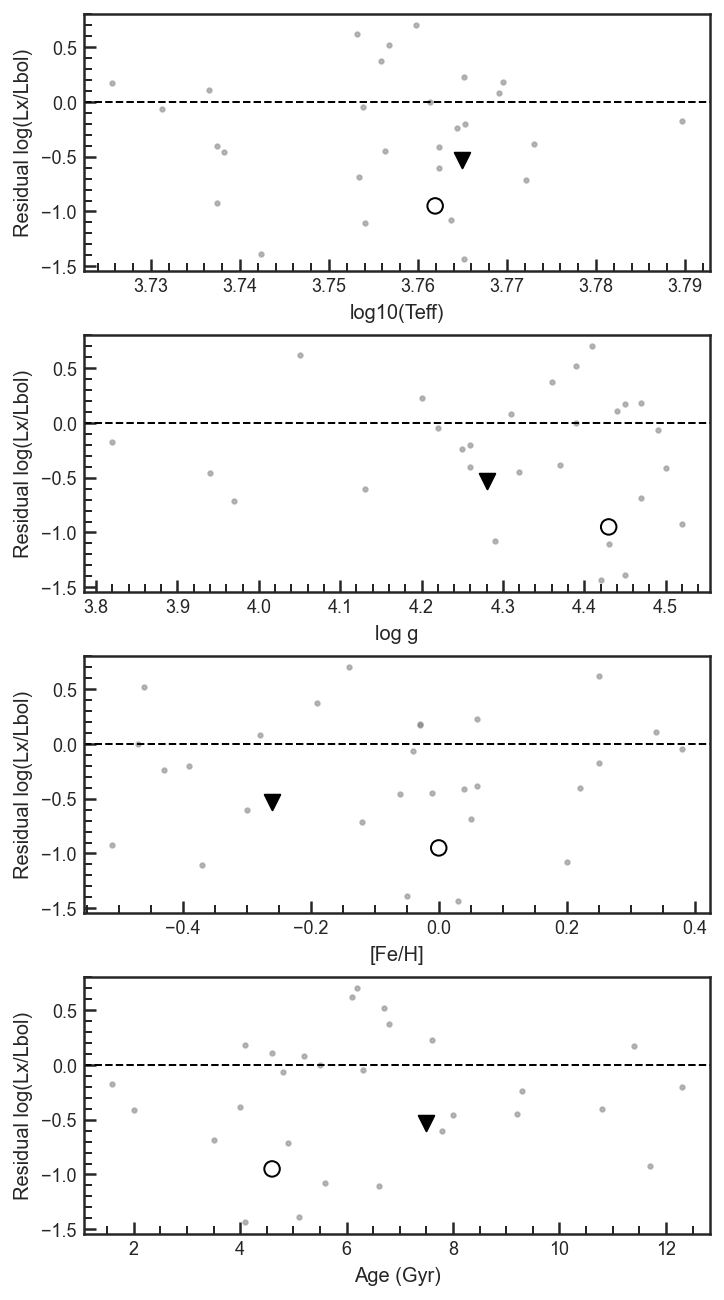}
\caption{Residual L$_{X}$/L$_{bol}$ values from regression fit using Equation~\eqref{eq:eq_X-ray} to the data from \citet{zhu25}. The triangle is an upper limit.}
\label{fig:X-ray_fig}
\end{figure}

\subsubsection{The {\it Kepler} field and {\it TESS}}
\label{subsubsec:phot_var_Kepler}
The original {\it Kepler} mission produced four years of nearly continuous photometric observations of over $10^5$ stars in a region of the constellation Cygnus. From these observations photometric variability has been determined for all the target stars. A number of indices have been defined to quantify variability on various timescales. These include the combined differential photometric precision (CDPP) noise metric at a timescale of 6.5 hours (half the time for an Earth analogue to transit its host star) and the astrophysical variability amplitude (S$_{\rm ph}$) over the timescale of the dominant periodic variability (usually at least a few days from rotational modulation).

Throughout this section we make extensive use of the {\it LAMOST} DR7 v2.0 LRS Stellar Parameter Catalog of A, F, G, and K stars.\footnote{\url{https://dr7.lamost.org/doc/lr-data-production-description\#S3.2}} We cross referenced this catalog with the high quality spectroscopic survey of {\it Kepler} field stars from \citet{brew18}. We restricted our subsample of their catalog to stars with $5500 <$ T$_{\rm eff} < 6000$ K and log g $ > 4.0$ dex; we also removed stars with signal-to-noise ratios below 30 and line broadening above 10.0 km/s. This resulted in a sample of 282 stars in common between the two catalogs. The mean differences ({\it LAMOST} - Brewer) and standard deviations are: $+8 \pm 79$ K, $-0.05 \pm 0.12$ dex, and $0.00 \pm 0.07$ dex, for T$_{\rm eff}$, log g, and [Fe/H], respectively. These differences are sufficiently small to be considered negligible with respect to the following analyses.

\citet{gill11} determined the sun's mean intrinsic CDPP in the {\it Kepler}-equivalent bandpass to be $11.0 \pm 1.5$ ppm (0.011 mmag) from solar {\it VIRGO} observations spanning a full solar cycle from 1996 to 2008. 

Following the procedure described in \citet{zhang24} we took the CDPP noise values published in the {\it Kepler} Stellar Properties Catalog for Q1–Q17 DR25 and corrected them for instrumental noise using their noise floor model (their equation 1 and table 1) with their equation 2; this quantity is called 'rrmscdpp$_{\rm intrinsic}$'. They estimate a typical measurement error of 12 ppm. We then cross-referenced this catalog with the {\rm LAMOST} DR7 stellar parameters catalog, the {\it Gaia} DR3 astrophysical parameters supplement catalog, along with the Kepler field {\it Gaia} DR3 cross-matched database.\footnote{\url{https://gaia-kepler.fun/}} The stellar parameters are from {\rm LAMOST}, and the ages are from the \texttt{Age-Flame} parameter in the {\it Gaia} supplement. We applied the following cuts to the resulting catalog: $5600 <$ T$_{\rm eff} < 5940$ K and log g $ > 4.2$ dex. We also required that the {\it Gaia} parameters \texttt{non\_single\_star} and the renormalized unit weight error (\texttt{ruwe}) be '0' and $<1.25$, respectively; these eliminate likely binaries.\footnote{Our adopted cut for the \texttt{ruwe} parameter is more restrictive than the value adopted by \citet{bud21}, 1.4, and slightly less restrictive than the value adopted by \citet{sant23}, 1.2.} These cuts resulted in a subsample of 2987 stars.

We fit the following equation to our subsample:
\begin{equation}
\begin{split}
\text{$\log_{10}$ rrmscdpp$_{\rm intrinsic}$} = (22.43 \pm 8.45) \\
+ (1.59 \pm 0.81) \times \log_{10}(\text{T$_{\rm eff}$}) - (13.16 \pm 3.75) \times \text{log g} \\
+ (1.60 \pm 0.43) \times (\text{log g})^{2} 
+ (0.18 \pm 0.03) \times [\text{Fe/H}] \\
+ (0.033 \pm 0.020) \times \log_{10}(\text{Age}) \\
\end{split}
\label{eq:eq_rrmscdpp}
\end{equation}
The adjusted R-squared of the fit is 0.12, and the standard deviation of the residuals is 0.31 dex. This is larger than the estimated error, implying a large contribution from cosmic variance. The strong dependence on log g confirms the well-known 'flicker' effect \citep{bast13,bast16,vankoot21} and the related FliPer metric \citep{bug18}. There is also a strong dependence on [Fe/H]. The predicted solar rrmscdpp$_{\rm intrinsic}$ from this model is $39.9 ^{+42}_{-20}$(se) $\pm~0.7$(sem) ppm. Only 175 stars out of the 2987 (6 percent) have smaller rrmscdpp$_{\rm intrinsic}$ values than the sun's 11.0 ppm; of these, 111 are more metal-poor, 161 have lower surface gravity than the sun, and 168 are either more metal-poor or lower gravity. This puts the sun at the lowest 0.2 percent for this metric when corrected for differences in these two parameters.

\begin{figure}
\includegraphics[width=3.2in]{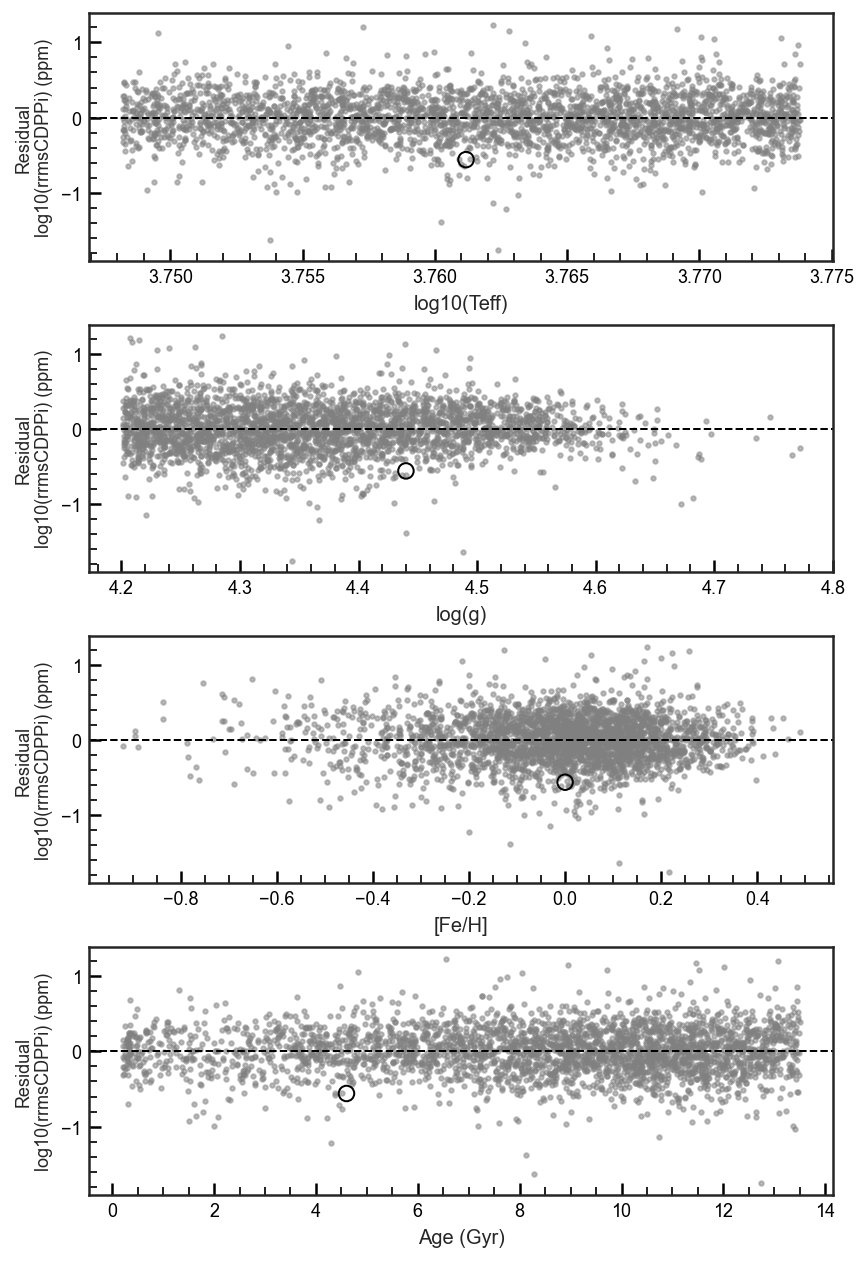}
\caption{Residual $\log_{10}$ rrmscdpp$_{\rm intrinsic}$ values from regression fit using Equation~\eqref{eq:eq_rrmscdpp} to the data from \citet{zhang24}.}
\label{fig:rrms_fig}
\end{figure}

\begin{figure}
\includegraphics[width=3.2in]{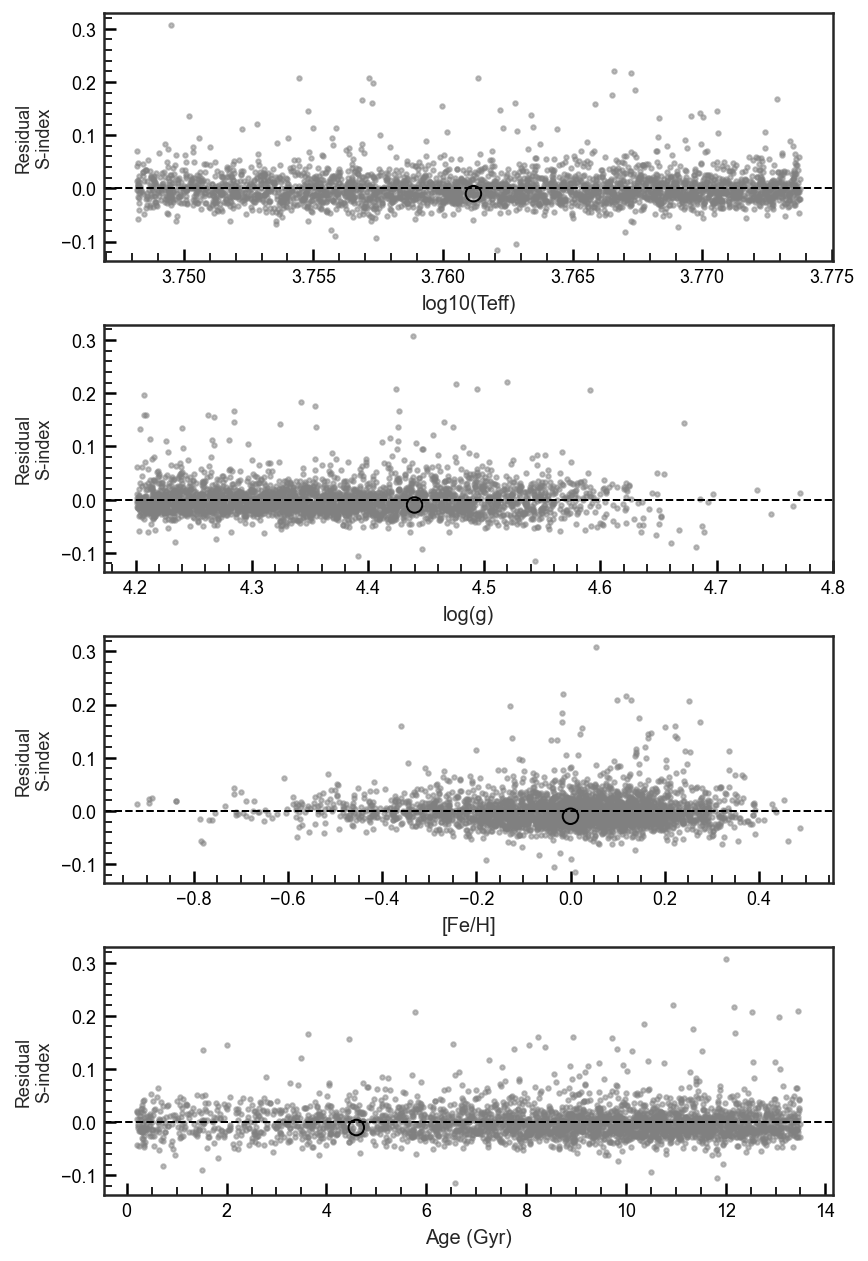}
\caption{Residual S$_{\rm L}$ values from the regression fit using Equation~\eqref{eq:eq_SL} to the data from \citet{zhang22}.}
\label{fig:SL_fig}
\end{figure}

In addition, chromospheric activity indices have been measured for many stars in the {\it Kepler} field. \citet{zhang22} calculated the chromospheric Ca II H and K lines S-index values for sun-like stars measured with the {\it LAMOST} low resolution spectra, denoted as S$_{\rm L}$. It is important to assemble a large sample in order to average out the intrinsic astrophysical S-index variations from individual stars; the authors estimate that the uncertainty in S$_{\rm L}$ resulting from activity cycle variability is about 0.047. We cross-referenced this catalog with the sample we prepared for the rrmscdpp$_{\rm intrinsic}$ analysis above and fit to the same stellar parameters. The results are as follows:
\begin{equation}
\begin{split}
\text{S$_{\rm L}$} = (4.150 \pm 0.823) \\
+ (0.237 \pm 0.078) \times \log_{10}(\text{T$_{\rm eff}$}) 
- (2.31 \pm 0.37) \times \text{log g} \\
+ (0.275 \pm 0.042) \times (\text{log g})^{2} 
+ (0.010 \pm 0.003) \times [\text{Fe/H}] \\
- (0.006 \pm 0.002) \times \log_{10}(\text{Age}) \\
\end{split}
\label{eq:eq_SL}
\end{equation}
The adjusted R-squared of the fit is 0.15, and the standard deviation of the residuals is 0.03. We show the residuals in Fig.~\ref{fig:SL_fig}. While this is smaller than the estimated uncertainty in S$_{\rm L}$, we will assume that the entirety of the residual variance is due to activity cycle variability. Again, the sensitivity to surface gravity is detected with high confidence, with some sensitivities to [Fe/H] and age as well. The predicted solar S$_{\rm L}$ from this model is $0.1961 \pm 0.03$(se) $\pm 0.0008$(sem). \citet{zhang20} calculated the value of S$_{\rm L}$ for the sun during cycles 15-24 to be 0.179 to 0.194 (mean, 0.1865). Thus, the observed mean solar S$_{\rm L}$ value is less than 1-$\sigma$ from the predicted value. Also, about half the stars in the sample have smaller S$_{\rm L}$ than the sun.

The chromospheric activity index R$^{\rm +}_{\rm HK}$ is based on the S-index. \citet{ye24} compiled a catalog of R$^{\rm +}_{\rm HK}$, mass, age, rotation period, and photometric activity index values from LAMOST DR7, {\it Kepler}, and {\it Gaia} DR3 (their Table 3). They only calculated R$^{\rm +}_{\rm HK}$ for stars in their catalog with at least two observations. Restricting their sample to $5600 <$ T$_{\rm eff} < 5900$ K, log g $> 4.2$, and $0.90 <$ Mass $< 1.1$ M$_{\odot}$, we obtain a subsample of 212 stars. The mean uncertainty in log R$^{\rm +}_{\rm HK}$ is 0.011 dex. We fit the following equation to log R$^{\rm +}_{\rm HK}$:
\begin{equation}
\begin{split}
\text{log R$^{\rm +}_{\rm HK}$ (dex)} = -(7.642 \pm 0.603) + (0.66 \pm 0.13) \times (\text{log g}) \\
- (0.275 \pm 0.072) \times [\text{Fe/H}] - (0.095 \pm 0.055) \times \log_{10}(\text{Age}) \\
\end{split}
\label{eq:eq_R_HK+}
\end{equation}
We removed one outlier following the first round (KIC 9154550). The adjusted R-squared of the final fit is 0.27, and the standard deviation of the residuals is 0.125 dex. We found R$^{\rm +}_{\rm HK}$ to be only weakly dependent on temperature in the first round of fits; we removed it from the final model fit, shown above. The predicted solar log R$^{\rm +}_{\rm HK}$ from this model is $-4.794 \pm 0.127$(se) $\pm 0.010$(sem) dex. Cosmic variance dominates the residuals. The range for the sun was -4.918 to -4.796 dex in cycles 15-24 (mean of -4.852 dex).

\begin{figure}
\includegraphics[width=3.2in]{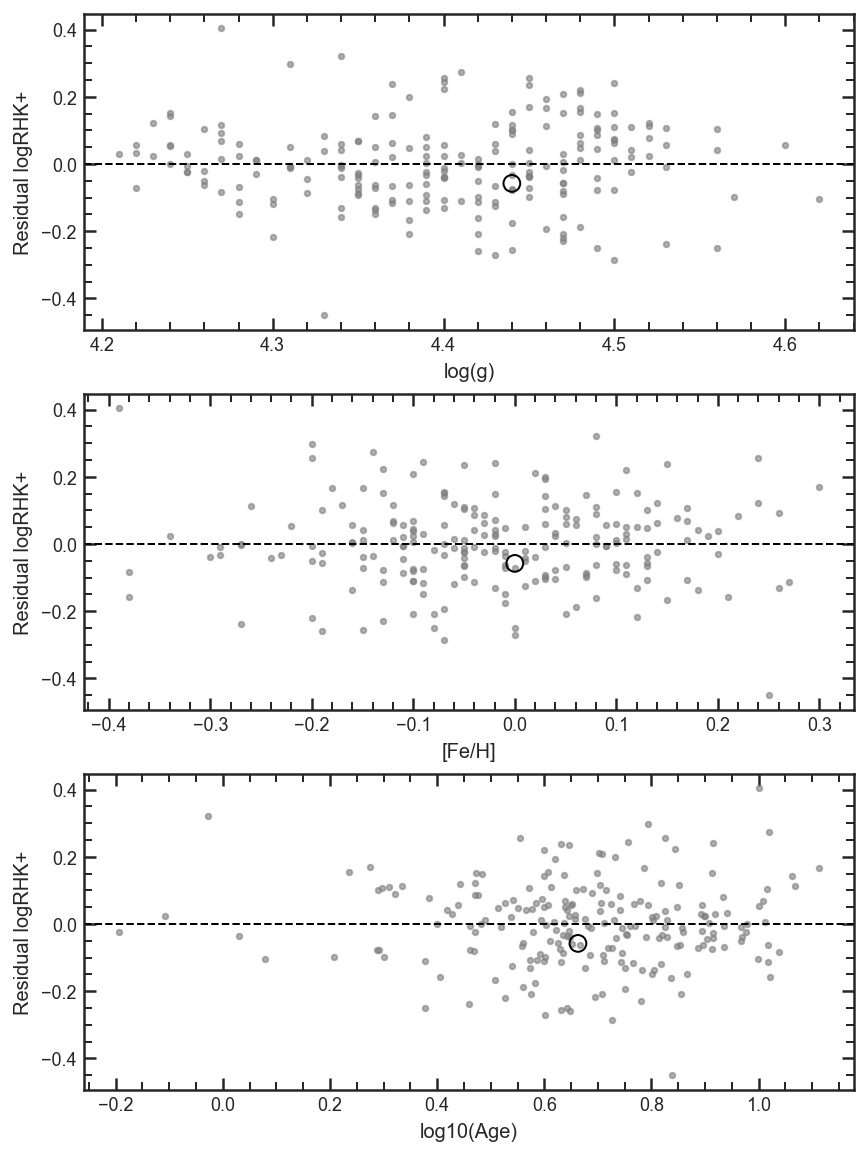}
\caption{Residual log R$^{\rm +}_{\rm HK}$ values from the regression fit using Equation~\eqref{eq:eq_R_HK+} to the data from \citet{ye24}.}
\label{fig:R_HK+_fig}
\end{figure}

Next, we analyze the photometric variability amplitude, S$_{\rm ph}$, from \citeauthor{ye24} (their Sample I); their S$_{\rm ph}$ and P$_{\rm rot}$ values are from \citet{sant19,sant21}. S$_{\rm ph}$ is considered a proxy for chromospheric/magnetic activity. It is important to note that S$_{\rm ph}$ is only measured for stars with detected rotational modulation. For this reason, it is biased in favor of stars with relatively large rotational modulation amplitude \citep{masuda22}. The average error in S$_{\rm ph}$ is 40 ppm. We obtain the following fit for our subsample of 214 stars:
\begin{equation}
\begin{split}
\text{S$_{\rm ph}$ (ppm)} = (8.693\pm2.17)\times10^{5}) \\
- (2.681\pm1.54)\times10^{4} \times \log_{10}(\text{T$_{\rm eff}$}) \\
- (3.605\pm0.978)\times10^{5} \times (\text{log g}) \\
+ (4.23\pm1.11)\times10^{4} \times (\text{log g})^{2}
+ (462 \pm 732) \times [\text{Fe/H}] \\
\end{split}
\label{eq:eq_S_ph}
\end{equation}
We did not include age in the fit, as S$_{\rm ph}$ was found to be very weakly dependent on it in the first round of fits. The adjusted R-squared of the fit is 0.35, and the standard deviation of the residuals is 1395 ppm. The predicted solar S$_{\rm ph}$ from this model is $1500 \pm 1417$(se) $\pm 123$(sem) ppm. \citet{sant23} calculated the mean solar value of S$_{\rm ph}$ to be about 300 ppm (see their Figure A.5). This confirms what had already been known, that compared to {\it Kepler} stars with detectable rotational modulation, the sun is less variable. But, in this dataset it is within 1-$\sigma$ of the predicted value.

\begin{figure}
\includegraphics[width=3.2in]{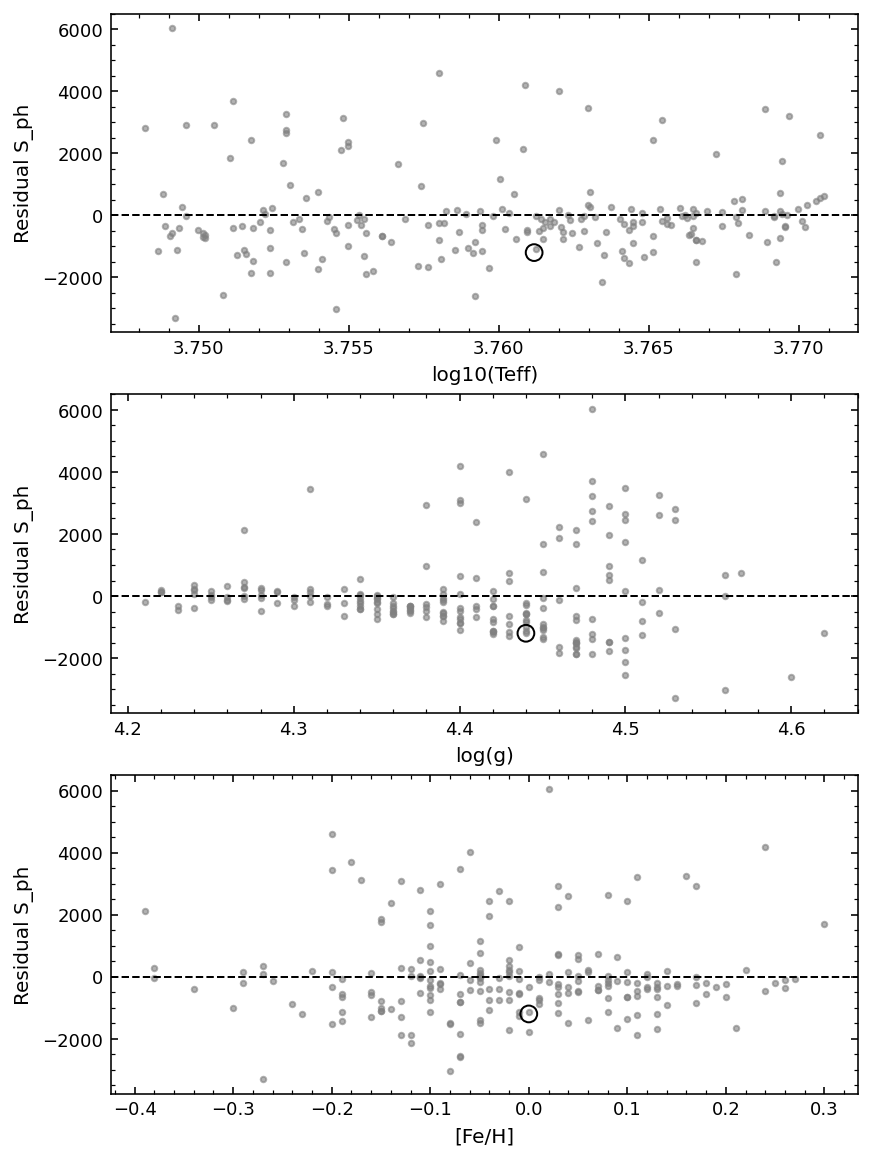}
\caption{Residual S$_{\rm ph}$ values from the regression fit using Equation~\eqref{eq:eq_S_ph} to the data from \citet{ye24}.}
\label{fig:S_ph_fig}
\end{figure}

\citet{ponte23} derived variability amplitude indices, {\it A}$_{\it TESS}$, from {\it TESS} photometry for 30 single solar twins from the first year of observations, all of which have small-uncertainty stellar parameters available from prior studies published by their group. The typical quoted error in log{\it A}$_{\it TESS}$ is $\pm 0.2$ dex. We obtain the following fit to the 20 stars in their sample with detected variability and the 10 stars with only upper limits for {\it A}$_{\it TESS}$\footnote{The upper limits were included in the analysis in the same way ans was done for the X-ray analysis above.}:
\begin{equation}
\begin{split}
\text{log{\it A}$_{\it TESS}$} = (5.03 \pm 3.86) - (0.31 \pm 0.87) \times (\text{log g}) \\
- (0.69 \pm 1.48) \times [\text{Fe/H}] - (0.90 \pm 0.12) \times \log_{10}(\text{Age}) \\
\end{split}
\end{equation}
The concordance parameter is 0.79. The standard deviation of the residuals is 0.3 dex; since this is larger than the measurement error, there is still some unmodelled error. The predicted solar log{\it A}$_{\it TESS}$ from this model is $3.05 \pm 0.08$(sem) dex. \citeauthor{ponte23} calculated the median solar{\it TESS} equivalent value to be $2.87^{+0.27}_{-0.40}$. Thus, these values are entirely consistent with being equivalent within the statistical errors.

\citet{rein20} calculated from {\it Kepler} data the photometric variability index, R$_{\rm var}$, which they define as ``...the difference between the 95th and 5th percentile of the sorted flux values (normalized by its median) in a light curve.'' This index has the advantage over the S$_{\rm ph}$ index in that it can be applied to any star, not just to a periodic variable. They prepared two subsamples of solar type stars, periodic and nonperiodic; the periodic subsample was restricted to periods between 20 and 30 days. It is worthwhile to revise their dataset and perform a new analysis with improved stellar parameters. Note that they also calculated a 'corrected' R$_{\rm var}$ index for the periodic subsample that removes the temperature, rotation period, and metallicity dependencies (based on their multiple regression fits); we will be using only the raw R$_{\rm var}$ index values in the following analysis.

We cross-matched the two \citeauthor{rein20} subsamples with the {\it LAMOST} DR7 and {\it Gaia} DR3 astrophysical parameters supplement catalogs. We adopted the {\it LAMOST} stellar parameters, (T$_{\rm eff}$, log g, [Fe/H]) and the {\it Gaia} \texttt{Age-Flame} age. We then applied the following cuts: $5600 <$ T$_{\rm eff} < 5940$ K, and log g $> 4.2$.  We also required that the {\it Gaia} parameters \texttt{non\_single\_star} and \texttt{ruwe} be '0' and $<1.25$, respectively. The resulting periodic subsample has 39 stars, and the nonperiodic one has 283 stars. We obtained the following fit for the {\it periodic} subsample: 
\begin{equation}
\begin{split}
\text{R$_{\rm var}$} = (54.26 \pm 23.07) - (13.56 \pm 5.90) \times \log_{10}(\text{T$_{\rm eff}$}) \\
- (0.69 \pm 0.37) \times (\text{log g}) + (0.08 \pm 0.21) \times [\text{Fe/H}] \\
+ (0.003 \pm 0.007) \times \text{Age} \\
\end{split}
\label{eq:eq_Rvar_per}
\end{equation}
The adjusted R-squared of the fit is 0.11, and the standard deviation of the residuals is 0.13 percent. We included the age parameter in a first round of fitting, but it was not significant. The predicted solar value of R$_{\rm var}$ from this model is $0.176 \pm 0.15$(se) $\pm 0.05$(sem) percent. \citeauthor{rein20} calculated the solar equivalent value of R$_{\rm var}$ to be $0.07$ percent from the VIRGO TSI measurements (1996--2018) and the reconstructed TSI (see the Supplementary Materials to their paper for details). Only 8 of the 39 stars in this subsample have more negative residuals than the sun.

\begin{figure}
\includegraphics[width=3.2in]{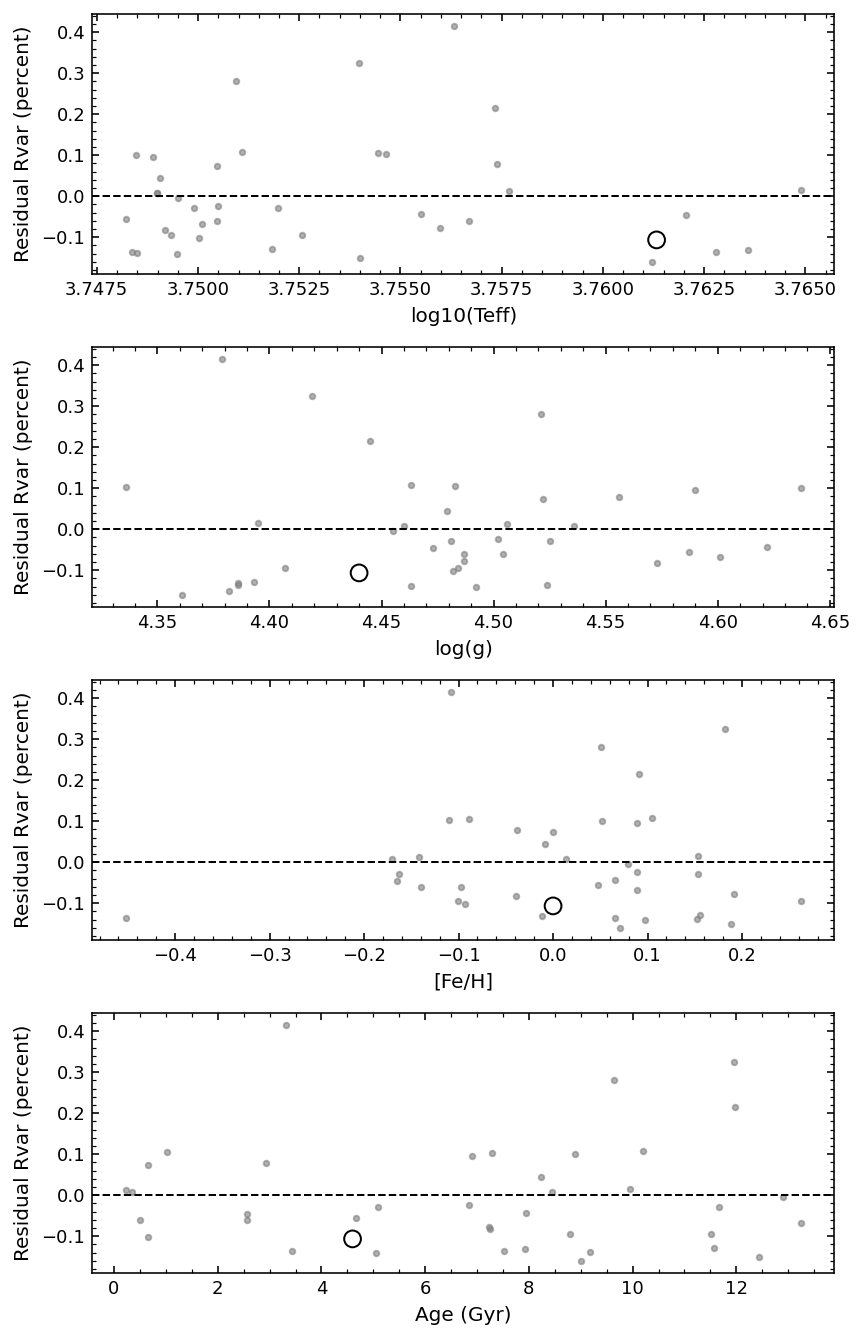}
\caption{Residual R$_{\rm var}$ values for the periodic stars from the regression fit using Equation~\eqref{eq:eq_Rvar_per} to the data from \citet{rein20}.}
\label{fig:Rvar_per_fig}
\end{figure}

Keeping the same regression parameters for consistency, the fit for the {\it nonperiodic} subsample is:
\begin{equation}
\begin{split}
\text{R$_{\rm var}$} = (1.589 \pm 1.750) - (0.50 \pm 0.46) \times \log_{10}(\text{T$_{\rm eff}$}) \\
+ (0.087 \pm 0.032) \times (\text{log g}) + (0.041 \pm 0.016) \times [\text{Fe/H}] \\
- (0.002 \pm 0.001) \times \text{Age} \\
\end{split}
\label{eq:eq_Rvar_nonper}
\end{equation}
The adjusted R-squared of the fit is 0.06, and the standard deviation of the residuals is 0.05 percent. The predicted solar R$_{\rm var}$ from this model is $0.096 \pm 0.05$(se) $\pm 0.005$(sem) percent. The sun's R$_{\rm var}$ value is 5-$\sigma$ smaller. Thirty-three percent of the stars in this subsample have more negative residuals (observed - fit) than the sun's.

\begin{figure}
\includegraphics[width=3.2in]{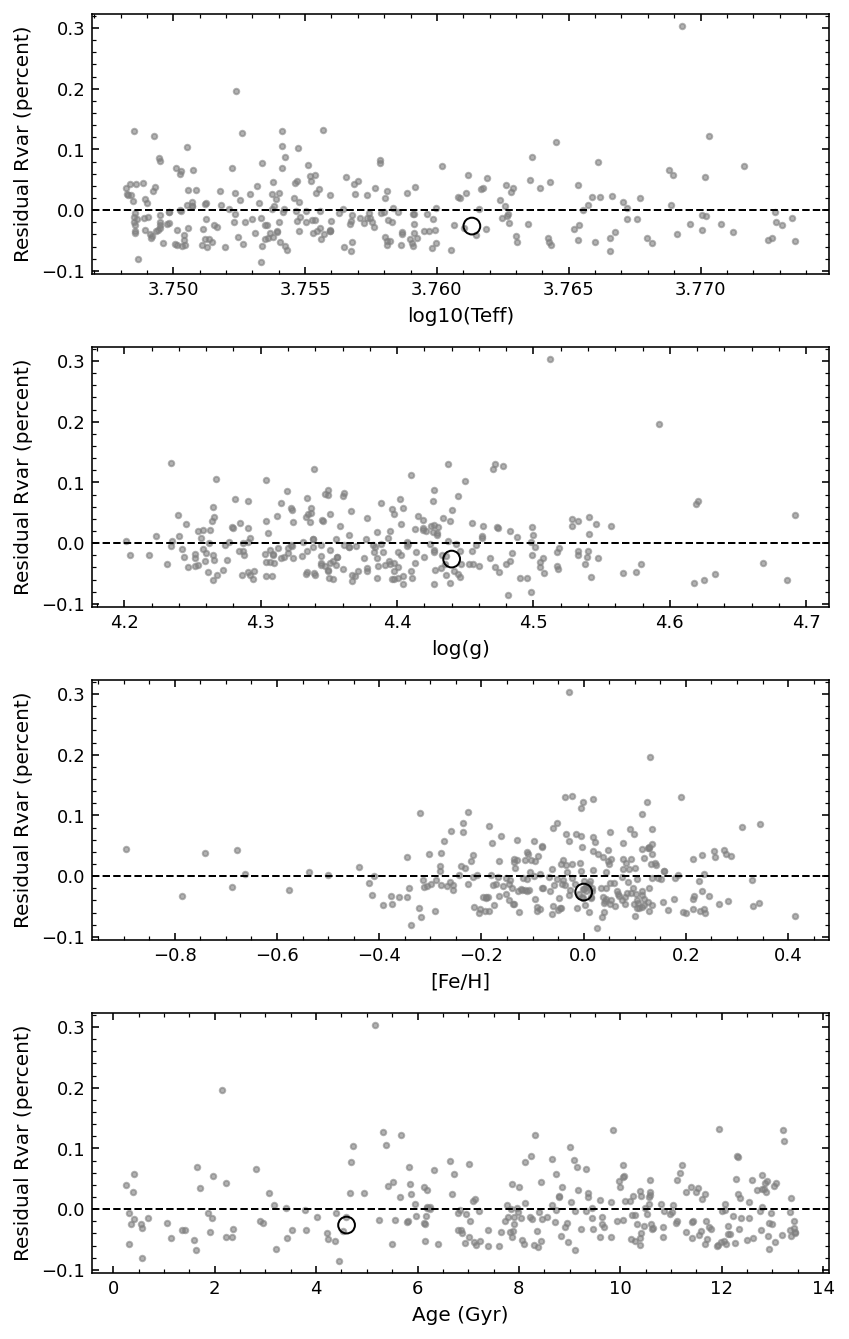}
\caption{Residual R$_{\rm var}$ values for the nonperiodic stars from the regression fit using Equation~\eqref{eq:eq_Rvar_nonper} to the data from \citet{rein20}.}
\label{fig:Rvar_nonper_fig}
\end{figure}

Our analysis of the {\it Kepler} R$_{\rm var}$ data generally confirms the conclusion of \citet{rein20} that the sun exhibits smaller photometric variations than the periodic subsample; however, we found that the value of R$_{\rm var}$ for a periodic solar twin to be half that quoted by them. This is also consistent with our findings for the S$_{\rm ph}$ index. We find, for the first time, tentative evidence that the sun is also less variable when compared to nonperiodic stars. Comparing the fit for the nonperiodic subsample to that for S$_{\rm ph}$ with the data from \citet{ye24} discussed earlier, we find that the dependencies on the stellar parameters are similar. In particular, the strong dependence on surface gravity has been a common theme in this section. Surface gravity dependence was ignored by \citeauthor{rein20}. On the other hand, the periodic stars display a different dependence on surface gravity. This might arise from the variability being related to different phenomena. In the periodic sample, the variability due to rotation dominates, while for the non-periodic sample other phenomena might dominate (e.g., perhaps granulation which is sensitive to log g).

The most energetic solar flare observed in the telescopic era was the Carrington flare of 1859, which is estimated to have had an energy of about $5\times$10$^{32}$ erg \citep{cliver22}. Flares have been observed in solar type stars exceeding the Carrington flare energy by several orders of magnitude. Since such strong flares are rare, it is necessary to monitor a large number of stars to detect a statistically significant number of flares. \citet{vas24} examined the statistics of superflares (energy $>$ $10^{34}$ erg) on sun-like {\it Kepler} stars (weaker flares suffer from incompleteness uncertainties in their sample). Within the four-year {\it Kepler} sample they identified 2889 flares on 2527 stars out of 56,450 stars selected to have $5000 <$ T$_{\rm eff} < 6500$ K and $4 <$ M$_{\rm G} < 6$ magnitudes (they did not select according to surface gravity or metallicity); they did not give age estimates. From this they estimate that superflares occur on sun-like stars at a rate of about one per century. We can improve on their analysis in two ways.

First, we cross-referenced their catalog with the {\it Gaia} DR3 astrophysical parameters supplement catalog using the {\it Kepler} field-{\it Gaia} DR3 cross-matched database. We adopted the stellar parameters from the {\it Gaia} \texttt{\_gspphot} (T$_{\rm eff}$, log g, [Fe/H]), \texttt{-Flame} age, and \texttt{ruwe} fields; we excluded stars with \texttt{ruwe} values $> 1.25$. The median temperature difference ({\it Gaia} - Vasilyev) is $-189$ K with a standard deviation of 184 K. Given this large difference, we repeat our analysis of the flare statistics with both temperature scales. The median age is high, at 9.2 Gyr among solar analogues (using T$_{\rm eff}$ from {\it Gaia}); this age bias results, in part, from \citeauthor{vas24} removing stars with known rotation periods less than 20 days.

Second, as we have done with other analyses above, we can correct the observed flare energy values for differences in stellar parameters relative to solar. We obtain the following fit for the stars with flares in \citeauthor{vas24} using the {\it Gaia} temperature scale:
\begin{equation}
\begin{split}
\text{E$_{\rm flare}$ (erg)} = (33.5 \pm 1.4) \\
+ (1.9 \pm 0.3) \times (\text{log T$_{\rm eff}$}) - (1.47 \pm 0.06) \times (\text{log g}) \\
- (0.08 \pm 0.03) \times [\text{Fe/H}] + (0.06 \pm 0.02) \times \text{Age} \\
\end{split}
\label{eq:eq_flares}
\end{equation}
The adjusted R-squared of the fit is 0.39. The dependency of flare energy on temperature and surface gravity are especially pronounced, with warmer lower gravity stars having more energetic flares.

\begin{figure}
\includegraphics[width=3.2in]{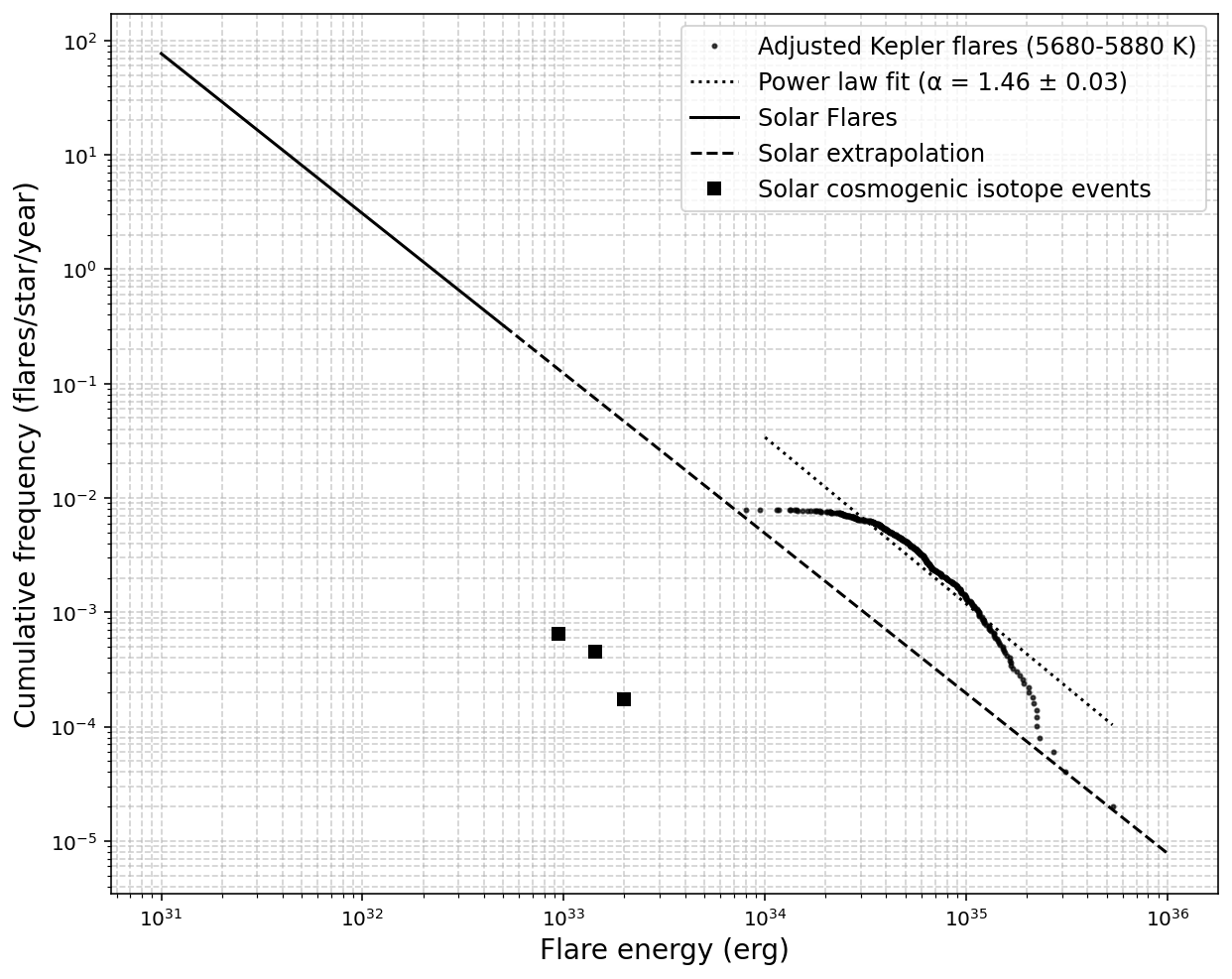}
\caption{Cumulative frequency of flares in the Sun and among the {\it Kepler} stars in the sample of \citet{vas24} within 100 K of the solar temperature (using the {\it Gaia} temperature scale). The solid and dashed lines are the observed and extrapolated solar flares, respectively (adapted from \citeauthor{vas24}). The dots are the adjusted frequencies for stars near the solar temperature adjusted for differences in stellar parameters from the solar values (see text for details). The power law fit is for flare energies greater than 10$^{34}$ erg. The squares are historical energetic solar flares inferred from the cosmogenic isotopes $^{14}$C and $^{10}$Be \citep{cliver22}. }
\label{fig:flares_fig}
\end{figure}

We applied this equation to the flares identified by \citeauthor{vas24} in order to to effectively adjust their energies to the solar parameters (i.e., as if all the stars were identical to the sun). The resulting flare frequency curve among the stars closest to the solar temperature in \citeauthor{vas24} and adjusted according to Equation~\eqref{eq:eq_flares} is shown in Fig.~\ref{fig:flares_fig}; the curve is based on 444 flaring stars out of 14,562 stars within 100 K of the solar temperature. The downturn in flare frequency for energy values below 10$^{34}$ erg is most likely due to inaccuracy in the correction for incompleteness. These results imply a rate of superflares among solar twins near one every 30 years; this is about 3 times more frequent than the estimate quoted by \citeauthor{vas24}. If we repeat this analysis with the \citeauthor{vas24} temperature scale (but all other stellar parameters from {\it Gaia}), we obtain a rate half as frequent but still 1.5 times their estimate based on a much broader temperature range. The power law fit constant is similar with each temperature scale, near 1.5; this is near the value for the sun at lower energies.

Combined with a lack of superflares seen in the historical cosmogenic isotope record, this reanalysis of \citeauthor{vas24} suggests that the superflare frequency in the sun is much smaller than it is among the {\it Kepler} stars.

\citet{herbst25} estimated the relative contributions of spot versus faculae to brightness variations among 48 {\it Kepler} stars within the temperature range 5600-6000 K (using the {\it Gaia} temperature scale). From this analysis they identified 4 stars that are in the transition from the spot- to the facula-dominated regimes (similar to the sun in this regard). Of these stars, only one has a rotation period comparable to the sun, KIC 11599385, but the {\it Gaia} Flame upper limit on its age is only 1.6 Gyr; interestingly, this star was not included in the sample of \citeauthor{vas24}. They note that \citet{okamoto21} had detected one flare from this star with an energy of $5.5\times10^{33}$ erg, which is just below the superflare cutoff as defined by \citeauthor{vas24}.

In summary, sun-like stars observed over the four-year {\it Kepler} mission have superflare rates anywhere between one and three per century. Given that even the strong AD 774/775 event discovered in tree rings falls below the superflare cutoff, the data suggest that the sun's superflare rate is far below that of the {\it Kepler} sample. Even though the tree ring record has yet to be fully analyzed at the required annual resolution to definitively rule out any superflares within the last millennium, it is unlikely that the sun has had a superflare within this time period.

\subsection{Rotation period from photometry}
\label{subsec:phot_var_rot}

We don't know the spot distributions on surfaces of single sun-like stars. We will assume that the most appropriate solar rotation period to employ in comparing to photometric observations of sun-like stars is the rotation period of the sun at the average heliographic latitude of sunspots. As seen by a distant observer making photometric measurements of the sun, this would be the typical rotation period they would determine for it. To arrive at this value, we employ two studies of long-term sunspot observations. \citet{jha21} determined the A and B constants for the solar rotation profile from nearly a century of sunspot observations:
\begin{equation}
\Omega = A + B \sin^{2} \theta
\label{eq:eq_omega_rot}
\end{equation}
where $\Omega$ is the sidereal angular rotation rate in deg/day, $\theta$ is the heliographic latitude, $A = 14.381 \pm 0.004$ deg/day, and $B = -2.72 \pm 0.04$ deg/day. The value of the $A$ constant implies a solar equatorial sidereal rotation period of $25.033 \pm 0.006$ days. Next, we make use of the latitude migration of sunspots measured by \citet{tak24}. Taking 65 months as the mid-point of a cycle, and averaging the latitude from equations 1 and 2 of \citeauthor{tak24}, we arrive at a mean latitude of 13.45 degrees. Inserting this value into Equation~\eqref{eq:eq_omega_rot} yields a rotation period at the average sunspot latitude of $25.29 \pm 0.008$ days. We will adopt this value for our comparisons below. It is close to the traditional sidereal Carrington rotation period of 25.38 days.

\subsubsection{Santos and Mathur dataset}
\citet{sant21} published a catalog of rotation periods and photometric variability measurements for 39,592 {\it Kepler} F and G main sequence and late subgiant stars. \citet{math23} added age and mass estimates for these stars using two methods. First, they inferred them using the usual fundamental stellar parameters and stellar isochrones along with the observed rotation period using the code \texttt{kiauhoku}\footnote{\url{https://github.com/zclaytor/kiauhoku}}; the code makes use of the observed rotation period to constrain stellar evolution models according to angular momentum evolution. Second, they inferred the ages and masses from the \texttt{STAREVOL} stellar isochrones without using individual observed rotation periods.\footnote{\url{https://obswww.unige.ch/Research/evol/starevol/starevol.php}} The two sets of ages differ mostly for stars younger than 1 Gyr and older than 5 Gyr. We perform our analysis below with both sets of stellar age inferences.

First, we restricted the dataset as described in \citeauthor{math23} to the stars with \texttt{flag} equal to 0 in their Table 1; this eliminates the stars most likely to be binaries. In addition, we applied the following cuts to their sample: $5600 <$ T$_{\rm eff} < 5940$ K, log g $> 4.1$, and $0.90 <$ mass $< 1.1$ M$_{\odot}$, and age $> 0.5$ Gyr. The mean error of P$_{\rm rot}$ for this sample is 2.17 d. These cuts resulted in 5,571 stars with the \texttt{kiauhoku} age inferences, which we fit to the following model using the usual procedure described in earlier sections:
\begin{equation}
\begin{split}
\text{P$_{\rm rot}$} = -(136.72 \pm 29.75) 
- (72.37 \pm 3.24) \times (\text{log T$_{\rm eff}$}) \\
+ (182.933 \pm 12.68) \times (\text{log g})
- (19.99 \pm 1.44) \times (\text{log g})^{2} \\
+ (6.62 \pm 0.19) \times [\text{Fe/H}] + (16.019 \pm 0.253) \times \log_{10}(\text{Age}) \\ 
+ (20.014 \pm 1.285) \times (\log_{10}(\text{Age}))^{2} \\
- (36.506 \pm 2.764) \times (\log_{10}(\text{Age}))^{3} \\
+ (29.151 \pm 1.679) \times (\log_{10}(\text{Age}))^{4}
\end{split}
\label{eq:Prot_Mathur_kia}
\end{equation}
The adjusted R-squared of the fit is 0.94, and the standard deviation of the residuals is 1.65 d. Since this is smaller than the mean estimated error, we assume that cosmic variance is negligible. The predicted solar P$_{\rm rot}$ from this model is $23.55 \pm 1.65$(se) $\pm~0.04$(sem) d, which is just over 40 $\sigma$ smaller than the solar sidereal rotation period of 25.29 d.

The fit to the rotation periods with the age inferences from the \texttt{STAREVOL} models instead of \texttt{kiauhoku} ages is (with 5,260 stars):
\begin{equation}
\begin{split}
\text{P$_{\rm rot}$} = -(364.49 \pm 32.92) 
- (109.65 \pm 3.62) \times (\text{log T$_{\rm eff}$}) \\
+ (17.74 \pm 14.11) \times (\text{log g})
- (0.98 \pm 1.61) \times (\text{log g})^{2} \\
+ (5.99 \pm 0.22) \times [\text{Fe/H}] + (15.661 \pm 0.278) \times \log_{10}(\text{Age}) \\ 
+ (11.657 \pm 1.423) \times (\log_{10}(\text{Age}))^{2} \\
- (45.070 \pm 3.546) \times (\log_{10}(\text{Age}))^{3} \\
+ (51.727 \pm 2.430) \times (\log_{10}(\text{Age}))^{4}
\end{split}
\label{eq:Prot_Mathur_iso}
\end{equation}
The adjusted R-squared of the fit is 0.92, and the standard deviation of the residuals is 1.79 d. The mean measurement P$_{\rm rot}$ error for this sample is 2.26 d. The predicted solar P$_{\rm rot}$ from this model is $23.82 \pm 1.79$(se) $\pm~0.05$(sem) d. It is interesting that the rotation period is so sensitive to metallicity; this confirms the modeling of \citet{amard20} and the analysis of {\it Kepler} rotation periods of \citet{amard20b} and \citet{see24}.

\begin{figure}
\includegraphics[width=3.2in]{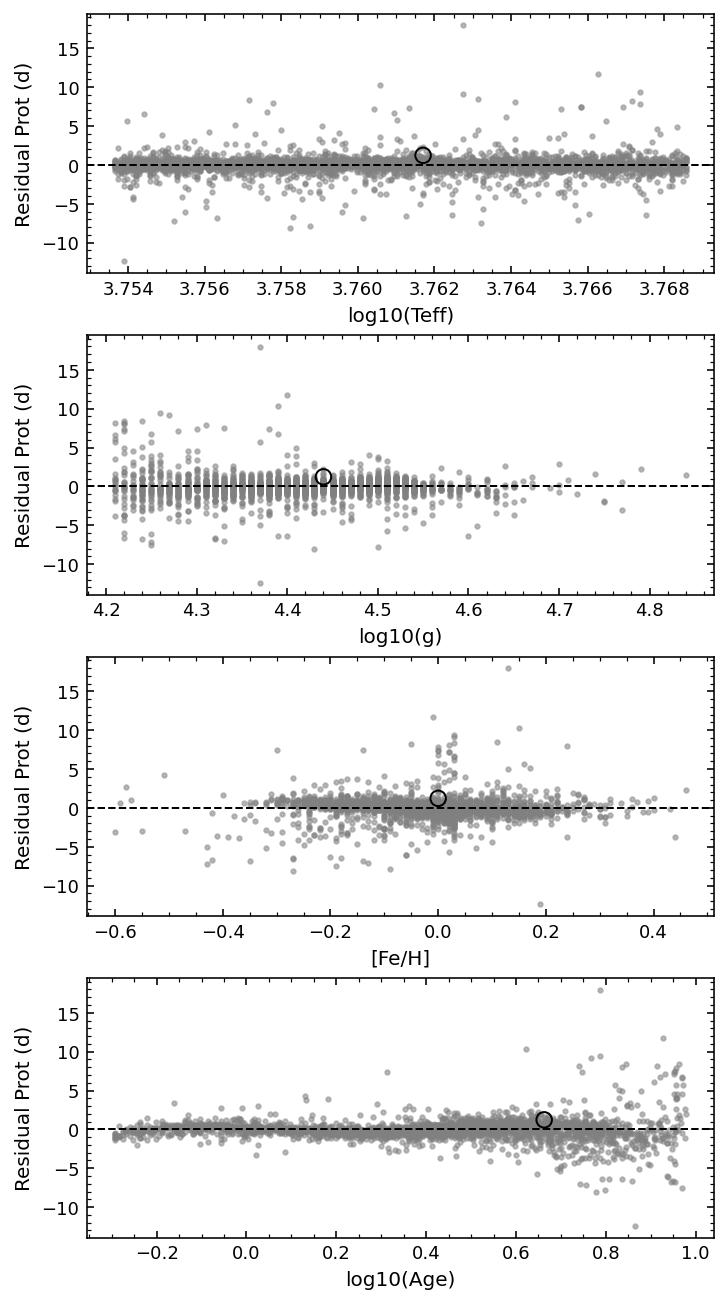}
\caption{Residual P$_{\rm rot}$ values from the regression fit using Equation~\eqref{eq:Prot_Mathur_solar} to the data from \citet{math23}.}
\label{fig:Prot_solar_iso_fig}
\end{figure}

In order to test the robustness of this result to changes in the sample selection, we produced a new sample with the \texttt{STAREVOL} models, narrowing the parameter ranges to be closer to solar: $5670 <$ T$_{\rm eff} < 5870$ K, log g $> 4.2$, and $0.90 <$ Mass $< 1.1$ M$_{\odot}$, and age $> 0.5$ Gyr. This yields a sample size of 3,129 stars. The mean estimated error of the P$_{\rm rot}$ values in this sample is 2.23 d. The fit is:
\begin{equation}
\begin{split}
\text{P$_{\rm rot}$} = -(266.85 \pm 45.76) 
- (114.10 \pm 5.72) \times (\text{log T$_{\rm eff}$}) \\
+ (308.61 \pm 18.52) \times (\text{log g})
- (33.60 \pm 2.09) \times (\text{log g})^{2} \\
+ (5.26 \pm 0.22) \times [\text{Fe/H}] + (15.322 \pm 0.267) \times \log_{10}(\text{Age}) \\ 
+ (12.272 \pm 1.346) \times (\log_{10}(\text{Age}))^{2} \\
- (44.285 \pm 3.335) \times (\log_{10}(\text{Age}))^{3} \\
+ (50.603 \pm 2.311) \times (\log_{10}(\text{Age}))^{4}
\end{split}
\label{eq:Prot_Mathur_solar}
\end{equation}
The adjusted R-squared of the fit is 0.95, and the standard deviation of the residuals is 1.33 d. We show the residuals in Fig.~\ref{fig:Prot_solar_iso_fig}. Since this is less than the mean estimated error in this sample, we can assume that the cosmic variance is negligible. The reduction in the standard deviation of the residuals relative to the previous fit is most likely due to a group of old low gravity stars (likely leaving the main sequence), which are more common in the sample stars with log g $< 4.2$. The predicted solar rotation period from this fit is $24.06 \pm 1.33$(se) $\pm~0.05$(sem) d, with the actual solar rotation period being nearly 25-$\sigma$ larger. This period is only 0.2 d higher than our previous estimate. This shows that the fit is robust to the precise criteria used in the construction of the sample.

It is worth noting the implicit assumption in the above analyses that a star with the same physical properties as the sun (including rotation period) will have the same spot distribution on its surface. If the spot distribution is fundamentally different, say in latitude, then there will be a systematic error in the period determined from the light variations. 

We also note that it is not the case that the age is an independent variable in relation to the rotation period in Equation~\eqref{eq:Prot_Mathur_kia}, since the rotation period measurements went into constraining the stellar ages with \texttt{kiauhoku}. For this reason, we strongly prefer the results of the fits that use the \texttt{STAREVOL} stellar isochrone age inferences.\footnote{\texttt{STAREVOL} also uses stellar rotation information by modelling the angular momentum transport, but this is done in a general way rather than requiring the input of a specific rotation period for a given star.} From these, we tentatively conclude that the sun's rotation period is about 0.2 d faster than {\it Kepler} field solar analogues. As period-estimation methods from photometry of old sun-like stars improve, this analysis should be revisited.

\subsubsection{M67}
\label{subsubsec:phot_var_M67}

Another approach is to compare the sun's rotation period to the member stars of the open cluster M67, which is similar in age and metallicity to the sun. \citet{reyes24} estimated its age to be $3.95\pm0.16$ Gyr, and \citet{casa19} determined that its mean [Fe/H] is $+0.03$ dex. 

\citet{grun23} is the largest detailed study of rotation among M67 member stars. They measured the rotation periods of 47 single main sequence stars from early G to M spectral types; to these stars we added 27 single stars from \citet{gon16b} listed in their Table C.3. The periods of these stars were fit to the colors of the 20 stars in the range $0.40 \le$ \( (G-G_{\text{RP}})_{0} \) $\le 0.60$ mag with a quadratic polynomial weighted by the period errors. From this fit, we calculated that the rotation period in M67 at the solar color (see Table~\ref{tab:sun_params}) is $24.3\pm1.3$ days.

If we apply Equation~\eqref{eq:Prot_Mathur_solar} from the Santos data to the known age and [Fe/H] values of M67, we calculate that the rotation period at the solar temperature and surface gravity is $22.40\pm0.04$ days. This is about 0.6 d (1.5-$\sigma$) smaller than our estimate above. Thus, while the M67 data do not constrain the rotation period at the solar temperature any better than the {\it Kepler} field stars do, they are approximately consistent with them.

\subsubsection{Reinhold dataset}

\citet{rein23} derived rotation periods for 67,163 {\it Kepler} stars using a new technique called gradient of the power spectrum (GPS). GPS shows promise in that it can be used to determine rotation periods for stars missed by other methods. 

\citeauthor{rein23} do not provide the fundamental parameters for the stars in their sample. Given this, we cross-referenced their catalog with the {\it Gaia} DR3 astrophysical parameters supplement catalog using the {\it Kepler} field-{\it Gaia} DR3 cross-matched database. We adopted the rotation periods from the P$_{\rm rot, fin}$ column of Table C.1 of  \citeauthor{rein23} and the stellar parameters from the {\it Gaia} \texttt{\_gspphot} (T$_{\rm eff}$, log g, [Fe/H]), \texttt{-Flame} (mass, age), \texttt{non\_single\_star}, and \texttt{ruwe} fields; these last two parameters were required they have values of 0 and $<1.25$, respectively.

Restricting the final sample to $5670 <$ T$_{\rm eff} < 5870$ K, log g $> 4.3$, and $3.5 <$ age $< 6.0$ Gyr, resulted in 654 solar analogue stars. Fitting a function to this sample like the one above for the Santos data produced a poor fit, not even justifying a quadratic term in log age. Simplifying the fit to be linear in log age, still produced a poor fit, with a reduced R-squared of 0.02. The log age coefficient was still not significant, and the sign of the [Fe/H] coefficient was the opposite of what was determined for the Santos data. The predicted rotation period of the sun from this fit is 15.6 days.

In order to check on the accuracy of the \citeauthor{rein23} rotation periods, we employed the stellar rotation isochrones model of \citet{long23}.\footnote{\url{https://github.com/longliu-git/gyroage.git}. The online code, which we employed in our calculations, is a slightly modified version of equations 9-12 and Table 6 in \citeauthor{long23}.} It was necessary to multiply the first term of their equation 9 by 1.015 to match the best-fit rotation-color curve of M67 from \citet{grun23} for the G dwarfs. We calculated the ratio of the predicted-to-measured period for each star. The mean was 0.6 days. This is consistent with the very low value we predicted for the sun from our fit. It implies that the GPS method is often not able to distinguish between the true period and the half-period of a star. For this reason we cannot use the estimated periods of \citeauthor{rein23}, even though they are more extensive than any other study.

\subsection{Rotation from {\it v} sin{\it i}}
\label{subsec:rot_vsini}
Measuring photometric or chromospheric variations is a direct way to determine rotation. A second, less direct, way is to projected rotational velocity ({\it v} sin{\it i}) via spectroscopy.

The sun's sidereal rotation period 25.4 d. Although the equatorial velocity of the sun is 2.0 km s$^{\rm -1}$, this is not the value appropriate for comparison with spectroscopically determined {\it v} sin{\it i} values. Most spectroscopic determinations of {\it v} sin{\it i} do not include differential rotation; fitting a spectrum of the sun as a star yields a value of 1.63 km s$^{\rm -1}$ \citep{vp96}. Furthermore, when comparing the sun's {\it v} sin{\it i} value to the mean of an ensemble of stars with randomly oriented rotation axes, it is necessary to employ the typical value of the solar {\it v} sin{\it i} that a distant random observer would measure, namely 1.28 km s$^{\rm -1}$ \citep{gonz10a}.

\subsubsection{Nearby field stars}
\label{subsubsec:rot_nearby}
\citet{lor19} studied the long term evolution of rotation periods of nearby solar twins. While only a few of the stars in their sample as old as the sun have well-determined ages and rotation periods, they found the sun's rotation period to be typical for its age. 

sun-like stars in star forming regions and young clusters display a wide range of rotation periods, termed slow, medium and fast rotators; the slow rotators make up the slowest 25 percent. Using an indirect method involving the sodium and potassium abundances in the lunar regolith, \citet{sax19} were able to place strong constraints on the sun's early rotation history. They concluded that the sun is ``highly likely'' to have been a slow rotator in its youth.

Stellar rotation can also be studied by measuring {\it v} sin{\it i}. Given the sin{\it i} ambiguity, {\it v} sin{\it i} must be studied statistically. In \citet{gonz10a} and \citet{gonz11} we compared the {\it v} sin{\it i} values of stars with Doppler-detected giant planets to those without. In \citet{gonz10a} we found that the sun's {\it v} sin{\it i} is smaller than stars without planets in the \citet{VF05} sample by 1.25 km s$^{\rm -1}$; in \citet{gonz11} we found it to be 0.90 km s$^{\rm -1}$ smaller. 

We revisit this comparison here using the large high quality spectroscopic sample of \citet{brew16a,brew16b}. It is an expansion of the SPOCS sample \citep{VF05}. The new sample consists of single F, G, and K dwarfs with small {\it v} sin{\it i} values that are RV planet search targets. \citet{brew16a,brew16b} determined the standard stellar parameters, {\it v} sin{\it i}, as well as ages and masses for many stars. They quote an average uncertainty of 0.5 km s$^{\rm -1}$ for {\it v} sin{\it i}; the typical uncertainty in the age is 50 percent. A large fraction of their {\it v} sin{\it i} determinations are equal to 0.1 km s$^{\rm -1}$, no doubt due to the relatively large measurement uncertainty for very low values. From the random distribution of rotation axes, the actual distribution of {\it v} sin{\it i} should have few values so close to zero. By including enough stars in the comparison sample, we can greatly reduce the effects of the random errors. Starting with the full sample of 1615 stars, we applied several cuts to restrict it to nearby single sun-like stars for comparison with the sun: 5600 K $<$ T$_{\rm eff} <$ 5940 K, log g $>$ 4.3 dex and $-0.3 <$ [M/H] $< 0.3$ dex. This leaves us with 244 stars. These cuts are a balance between selecting the most sun-like stars and still having a large enough sample to conduct the following analysis.

\begin{figure}
\includegraphics[width=3.2in]{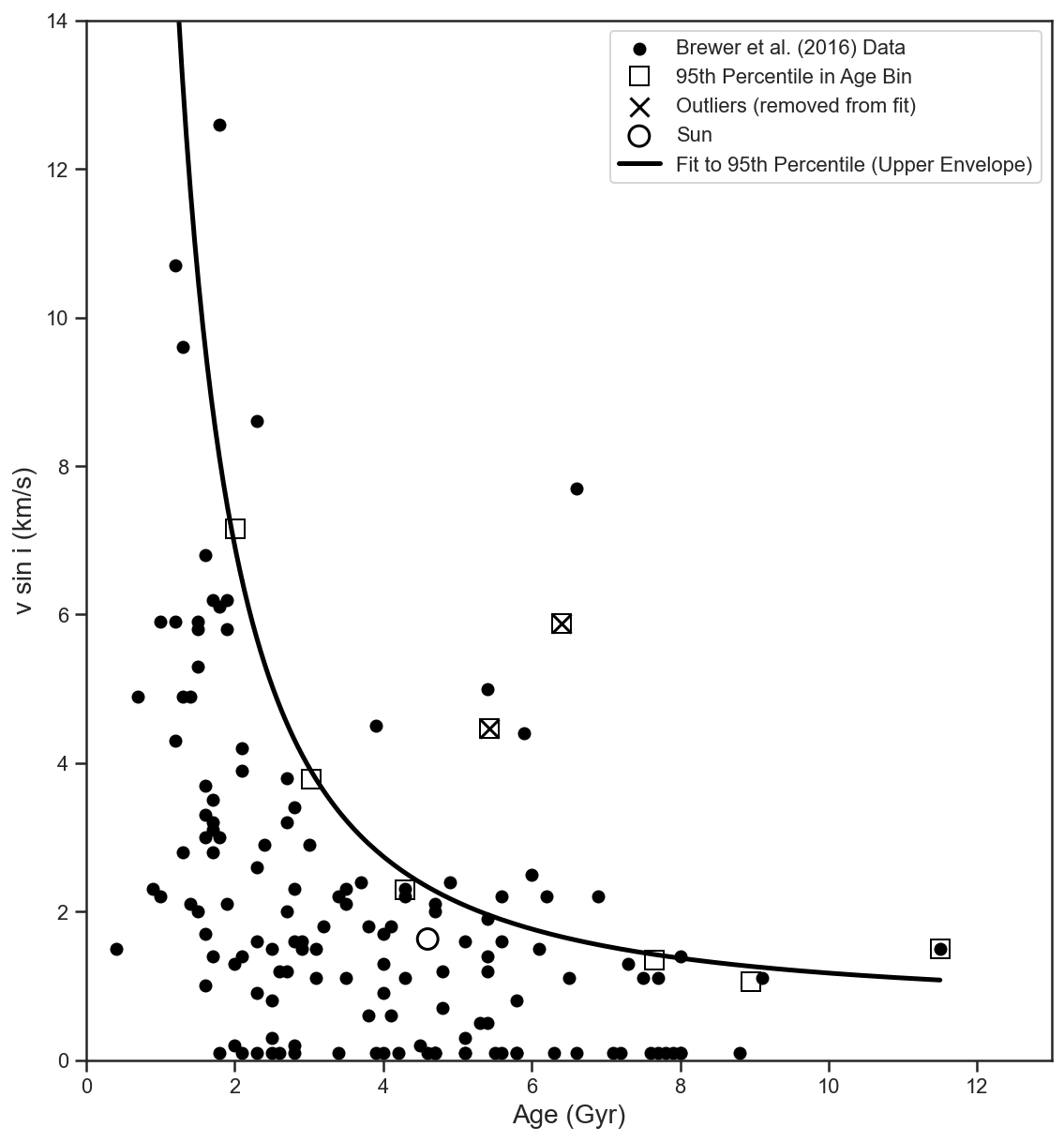}
\caption{The {\it v} sin{\it i} values of 244 sun-like stars from \citet{brew16a,brew16b}. The boxes are 95th percentile binned {\it v} sin{\it i} values; the two outliers are shown with x's. The solid curve is a power law function fit to the binned values and represents the ``upper envelope'' of the data.}
\label{fig:vsini_Brewer}
\end{figure}

We plot our subset of the \citet{brew16a,brew16b} {\it v} sin{\it i} values in Fig.~\ref{fig:vsini_Brewer}. Due to measurement error and the random orientation of the rotation axes, at a given age there should be a spread in {\it v} sin{\it i} values from zero to some maximum value, where the inclination angle relative to our line-of-sight is 90 degrees. Following \citet{dos16}, we seek to describe the location of this ``upper envelope'' with a simple power law, {\it v} sin{\it i} $=$ {\it v}$_{\rm f} + m t^{-b}$, where t is age. We fit this equation to the 95th percentile {\it v} sin{\it i} values binned in age with least-squares in order to locate the apparent upper envelope in Fig.~\ref{fig:vsini_Brewer}. The challenge in fitting the curve is to find where the upper limit of the real distribution is located at each age while allowing for measurement errors and outliers with excessively high {\it v} sin{\it i}. At least the location of the upper envelope is not affected by errors in locating the ``floor'' of the distribution.

There are only a few outliers for ages greater than about 2 Gyr, but there are many among the younger stars. Some of the younger outliers could be explained by age errors, but they should have little impact on the location of the upper envelope for the older stars. In addition, there could still be a few spectroscopic binaries in the sample.

The appropriate value of the sun's {\it v} sin{\it i} to compare to the upper envelope in Fig.~\ref{fig:vsini_Brewer} is 1.63 km s$^{\rm -1}$. This is 0.7 km s$^{\rm -1}$ smaller than the location of the upper envelope for the sun's age in Fig.~\ref{fig:vsini_Brewer}.

To help check the robustness of this result, we can examine an independent dataset. \citet{dos16} performed high precision spectroscopic analyses of 82 solar twins, for which they also determined {\it v} sin{\it i}; their sample substantially overlaps with the lithium study of \citet{mart23}. The typical uncertainty in their {\it v} sin{\it i} estimates is only 0.12 km s$^{\rm -1}$. We cross referenced their sample with \citeauthor{mart23}, adopting the more recent stellar parameters. The age estimates have much smaller errors than the \citet{brew16a,brew16b} sample. Our final selection includes 61 single stars.

\begin{figure}
\includegraphics[width=3.2in]{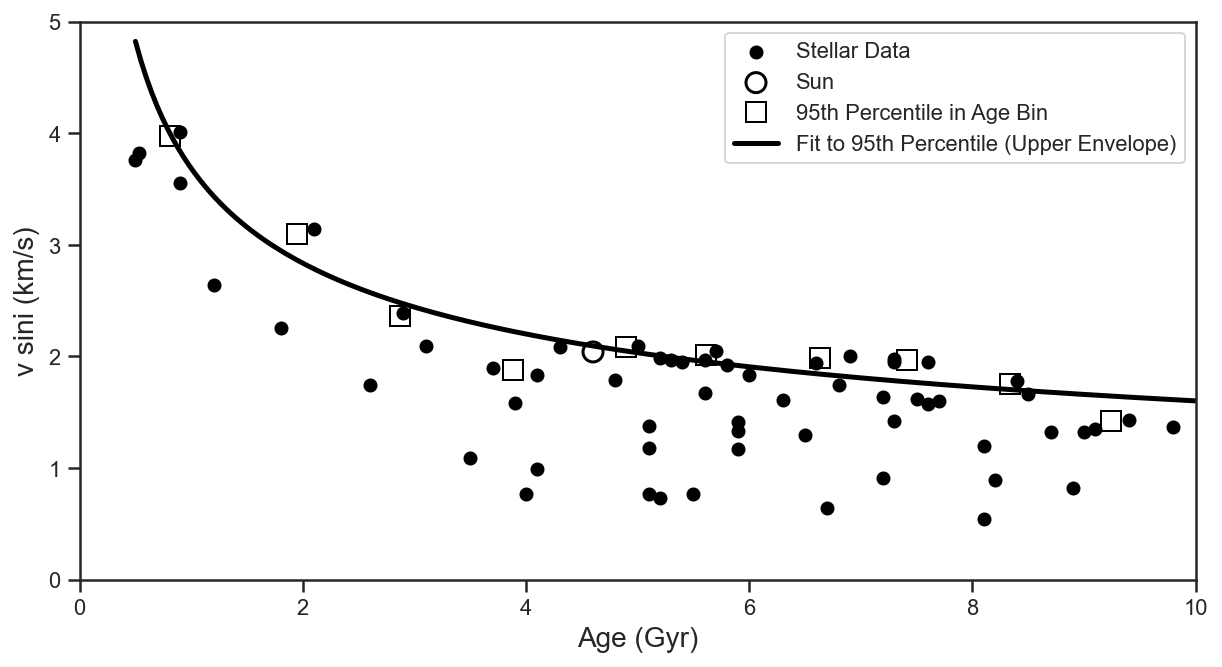}
\caption{{\it v} sin{\it i} values from \citet{dos16} with revised ages from \citet{mart23}. Symbols are the same as Fig.~\ref{fig:vsini_Brewer}.}
\label{fig:vsini_dos}
\end{figure}

We show the resulting data in Fig.~\ref{fig:vsini_dos}. As above, the solid curve is a least-squares fit to 95th percentile {\it v} sin{\it i} values in each age bin, as was done for the data in Fig.~\ref{fig:vsini_Brewer}. There don't appear to be any outliers, allowing for greater confidence in defining the upper envelope. It is important to note that \citeauthor{dos16} determined the solar {\it v} sin{\it i} value to be $2.04 \pm 0.12$ km s$^{\rm -1}$. This is only about 0.05 km s$^{\rm -1}$ less than the upper envelope at the solar age. It is also notable that the floor of their {\it v} sin{\it i} values is near 0.5 km s$^{\rm -1}$, which is considerably greater than that of the \citet{brew16a,brew16b} data. No doubt this is due to the smaller uncertainties of the {\it v} sin{\it i} measurements of \citeauthor{dos16}.

While the individual measurements shown in Fig.~\ref{fig:vsini_dos} are higher quality than those in Fig.~\ref{fig:vsini_Brewer}, the smaller sample size is a weakness. Overall, the data from \citet{brew16a,brew16b} and \citet{dos16} are consistent with the sun having close to but just below the estimated maximum {\it v} sin{\it i} for nearby sun-like stars for its age. We would tend to favor the results shown in Fig.~\ref{fig:vsini_dos}, given the higher quality of the individual measurements. It is difficult to be more firm until additional high quality {\it v} sin{\it i} measurements of solar analogues are obtained.

\subsection{Binarity}
\label{subsec:binarity}
Single stars (excluding brown dwarfs) comprise 54 percent of the RECONS sample \citep{hen18}. \citet{moe19} confirmed earlier reports that the fraction of short-period binaries (P $< 10^{\rm 4}$ d, {\it a} $ < 10$ AU) is strongly dependent on metallicity, with the binary fraction for solar-type stars decreasing from 40 percent at [Fe/H] $= -1.0$ to 10 percent at [Fe/H] $= 0.5$. They also found that early-B stars have a much higher binary fraction, at 70 percent. 

From analysis of {\it Gaia} EDR3 data for G and K dwarfs within 0.95 kpc of the sun, \citet{niu21} estimate an average binary fraction of $0.42 \pm 0.01$. They confirm that binary fraction decreases with increasing metallicity and also find that thin disk stars have smaller binary fraction than thick disk stars. Thus, the sun appears to be a typical thin disk star in lacking a stellar companion for its metallicity. 

\section{Comparing the Solar System to Exoplanetary Systems}
\label{sec:exoplanets}
As of this writing there are over 7000 confirmed exoplanets\footnote{\url{http://exoplanet.eu} Accessed August 20, 2024.}, mostly discovered with the RV or photometric transit methods \citep{sant18,trifon24}. Our knowledge of exoplanets includes planet masses (or minimum masses), planet sizes, orbital periods, eccentricities {\it e}, and orbital inclinations {\it i} as well as the host star properties. \citet{mlivio15} and \citet{livio19} compared the properties of the Solar System planets to those of exoplanetary systems in order to try to answer the question, "How special is the Solar System?". They concluded that the Solar System differs in two ways from exoplanetary systems: 1) lack of super-earths, and 2) lack of close-in planets. We review and evaluate these and other aspects of the Solar System planets' properties in the following.

Of course, it is not proper to compare the properties of the full complement of planets in the Solar System to exoplanetary systems, given that only a fraction of its planets could be detected from afar with current methods and a baseline of nearly 30 years. It is likely that Venus would have been detected with the transit method with {\it Kepler}, but detection of Earth is much less likely; also, the detection of two transiting planets in the Solar System is about 10 percent as probable as detecting only one planet \citep{wells17}. Given this, the most likely situation for a random observer of the Solar System is not to see any transiting planets, and the next most likely is to see only one planet. We will designate as ``{\it Kepler}-equivalent'' the detections of Solar System planets and measurement of their properties with {\it Kepler}.

Similarly, it is also likely that Jupiter would have been detected with the Doppler/RV method by a random observer; we will designate as ``RV-equivalent'' the detections of Solar System planets and measurement of their properties with the RV method. When both detection methods are combined, the Solar System would be a two-planet system among those systems with at least one transiting planet, and a one-planet system for 97.5 percent of random observers.

\subsection{Relationships between Host Star and Planet Properties}
\label{subsec:exo_host}
In the previous section we compared the sun's properties to nearby sun-like stars. Prior to transitioning into comparisons of planet properties, it is instructive to compare how the properties of the stellar hosts relate to their complement of planets relative to the Solar System.

The age of the host star is an important parameter when comparing the properties of the Solar System planets to exoplanets, given that planet orbits change over time. In addition, older stars tend to me more metal-poor, which affects planet formation \citep{gonz14b, sant17}.

\citet{baner24} compared the [Fe/H] values and ages of stars hosting giant planets with different orbits: hot, warm, and cold giant planets (circular and eccentric orbits). First, they confirmed prior research that showed a correlation between the presence of giant planets and the metallicity of the host star. They also showed that giant planets in the hot, warm, and cold eccentric categories are more metal-rich than giant planets in cold circular orbits. They found that the median [Fe/H] for cold circular giant planet orbits is 0.03 dex. The Solar System is normal in this respect, since it contains a giant planet in a cold circular orbit and the sun has [Fe/H] $= 0$. 

\citeauthor{baner24} also compared the mean ages of the different categories of giant planets. They found that the giant planets in cold circular orbits have a median age of $6.07 \pm 0.79$ Gyrs; the median age for cold eccentric giant planets is only slightly greater. The median age for ``hot Jupiters'' is $3.97 \pm 0.51$ Gyrs. The age of the Solar System is almost two $\sigma$ younger than the cold circular giant planet systems. These results could be biased if the mass distributions of their samples differ (since more massive host stars would be younger on average). While they don't explicitly discuss mass distributions, they do compare the host star temperatures of their sub-samples, finding no significant differences. 

\citet{yang20} explored the dependence among {\it Kepler} stars of the fraction of main sequence stars with planets and planet multiplicity (the mean number of planets per system with planets) on host star temperature. They model the biases inherent in the {\it Kepler} sample, e.g., easier to detect larger planets in shorter periods around smaller stars. They found that the fraction of stars with planets (planet radii $> 0.4$ R$_{\oplus}$ and orbital periods $< 400$ d) declines from 75 percent for stars with later spectral types to about 35 percent for mid-F or earlier. Multiplicity drops from 2.8  planets per star for late type stars to 1.8 for early type stars. In each case the decline occurs over a small range of spectral types from about G1 to F8 (see their Figures 5 and 6). They also find that the mean number of planets per star drops from 2.1 to 0.5. Thus, the sun is one of the most massive main sequence stars within the high occurrence/multiplicity group. However, based on its {\it Kepler}-equivalent multiplicity alone ($= 1$ if we only include Venus), the Solar System is a better match to the low multiplicity group. This makes it anomalous in this context.

\subsection{Multiplicity, Eccentricity, and Size Scale}
\label{subsec:mult}
Detailed comparisons relying on either RV data alone \citep{lt15}, the large {\it Kepler} primary mission sample \citep{xie16}, or a combination \citep{zinzi17} imply that the mean {\it e} value of the full complement of Solar System planets is similar to that of exoplanet systems with high multiplicity, especially for $>$ 6 planets. \citet{zinzi17} noted that the exoplanetary system with the greatest number of planets with known {\it e}, TRAPPIST-1 with 7 planets, has a mean {\it e} value almost as small as that of the Solar System. 

\citet{bach21} examined multiple planet systems within the context of both {\it e} distribution and frequency for various multiplicities. They confirm that the mean {\it e} of the Solar System planets follows the trend with multiplicity among exoplanets and appears normal in this regard. They also estimate that systems with 8 planets only constitute 1 percent of all systems. Based on this, the Solar System may have a normal mean {\it e} value, but it is rare in having 8 planets. Of course, there is a severe bias in present detection methods against detection of planets with long orbital periods.

Jupiter has an {\it e} value 0.048, and the mean {\it e} of Venus plus Jupiter is 0.028. The mean {\it e} of two-planet exoplanetary systems according to \citet{bach21} is 0.24. If we add Earth and make it a three-planet system, the mean {\it e} becomes 0.024, which compares to 0.15 for three-planet exoplanetary systems. In any case, the Solar System is far from typical in this respect (we perform a more detailed analysis of {\it e} below).

\begin{figure}
\includegraphics[width=3.2in]{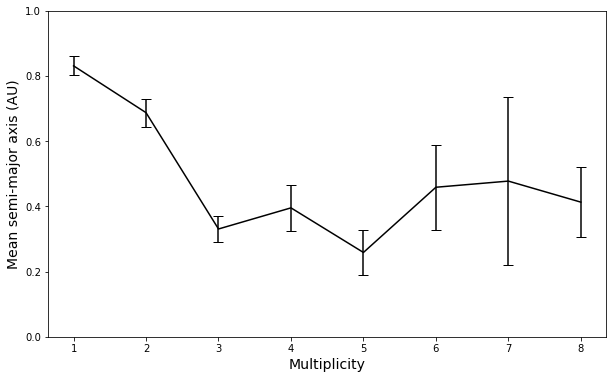}
\caption{Mean semi-major axis versus multiplicity. Error bars are the standard deviations of the mean. The error bars are larger for higher multiplicities due to the smaller sample size.}
\label{fig:planet_a_mult}
\end{figure}

How does the size scale of the Solar System compare to exoplanetary systems? The mean of the semi-major axes of the planets in the Solar System is 8.45 AU. Restricting the planets to just Venus and Jupiter, as noted above, the mean size is 2.95 AU. We prepared a sample of exoplanetary systems to compare to this value using the following procedure. We downloaded the {\it Catalogue of Exoplanets}\footnote{\url{http://exoplanet.eu} Accessed August 20, 2024.} and limited it to confirmed planets with semi-major axes less than 10 AU. This resulted in a sample of 4610 planets. We plot the mean semi-major axis values against multiplicity in Fig.~\ref{fig:planet_a_mult}. There is a declining trend of mean orbit size with multiplicity; in other words, the greater the multiplicity, the more compact the system. The planets in the two-planet exoplanetary systems have a mean semi-major axis of only 0.7 AU with a standard deviation of 1.4 AU, more than a factor of four smaller than the two-planet Solar System. Only 34 of 543 (or 6 percent) of two-planet systems have mean semi-major axes greater than 2.95 AU. Therefore, although there is bias favoring detection of compact systems, the current data indicate that the detectable two-planet Solar System is not nearly as compact as the typical two-planet exoplanetary system; the discrepancy is even greater if we limit the Solar System to just Jupiter.

\subsection{System Architecture}
\label{subsec:arch}
\subsubsection{Structure}
\label{subsubsec:struct}
How does the architecture of the Solar System compare to exoplanetary systems? The Solar System is characterized by multiple low mass inner planets and gas giant outer planets on nearly circular and coplanar orbits. Only two exoplanetary systems, GJ 676A \citep{escude12} and Kepler-90 \citep{shall18}, are Solar System analogues in the sense of their overall architectures, but they each have at least one super-earth ($\sim$1--10 M$_{\oplus}$ or $\sim$1--4 R$_{\oplus}$).\footnote{We might also add HD\,160691 as a Solar System analogue. It has four planets, three of which are giants. The orbits of planets c and d are similar to Earth and Mars, but they are probably gas giants. Planet e has very similar orbital properties to Jupiter with {\it P} $= 11.1$ y and {\it e} $= 0.045$, but it is at least twice as massive \citep{witt20}; the corresponding orbital properties for Jupiter are 11.9 y and 0.049, respectively.} While Kepler-90 is the only exoplanetary system known to have the same number of planets as the Solar System, it is a compact system with all its planets just fitting within the orbit of Earth, and several of its planets are in mean motion resonances; the planets in the Solar System lack such resonances.

\citet{gajen24} examined the conditional occurrences of super-Earths and giant planets using a sample of planets discovered with the RV method. They define a super-Earth as having $m \sin i < 10$ M$_{\oplus}$ and P $< 100$ d, and they define a cold Jupiter as having mass $> 0.4$ M$_{\rm J}$ and orbit larger than 1 AU. They calculate a conditional probability of no super-Earth given a cold Jupiter, P(no-SE$\vert$CJ), of about 3 percent. Since the Solar System lacks a super-Earth, this makes it rare in this aspect of its architecture.\footnote{A super-Earth in the Solar System inside the orbit of Earth would be detectable with the RV method by a distant observer.} Even this may be an overestimate, since the probability of hosting a cold Jupiter that \citeauthor{gajen24} assumed (10 percent) may be too high (see below).

Another way to compare system architectures is to examine the relative spacings and sizes (or masses) of planets in a given system. \citet{weiss23} created two simple metrics to measure planet size dispersion ($\sigma_{\mathcal{R}}$) and orbit spacing dispersion ($\sigma_{\mathcal{P}}$); each metric is based on the standard deviation of the log of the respective observed quantity (in the case of the spacing dispersion, it is based on ratios of consecutive periods). They find when comparing the inner four planets of the Solar System ($\sigma_{\mathcal{R}} = 0.20$ dex, $\sigma_{\mathcal{P}} = 0.10$ dex) to {\it Kepler} systems with four or more transiting planets within 1.52 AU, that the Solar System is less regular (larger dispersion indices) in the relative planet sizes and spacings.

\begin{figure}
\includegraphics[width=3.2in]{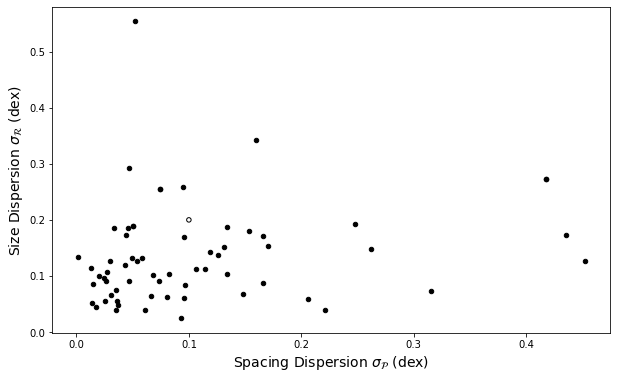}
\caption{Size dispersion versus spacing dispersion for compact systems with four or more planets. The sun is shown as an open circle.}
\label{fig:s_p_disps}
\end{figure}

We repeat the analysis of \citet{weiss23} using the latest version of the {\it Catalogue of Exoplanets}. We included both {\it Kepler} and non-{\it Kepler} planets in our analysis. We plot the resulting $\sigma_{\mathcal{R}}$ and $\sigma_{\mathcal{P}}$ values of 81 systems in Fig.~\ref{fig:s_p_disps}. The sun has larger $\sigma_{\mathcal{R}}$ and $\sigma_{\mathcal{P}}$ values than 68 and 59 percent of the exoplanets, respectively. We can conclude from this comparison that the Solar System planet sizes and spacings are moderately less regular than the exoplanetary systems with the caveat that the four inner planets of the Solar System would not be detectable from afar with current methods.

It is important to note that {\it Kepler} sampled a different population of planets than the RV surveys of nearby stars. The {\it Kepler} four year primary mission was more sensitive to smaller, shorter period planets. In contrast, some of the precision RV programs have been monitoring nearby stars for nearly 30 years, e.g. \citet{rick19}. \citet{xie16} suggested that this difference in the samples could explain why there is a steep decline in the mean value of {\it e} for increasing multiplicity in the {\it Kepler} sample but only a gradual one for the RV sample. Values of {\it e} for individual planets can only be obtained from RV and transit timing variations observations.

Employing stellar parameters from {\it LAMOST} DR8, \citet{an23} explored statistical trends between ${\it e}$ and ${\it i}$ and [Fe/H] for ensembles of {\it Kepler} systems. They confirm prior studies \citep[e.g., ][]{mills19} that suggested increasing ${\it e}$ with [Fe/H] for both single and multiple transiting systems, with the trend being steeper for the former. At solar [Fe/H] the mean value of ${\it e}$ for one-planet transiting systems is 0.09; this is over 10 times larger than the value of {\it e} of Venus, which would be the easiest planet to find orbiting the sun with {\it Kepler}. For multiple-planet transiting systems, they find that the mean value of ${\it e}$ at solar [Fe/H] to be about 0.03; it is not possible to compare the {\it Kepler}-equivalent Solar System to this estimate. They also find that the mean mutual ${\it i}$ increases with [Fe/H]; at solar [Fe/H] it is about 1.6 degrees. This is about half of the two-planet (Earth and Venus) Solar System mutual inclination (which would have a low probability of detection). The mean mutual ${\it i}$ of the two-planet Solar System, then, is larger, irrespective of which two planets are selected. Following the suggestion of \citet{an23} that the mean ${\it e}$ correlates with the mean ${\it i}$, we could tentatively conclude that the {\it Kepler} equivalent of the Solar System mean ${\it e}$ is larger than at the solar [Fe/H] {\it Kepler} multiple planet systems.

\citet{knud24} estimated host star obliquities from observations of the Rossiter-McLaughlin effect.  In their Figure 8 they plot the measured obliquities of hot Jupiter systems, finding a dispersion of 1.4 degrees, which they attribute to tidal dissipation. Among multiple planet systems, they found that the mean and intrinsic dispersion of the obliquity are zero and 2 degrees (each with uncertainty near $\pm1.6$ degrees), respectively. The obliquity of the sun relative to its invariable plane is 6.2 degrees \citep{gomes17}. Thus, the Solar System is an outlier on the high side at the few $\sigma$ level in this parameter. The authors state, "The low obliquities of the multis might therefore be primordial, namely, due to the flatness of the protoplanetary disk and its strong coupling to the accreting young star. From this point of view, the solar obliquity of $6^{\circ}$ may be unusually high."

\citet{albr22} showed that cool stars (with effective temperatures below approximately 6250 K) tend to have low obliquities, and hotter stars display a broader range of obliquities. While this shows a temperature/mass dependence, it does not affect our comparison to the sun's obliquity, since it falls among the cooler stars in their sample. Still, it is worthwhile revisiting the possible temperature dependence of obliquity among G dwarfs in a future study as the number of obliquity measurements increases.

\subsubsection{Planet Occurrence}
\label{subsubsec:occur}
Estimates of the fraction of FGK dwarf stars with at least one planet range from 0.30 for planet radii $> 1$ R$_{\oplus}$ and periods $< 400$ d \citep{zhu18} to 0.72 for planet radii between 0.5 and 16 R$_{\oplus}$ and periods between 0.5 and 500 d \citep{zinc19}. Intermediate between these is the estimate of \citet{he19} of 0.56 for planet radii between 0.5 and 10 R$_{\oplus}$ and periods between 3 and 300 d; they determine an average rate of 3.5 planets per system. The Solar System only has one planet in this size and period range (Venus), confirming \citet{mlivio15} and \citet{livio19} that the Solar System is an outlier in these respects among stars with at least one planet; among stars with planets, the Solar System ranks near the 5 percent level.

\citet{sand19} explore a range of statistical distributions to estimate the relative incidence of {\it Kepler} planets with multiplicity. Their cuts include: $0.5$ R$_{\oplus} <$ planet radius $< 32$ R$_{\oplus}$, 6.25 d $<$ period $< 400$ d, $0.8$ M$_{\odot} <$ star mass $< 1.2$ M$_{\odot}$. In their analysis the Solar System would count as a 2-planet system, which they estimate constitutes about 20 percent of the {\it Kepler} stars with planets.

We don't expect precise agreement among exoplanet studies given the different ways they count planets as well as the different modeling assumptions. For example, \citet{he19} include period and period plus size clustering models in their analysis of the {\it Kepler} planets and find evidence of clustering in the dataset. They also find that the non-clustering models give a dramatically larger incidence of planets. What's more, \citet{moe20} have shown that binary systems bias RV and {\it Kepler} surveys in important ways. For example, they calculate that the occurrence rate estimates of small planets around {\it Kepler} G dwarfs should be increased by a factor of 2.1 when accounting for the presence of binaries. Given this, the planet occurrence values we quoted above for {\it Kepler} data are underestimates. They also find that RV surveys, by excluding binaries, overestimate their giant planet detection rates by a factor of 1.8 relative to transit surveys; for example, \citet{witt20} would have to reduce their estimated rates of cool/cold Jupiters (P $> 100$ d, 6.7 percent) and hot Jupiters (0.84 percent). From this we infer that only about 4 percent of sun-like stars have cool/cold Jupiters.

\subsubsection{Eccentricity distributions}
\label{subsubsec:eccensec}
We performed the following cuts on the {\it Catalogue of Exoplanets} to produce a uniform well-defined sample:

\begin{itemize}
\item[(i)]   \texttt{Orbital Period} $\leq$ 30,000 d. This cut retains only the planets with orbital periods comparable to those of the Solar System planets;
\item[(ii)]  0.08 $\leq$ \texttt{Star mass} $\leq$ 2.5 M$_{\odot}$. This criterion restricts the host stars to have similar mass to the sun;
\item[(iii)] \texttt{luminosity class} V or IV/V only. This cut excludes subgiants, giants, bright giants, white dwarfs and pulsars;
\item[(iv)] \texttt{stellar radius} $\leq$ 2.5 R$_{\odot}$. This criterion excludes any remaining evolved stars that lack a spectral type;
\item[(v)] \texttt{microlensing} planets excluded. This criterion excludes planets that don't have precise orbital parameters;
\item[(vi)] \texttt{planet mass or msini} $\leq$ 15 M$_{\rm J}$. This criterion excludes objects that are more likely to be brown dwarfs or stars;
\item[(vii)] \texttt{eccentricity measured}. This criterion only includes planets with a measured eccentricity.
\end{itemize}

The resulting {\it EXOCAT} sample includes 1947 planets. We used it to explore the statistics of the {\it e} distribution in 30 cases, the results of which we list in Table~\ref{tab:eccentab}. In some cases we applied a period cut of 10 d to exclude close-in planets likely affected by tidal circularization; this cut was not done in most of the published studies cited above. The period cut of 88 d restricts planets to have periods at least as long as Mercury, allowing for a more fair comparison with the Solar System planets; the 300 d period cut mostly restricts the sample to planets discovered with the RV method. From the numbers of planets in the first four cases, we calculate that planets with periods $<$ 10 d make up 44 percent of the sample (here we are ignoring the binaries biases). 

The last three columns in Table~\ref{tab:eccentab} list the $\lambda$ parameter in the Box-Cox transformation \citep{mlivio15} and the number of sigma that Jupiter's and Earth's transformed eccentricity values deviate from the transformed mean, respectively. For details see equations 1-4 in their paper. Their sample size was only 28 percent as large as ours, and there's also the benefit of about 10 years worth of additional observations relative to their dataset, which further reduces the uncertainties of {\it e} determinations for the longer period planets. \citet{mlivio15} quote deviations of the transformed {\it e} values from the mean for Jupiter and Earth of $-0.97\sigma$ and $-1.60\sigma$, respectively; they only considered the one case over the full dataset available to them. 

Our results are generally consistent with the analyses of \citet{mlivio15} and \citet{lt15} with respect to the dependence of {\it e} on multiplicity (i.e., declining {\it e} with increasing multiplicity). Restricting our sample to host stars with masses $>$0.8 M$_{\odot}$ results in only a small increase in the mean {\it e} value. Restricting the planets to periods $>$ 88 d, $>$ 300 d, and $>$ 3600 d leads to gradual increases in mean {\it e}. The final three cases in Table~\ref{tab:eccentab} are especially relevant to comparison with Jupiter; Jupiter's {\it e} value is about one sigma smaller than the mean compared to the multiple planet systems. As expected, in all cases Earth deviates much more than Jupiter. Overall, our results for the longer period cuts are similar to those of \citet{mlivio15}. Adopting $-1.0\sigma$ and $-1.7\sigma$ (case 24) for Jupiter and Earth, respectively, implies a joint probability of about 2 percent. Again, we have to remember the caveat that Earth would not be detectable in extant surveys; substituting Venus, however, would make the probabilities even lower, given its very small {\it e}. For the same reason, we did not include Saturn.

Our results in Table~\ref{tab:eccentab} are also consistent with \citet{kane24}, who calculated that the mean and median values of {\it e} for planets more massive than Saturn and beyond their respective snow lines are 0.29 and 0.23, respectively (sample size 846 planets). The main difference between our sample preparations methods is that we selected according to period, while \citet{kane24} selected according to snow line location, which depends on the host star mass.

\begin{table*}
\centering
\begin{minipage}{160mm}
\caption{Exoplanet eccentricity distribution analysis results.}
\label{tab:eccentab}
\begin{tabular}{ccccccccc}
\hline
Case & Period cut & Planets cut & Star mass cut & N & mean {\it e} & $\lambda$ & Jupiter & Earth\\
 & (d) & (planets) & (M$_{\odot}$) & (planets) & & & ($\sigma$) & ($\sigma$)\\
\hline
1 & all & 1 & all & 1947 & 0.151 & 0.20 & -0.06 & -0.48\\
2 & all & $>$1 & all & 766 & 0.155 & 0.22 & -0.14 & -0.61\\
3 & $>$10 & 1 & all & 1030 & 0.178 & 0.26 & -0.34 & -0.85\\
4 & $>$10 & $>$1 & all & 496 & 0.177 & 0.27 & -0.39 & -0.96\\
5 & $>$10 & 2 & all & 265 & 0.186 & 0.27 & -0.41 & -0.94\\
6 & $>$10 & 3 & all & 98 & 0.168 & 0.22 & -0.24 & -0.76\\
7 & $>$10 & 4 & all & 69 & 0.159 & 0.26 & -0.44 & -1.12\\
8 & $>$10 & 1 & $>$0.8 & 403 & 0.177 & 0.24 & -0.21 & -0.65\\
9 & $>$10 & $>$1 & $>$0.8 & 320 & 0.183 & 0.25 & -0.33 & -0.84\\
10 & $>$10 & 2 & $>$0.8 & 181 & 0.196 & 0.26 & -0.40 & -0.91\\
11 & $>$10 & 3 & $>$0.8 & 66 & 0.172 & 0.21 & -0.21 & -0.68\\
12 & $>$10 & 4 & $>$0.8 & 43 & 0.143 & 0.20 & -0.14 & -0.67\\
13 & $>$88 & 1 & all & 291 & 0.217 & 0.38 & -1.50 & -2.41\\
14 & $>$88 & $>$1 & all & 236 & 0.206 & 0.31 & -0.78 & -1.45\\
15 & $>$88 & 2 & all & 151 & 0.212 & 0.29 & -0.60 & -1.16\\
16 & $>$88 & 3 & all & 37 & 0.219 & 0.24 & -0.73 & -1.39\\
17 & $>$88 & 4 & all & 24 & 0.142 & 0.32 & -1.57 & -3.15\\
18 & $>$88 & 1 & $>$0.8 & 241 & 0.220 & 0.38 & -1.53 & -2.44\\
19 & $>$88 & $>$1 & $>$0.8 & 181 & 0.220 & 0.31 & -0.86 & -1.52\\
20 & $>$88 & 2 & $>$0.8 & 123 & 0.223 & 0.31 & -0.71 & -1.30\\
21 & $>$88 & 3 & $>$0.8 & 26 & 0.244 & 0.25 & -0.81 & -1.44\\
22 & $>$88 & 4 & $>$0.8 & 18 & 0.149 & 0.29 & -1.06 & -2.24\\
23 & $>$300 & 1 & all & 213 & 0.229 & 0.37 & -1.96 & -3.03\\
24 & $>$300 & $>$1 & all & 156 & 0.225 & 0.31 & -0.99 & -1.70\\
25 & $>$300 & 2 & all & 112 & 0.230 & 0.32 & -0.95 & -1.63\\
26 & $>$300 & 3 & all & 19 & 0.259 & 0.20 & -0.71 & -1.31\\
27 & $>$300 & 4 & all & 14 & 0.109 & 0.48 & -2.14 & -4.63\\
28 & $>$3600 & 1 & all & 33 & 0.232 & 0.34 & -1.97 & -3.08\\
29 & $>$3600 & $>$1 & all & 36 & 0.204 & 0.33 & -1.00 & -1.77\\
30 & $>$3600 & 2 & all & 24 & 0.220 & 0.33 & -0.95 & -1.63\\
\hline
\end{tabular}

\end{minipage}
\end{table*}

\subsubsection{Small Planets}
\label{subsubsec:small}
We prepared another sample similar to {\it EXOCAT} ({\it EXOCAT2}) that differs only in final selection criterion; instead of requiring an {\it e} measurement, we require either a valid mass or $m \sin i$ measurement. We employ the {\it EXOCAT2} sample of 2262 planets to examine the distribution of ``small planets'' (SPs, 1--20 M$_{\oplus}$),\footnote{This category is defined in terms of mass, rather than size, for the benefit of including planets not observed with the transit method. SPs encompass super-earths and sub-Neptunes, which are often defined in terms of size.} finding that SPs make up 33 percent of {\it EXOCAT2}. However, their distributions are very different when we take multiplicity into account; 17 percent of single-planet systems have a SP, while 56 percent of the planets found among multiple planet systems are SPs. For multiple planet systems with more than 5 planets the fraction is up to 74 percent. Among two-planet systems, the fraction is 44 percent. In the context of the absence of SPs, then, the Solar System (as a two-planet system) is not atypical.

Summarizing this section, the {\it detectable} Solar System would be most likely a two-planet system with the transit and/or RV detection methods. We can conclude that the Solar System differs most significantly ($ <10$ percent) in the following ways from the typical exoplanetary system: 1) the occurrence of {\it Kepler}-detectable planets, 2) the size-scale of the detectable two-planet Solar System, 3) the occurrence of an RV-detectable cold Jupiter, 4) the low-{\it e} of detectable Solar System (Venus + Jupiter), and 5) lack of super-earths + presence of cold Jupiter. The Solar System is likely typical in the orbit of Jupiter in relation to the metallicity of the sun. The full list of the characteristics we examined in this section is summarized in the second section of Table~\ref{tab:summary}.

\section{Galactic parameters}
\label{sec:galactic}
Our knowledge of the Solar System's location and motion within the Milky Way galaxy has improved dramatically in recent years. In \citet{gon99a} we argued that the sun's location relative to the mid-plane and its motion in the disk are anomalous. We revisit these questions and more below in the light of recent observations, beginning with the Solar System's location in the Milky Way.

The height of the Solar System relative to the mid-plane of the Milky Way, z$_{0}$, is determined by comparing star counts at the galactic north and south poles or tracing the distributions of young objects. Recent determinations span the range from about 5 to 25 pc (see Table 2 of \citet{griv21} for a listing) and seem to depend on the method of analysis and type of sample, but they tend to cluster around ``low'' values near 5 pc and ``high'' values near 16 pc (thus the listing of two values for z$_{0}$ in Table~\ref{tab:sun_params}). \citet{reid19} suggested that these differences are due to extinction and galactic disk warping. Rather than trying to settle the debate here and choosing one value, we base our comparison of the sun's motion to field stars on the two values of z$_{0}$ given in Table~\ref{tab:sun_params}.

As suggested above, neglect of the galactic disk warp can bias z$_{\odot}$ estimates. \citet{chen19} were able to map the disk warp and model it with a simple analytic expression from observations of Classical Cepheids. Two aspects of the warp are relevant to our present discussion. First, the warp is most obvious outside the solar circle. From their best-fit power law disk warp model, we calculate that the deviation in z from a flat plane at our location is only about 3 parsecs. There are two reasons for this: the warp is very small inside the solar circle, and the line of nodes of the warp is only 17.5 degrees from the solar position. Over the course of its galactic orbit the solar system spends most of its time farther from the galactic center and farther from the mid-plane than the present position (see below). Thus, the Solar System is currently about as close as it ever gets to the plane defined by the inner galactic disk.

\citet{kordo23} calculated galactocentric positions, velocities and orbital parameters for the {\it Gaia} targets with radial velocities. To do this they adopted R$_{0}$ and z$_{0}$ values of 8.249 kpc and 20.8 pc, respectively. For the velocities, they assumed (V$_{R}$, V$_{\phi}$, V$_{Z}$)$_{\odot}$ $= (-9.5, 250.7, 8.56)$ km~s$^{-1}$. Using the code \texttt{Galpy}\footnote{\url{https://github.com/jobovy/galpy}} \citep{bovy15} with an axisymmetric galactic model, they also calculated eccentricity and maximum distance from the midpane, Z$_{max}$. However, as we will show below, these are probably not the best values to use to calculate galactic orbits.

The relationship among the circular rotation speed at the sun's position, $\Theta_{0}$, the full circular speed of the sun, (V$_{\phi}$)$_{\odot}$, and the sun's peculiar azimuthal velocity, V$_{\odot}$, is:
\begin{equation}
(V_{\phi})_{\odot} = \Theta_{0} + V_{\odot}
\end{equation}
\citet{reid20} measured the proper motion of Sgr A* (presumably at the galactic center) in Galactic longitude and latitude to be $-6.411 \pm 0.008$ and $-0.219 \pm 0.007$ mas yr$^{\rm -1}$, respectively. Multiplying these by R$_{0} = 8.23$ kpc and reversing the sign yields the Solar System's velocity components relative to Sgr A*: (V$_{\phi}$)$_{\odot} = 250.1$ km~s$^{\rm -1} \pm 1.6$ km~s$^{\rm -1}$ and W$_{\odot} = 8.5 \pm 0.3$ km~s$^{\rm -1}$. It is not known why the value of W$_{\odot}$ is larger than the typically quoted value (e.g., Table~\ref{tab:sun_params}).

\citet{reid19} calculated (V$_{\phi}$)$_{\odot}$ to be $247\pm4$ km~s$^{\rm -1}$, R$_{0}$ to be $8.15\pm0.15$ kpc, and W$_{\odot}$ to be $7.6\pm0.7$ km~s$^{\rm -1}$ from observations of masers. From our adopted value of 11 km~s$^{\rm -1}$ for V$_{\odot}$ from Table~\ref{tab:sun_params}, this requires $\Theta_{0}$ to be 236 km~s$^{\rm -1}$. This is consistent with \citet{kaw19}, who employed {\it Gaia} DR2 observations of classical cepheids to derive $\Theta_{0} = 236 \pm 3$ km~s$^{\rm -1}$; it is notable that they adopted a prior on R$_{0}$ of $8.2 \pm 0.1$ kpc, which is consistent with the value in Table~\ref{tab:sun_params}.

We can calculate an independent value of $\Theta_{0}$ from:
\begin{equation}
\Theta_{0} = R_{0} \Omega = R_{0} \lvert A-B \rvert
\end{equation}
where {\it A} and {\it B} are the Oort constants. \citet{li19} determined $\Omega$ to be $28.5 \pm 0.1$ km s$^{\rm -1}$ kpc$^{\rm -1}$ and \citet{wang21b} derive $28.3 \pm 1.2$ km s$^{\rm -1}$ kpc$^{\rm -1}$. Adopting $28.4 \pm 0.6$ km s$^{\rm -1}$ kpc$^{\rm -1}$ and multiplying by R$_{0}$ yields $233.1 \pm 5.1$ km s$^{\rm -1}$. This is consistent with the value derived by \citet{kaw19}.

In the following analysis we adopt the galactic constants listed in Table~\ref{tab:sun_params},  (V$_{\phi}$)$_{\odot}$ = 247 km s$^{\rm -1}$, and $\Theta_{0} = 236$ km~s$^{\rm -1}$. These differ slightly from values used by \citet{kordo23} to calculate galactocentric positions, velocities and orbital parameters. We have applied the small corrections to their velocities to bring them into agreement with our assumptions and used our adopted galactic parameters with \texttt{Galpy} to calculate a new set of eccentricity and Z$_{max}$ values for the stars in K-sample 3 (Section~\ref{subsec:comp}).

With these adopted parameters (and assuming the sun's height above the midplane is 16 pc), the sun's Z$_{\max}$ and eccentricity values are found to be 103 pc and 0.061, respectively. The sun has a smaller Z$_{\max}$ value than 76 percent of the K-sample 3 stars and a smaller eccentricity than 82 percent. Together, only 5 percent of the sample stars have smaller values of both quantities. This implies that the sun's orbit is colder than 95 percent of otherwise similar nearby sun-like stars.

We can repeat these comparisons for the smaller {\it GALAH} Solar Sample 1 from Section~\ref{subsec:comp}. Instead of adopting the galactic parameters from the {\it GALAH} dynamics VAC\footnote{\url{https://www.galah-survey.org/dr3/the_catalogues/}}, we calculated them from the astrometric data. We find that the sun has a smaller Z$_{\max}$ value than 94 percent of the {\it GALAH} Solar Sample 1 stars and a smaller eccentricity than 79 percent. Taken together, only 2 percent of the sample stars have smaller values of both quantities. While both these samples are restricted to nearby solar twins, the {\it GALAH} sample includes a larger volume of space, and it is not as complete as K-sample 3 for its sampled volume. Still, it can be concluded that the sun's orbit is colder than 2 to 5 percent of nearby solar twins.

The observed number density profile in the z direction is well described by the function,
\begin{equation}
n(z) = A_{0} \sech^2 \left(\frac{z}{z_{h}}\right)
\end{equation}
This functional form is frequently used in studies of the galactic matter distribution \citep[e.g., ][]{buch19}. Fitting this equation to the number density counts in K-sample 3 between z = 0 and 150 pc yields z$_{h}$ = 191 pc; the stars with negative z values were not included in the fit, since they display a significant deficit compared to the stars with positive z (presumably due to the sun's location above the midplane and dust in the midplane). Integrating the equation over the range $-1000$ to $+1000$ pc in z with the best-fit z$_{h}$ value gives a total of 18,991 expected stars, which compares to 12,496 stars actually in this sample. We take this difference to be a measure of the completeness of K-sample 3 (that is, completeness in the z-direction).

The raw percentage of stars in K-sample 3 within 5 pc of the midplane is 4, while 12 percent are within 16 pc (this covers the range of most likely values of z$_{\odot}$). When corrected for completeness, and assuming that K-sample 3 is complete within 16 pc of the midplane, these percentages become 2.6 and 7.9 percent, respectively. In other words, anywhere between about 3 and 8 percent of sun-like solar-age stars in the solar neighborhood have z values smaller than the sun's.

Next, using the velocities we compare the Solar System's total velocity relative to the LSR, V$^{\rm tot}_{\rm LSR} (= 16.6$ km s$^{\rm -1}$), to the stars of K-sample 3. We calculate that 12 percent of the sample has smaller velocities relative to the LSR than the sun. This consistent with our finding that the sun has a colder orbit than most sun-like stars in the solar neighborhood.

\begin{figure}
\includegraphics[width=3.2in]{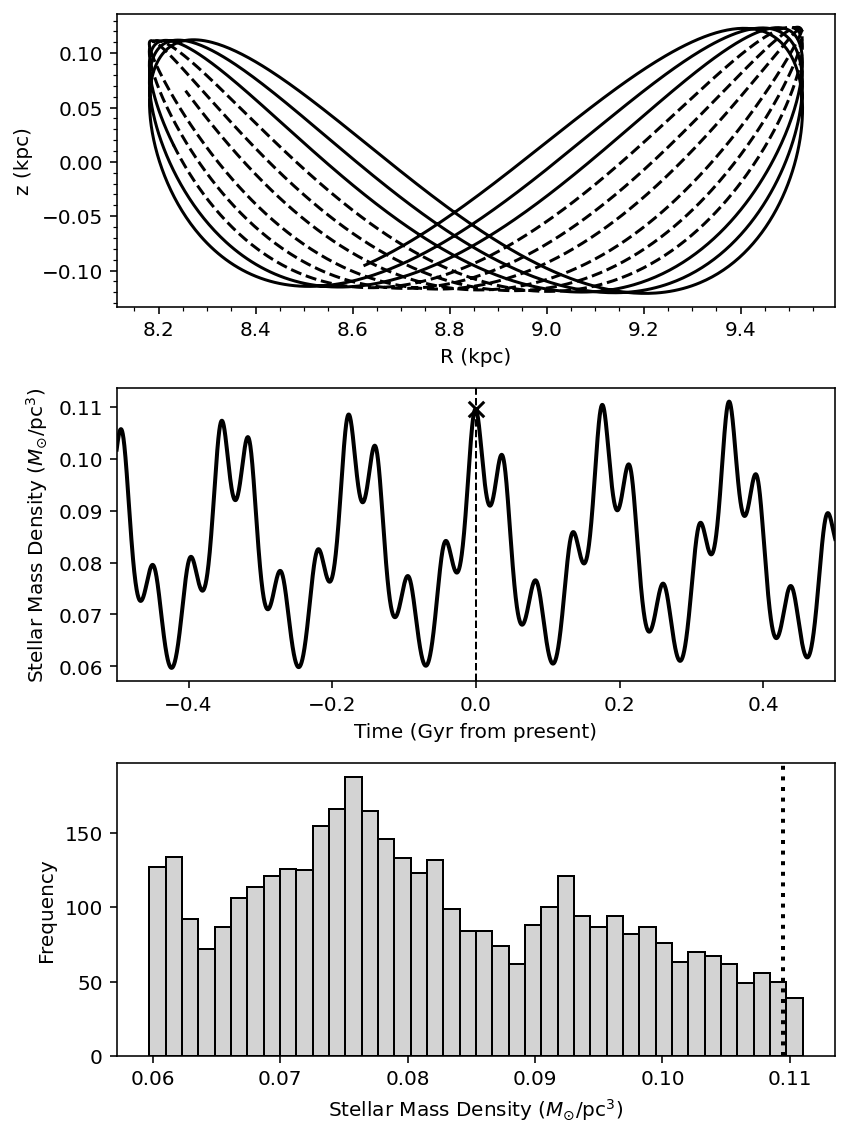}
\caption{The sun's galactic orbit integrated from $-0.5$ to $+0.5$ Gyrs; future orbit is shown as a solid curve and past orbit as a dashed curve (top panel). Stellar density near the sun over the same time interval (middle panel). Histogram of the stellar density values from the points plotted above (bottom panel).}
\label{fig:galpy_plots}
\end{figure}

We can also ask if we are living during a special time in the sun's galactic orbit. We integrated its orbit using \texttt{Galpy} with 4000 steps from $-0.5$ to $+0.5$ Gyr with the parameters described above and with a current midplane distance of 16 pc. We show the output from the simulation in Fig.~\ref{fig:galpy_plots}. We already showed that the sun is unusually close to the midplane. The simulated orbit also indicates that it is close to perigalacticon. The effect of the sun being so close to both these special places is for it to be near the maximum of the disk mass density it encounters during its full galactic orbit. From this simulation, we calculate that the mass density near the sun is greater than the current value during only 1.25 percent of this time interval. Changing the midplane distance makes very little difference. This is because stars spend most of their time near \( \left| Z_{\max} \right| \).

We can make a more global comparison of the location of the Solar System in the Milky Way by asking the question, "What fraction of the stars in the galaxy are inside the current/mean galactocentric distance?" The answer will give us an idea whether the Solar System is at a typical value of {\it R} in the galaxy. From the same \texttt{Galpy} Milky Way potential we calculate that 76 percent of the stellar mass is within {\it R$_{0}$}; it is 78 percent if, instead, we calculate the stellar mass interior to the Solar System's average value of {\it R}.

We can also compare the interstellar gas density in the solar neighborhood to other regions of the Milky Way. Both atomic (H I) and molecular gas (H$_{\rm 2}$) have been mapped throughout the galaxy. \citet{md17} plot the surface mass densities of H I and H$_{\rm 2}$ in their Figure 9. The H$_{\rm 2}$ density is well described with an exponential between $R =$ 4 and 17 kpc with a scale length of 2.0 kpc. The H$_{\rm 2}$ mass fraction in the solar neighborhood is about 10 percent. About 88 percent of the H$_{\rm 2}$ mass in the galaxy resides within the solar circle. In addition, the plot on the upper right of their Figure 9 shows a spike just inside the solar circle (it is a plot of the total mass of H$_{\rm 2}$ in galactocentric rings). On a smaller scale, the Solar System is located within the ``local bubble'', which is a region about 200 pc in diameter containing very little interstellar material (see their Figure 11).

Over the past several decades several studies have concluded that the Solar System is near the Milky Way's corotation resonance. For instance, \citet{mont21} determined that the corotation radius, $R_{\rm c}$, is $1.01 \pm 0.08$ R$_{0}$ from {\it Gaia} DR2 observations of young galactic clusters. \citet{boby23a} find that $R_{\rm c}$ is $1.14 \pm 0.06$ R$_{0}$ from three different types of objects; \citet{boby23b} derive the same result from young open clusters using {\it Gaia} DR3 data. However, this topic remains controversial (e.g., \citet{vall21} finds $R_{\rm c}$ to be about 16 kpc).

In summary, the sun's galactic position and motion are anomalous in several respects. The present position of the sun in its galactic orbit is near the maximum of the disk stellar density and the minimum in the z distance from the plane defined by the inner disk. The motion of the sun is colder than most local sun-like stars of solar age. It's location in the disk is outside most of the stellar and molecular mass of the galaxy. The full list of the characteristics we examined in this section is summarized in the third section of Table \ref{tab:summary}.

\section{Discussion}
\label{sec:disc}
We summarize the findings of our analyses from previous sections in Table \ref{tab:summary}, where we list for each examined property, whether it is anomalous (``yes'' or ``no''), and the percentage (i.e., 10 percent means that a property is different from 90 percent of the relevant comparison instances). Uncertain cases are enclosed in parentheses. The ``anomalous?'' column is labeled as ``yes'' if the ``percent'' column is less than 50. 

What could account for the anomalous properties of the Solar System? Possible explanations include anthropic self-selection bias \citep{gon99a,gon99b,livio19,wal19}, inadequate comparison samples, systematic errors, and chance. We will consider these in the following, with most of the discussion focused on anthropic explanations.

\begin{table*}
\centering
\begin{minipage}{200mm}
\caption{Summary of the aspects of the Solar System considered and how anomalous they are.}
\label{xmm}
\begin{tabular}{lccccr}
\hline
Category: property & Section & observed value & fit value & anom.? & \% \\
\hline
Sun: & & & &\\
~~~mass & \ref{subsec:mass} & 1 M$_{\odot}$ & -- & yes & 8\\
~~~age & \ref{subsec:comp} & 4.6 Gyr & various & no & --\\
~~~binarity & \ref{subsec:binarity} & single & -- & no & --\\
~~~photometric variability: & & & & &\\
~~~~~~rrmscdpp$_{\rm intrinsic}$ & \ref{subsubsec:phot_var_Kepler} & 11.0 ppm & 40 ppm & yes & 0.2\\
~~~~~~S$_{\rm ph}$ (compared to periodic variable stars) & \ref{subsubsec:phot_var_nearby} & 300 ppm & 1500 ppm & (yes) & 30\\
~~~~~~R$_{\rm var}$ (compared to periodic variable stars) & \ref{subsubsec:phot_var_nearby} & 0.07 percent & 0.176 percent & yes & 20\\
~~~~~~R$_{\rm var}$ (compared to non-periodic variable stars) & \ref{subsubsec:phot_var_nearby} & 0.07 percent & 0.096 percent & (yes) & 30\\
~~~~~~log{\it A}$_{\it TESS}$ & \ref{subsubsec:phot_var_Kepler} & 2.87 & 2.84 & no & --\\
~~~chromospheric activity: & & & & &\\
~~~~~~S$_{\rm L}$ & \ref{subsubsec:phot_var_Kepler} & 0.1865 & 0.1961 & no & --\\
~~~~~~log R$^{\rm '}_{\rm HK}$(T$_{\rm eff}$) & \ref{subsubsec:phot_var_nearby} & $-5.020$ & $-4.986$ & no & --\\
~~~~~~log R$^{\rm +}_{\rm HK}$ & \ref{subsubsec:phot_var_nearby} & $-4.852$ & $-4.794$ & no & --\\
~~~~~~H$\alpha$ F$_{\rm Total}$ (in units of 10$^6$ erg cm$^{-2}$ s$^{-1}$) & \ref{subsubsec:phot_var_nearby} & 5.120 & 5.121 & no & --\\
~~~~~~superflare frequency (per century) & \ref{subsubsec:phot_var_Kepler} & $<0.1$ & 1.5-3.0 & yes & $<<$1\\
~~~~~~X-ray to bolometric luminosity ratio, $\log_{10}$ (L$_{X}$/L$_{bol}$) & \ref{subsubsec:phot_var_nearby} & $-5.95$ & $-5.00$ & (yes) & 5\\
~~~rotation period & \ref{subsec:phot_var_rot} & 25.29 d & 24.06 d & yes & $<<$1\\
~~~{\it v} sin{\it i} & & 2.04 km s$^{\rm -1}$ & 2.09 km s$^{\rm -1}$ & (no) & --\\
~~~composition: & & & & &\\
~~~~~~metallicity, [M/H] & \ref{subsec:comp} & 0.00 & various & yes & 10\\
~~~~~~[$\alpha$/Fe] & \ref{subsec:comp} & 0.00 & 0.01 & no & --\\
~~~~~~abundance-T$_{\rm c}$ slope & \ref{subsec:other_elem} & -- & -- & yes & 10\\
~~~~~~C,N,O abundances and ratios and C isotopes & \ref{subsec:other_elem} & Table~\ref{tab:CNO} & Table~\ref{tab:CNO} & yes & 5\\
~~~~~~lithium abundance & \ref{subsec:other_elem} & 1.07 & 1.47 & yes & 5\\
~~~~~~beryllium abundance & \ref{subsec:other_elem} & 1.16 & 1.11 & yes & $<$1\\
~~~~~~phosphorus abundance, [P/Fe] & \ref{subsec:other_elem} & 0.00 & 0.00 & no & --\\
~~~~~~[Y/Mg] & \ref{subsec:other_elem} & 0.00 & 0.00 & no & --\\
~~~~~~heavy neutron-capture element abundances (Eu, Nd, Pr, Th, Pb) & \ref{subsec:other_elem} & various & various & yes & $<$1\\
Solar System planets: & & & & &\\
~~~median [Fe/H] of host stars for circular cold Jupiters & \ref{subsec:exo_host} & 0.00 & 0.03 & no & --\\
~~~median age of host stars for circular cold Jupiters & \ref{subsec:exo_host} & 4.6 & 6.07 & yes & 10\\
~~~{\it Kepler}-detectable planet occurrence/multiplicity with host star mass & \ref{subsec:exo_host} & -- & -- & yes & 5-20\\
~~~full complement of Solar System planets -- multiplicity & \ref{subsec:mult} & -- & -- & yes & 1\\
~~~full complement of Solar System planets -- eccentricity & \ref{subsec:mult} & -- & -- & no & --\\
~~~{\it Kepler}-detectable planet occurrence for P $<$ 300 d, 0.5 $<$ R $<$ 10 R$_{\oplus}$ & \ref{subsubsec:occur} & 1 & 3.5 & yes & 5\\
~~~size scale of detectable two-planet Solar System & \ref{subsec:mult} & 2.95 AU & 0.7 AU & yes & 6\\
~~~four Solar System inner planets size and spacing dispersions & \ref{subsubsec:struct} & -- & -- & (yes) & 40\\
~~~lack of super-earth(s) conditioned on presence of cold Jupiter & \ref{subsubsec:struct} & -- & -- & yes & 3\\
~~~occurrence of RV detectable cold Jupiter & \ref{subsubsec:occur} & -- & -- & yes & 4\\
~~~occurrence of ``small planets'' (super-Earths and sub-Neptunes) & \ref{subsubsec:small} & -- & -- & (no) & --\\
~~~mean {\it e} of Venus+Jupiter & \ref{subsubsec:eccensec} & Table~\ref{tab:eccentab} & Table~\ref{tab:eccentab} & yes & $<$2\\
~~~mean {\it e} of {\it Kepler}-equivalent planets at solar [Fe/H] & \ref{subsubsec:struct} & -- & -- & (no) & --\\
~~~mutual inclination of {\it Kepler}-equivalent planets & \ref{subsubsec:struct} & -- & -- & (no) & --\\
~~~sun's obliquity relative to invariant plane & \ref{subsubsec:struct} & 6.2 deg & $<2$ deg & yes & 10\\
Solar System galactic properties: & & & & &\\
~~~z$_{\odot}$ & \ref{sec:galactic} & 5, 16 pc & -- & yes & 3, 8\\
~~~Z$_{max}$ & \ref{sec:galactic} & 103 pc & -- & yes & 10\\
~~~{\it e} & \ref{sec:galactic} & 0.061 & -- & yes & 20\\
~~~Z$_{max}$ and {\it e} combined & \ref{sec:galactic} & -- & -- & yes & 2-5\\
~~~V$^{\rm tot}_{\rm LSR}$ & \ref{sec:galactic} & 16.6 km s$^{\rm -1}$ & -- & yes & 12\\
~~~temporal galactic star density near sun over 1 Gyr & \ref{sec:galactic} & -- & -- & yes & 1.25\\
~~~stellar mass $>$ {\it $\overline R_{\odot}$} & \ref{sec:galactic} & -- & -- & yes & 22\\
~~~H$_{\rm 2}$ mass $>$ {\it R$_{0}$} & \ref{sec:galactic} & -- & -- & yes & 12\\
~~~R$_{\rm c} =$  {\it $\overline R_{\odot}$} & \ref{sec:galactic} & -- & -- & (yes) & 10\\

\hline
\end{tabular}

\end{minipage}
\label{tab:summary}
\end{table*}

\subsection{Adequacy of Comparison samples}

For each property we examined, we selected the best available comparison sample(s). In most instances, a comparison dataset is based on the most recent observations and/or analyses. Many incorporate measurements from {\it Gaia} DR2 or DR3. The comparison samples for the sun's mass and binarity are among the most robust, but even for these we expect improvement among the latest spectral types following the next {\it Gaia} data release. Comparison samples requiring more attention include, for example, rotation periods for photometrically ``quiet'' stars, more sun-like stars with precise {\it v} sin{\it i}, and longer-baseline planet transit and RV surveys. This situation can be improved easily with additional observations. Continued surveys of exoplanets will only improve comparison of their properties to the sun's planets.

\subsection{Anthropic ``explanations''}

Anthropic explanations relate the particular properties of an environment to its habitability for observers. In the present application, anthropic considerations apply to the Earth's geophysical properties, the sun's properties, the overall properties of the Solar System, and the galactic context. As intelligent life,\footnote{We adopt the definition of intelligent life of \citet{gleiser12}: ``capable of self-awareness and of developing technology through the directed manipulation of energy and materials.''} we are the observers which this setting must be compatible with. In other words, anthropic self selection necessarily restricts our setting to one that is permits us to exist \citep{carter74,carter83}.

\subsubsection{Properties of the sun}

The mass of the host star in a planetary system has long been considered to be an important habitability factor \citep{wal11,wal17}. On the high end, the lifetime on the main sequence and astrosphere descreening \citep{smith09} have been considered as factors, while on the low end several other factors have been considered \citep{schw19,yama19,rodmoz19,atri20,cook24}. The low probability of the sun's high mass tends to argue against the habitability of low mass stars for intelligent life.

The overall metallicity of the protoplanetary disk around the early sun is reflected in the sun's photospheric metallicity, excepting for corrections due to element diffusion and gravitational settling in its envelope and possibly fractionation of accreted material \citep{gonz14a}. The probability of the presence of a giant planet is a sensitive function of the metallicity of the host star \citep{santos2001,santos2004,gonz14b}; smaller planets display less sensitivity to metallicity. Furthermore, \citet{baner24} showed that cold Jupiters in circular orbits tend to orbit solar metallicity stars, while hot Jupiters, warm Jupiters, and cold Jupiters on eccentric orbits tend to have metal-rich hosts; the latter three classes of planets have been considered to be detrimental to the formation of habitable planets \citep{ray04,mand07,must17}. While the sun's metallicity is not anomalous with respect to either the local interstellar medium or nearby stars, it is slightly anomalous with respect to nearby solar age stars. The presence of Jupiter, then, could be anthropically selected via the sun's metallicity. The habitability of a planetary system, in turn, depends on the presence (or absence) of giant planets (see below).

The sun's anomalous metallicity compared to nearby sun-like stars of solar-age has been cited as evidence for its birth location in the inner galactic disk \citep{wiel96}. The basic idea is that stars in the inner disk of the galaxy are born form more metal-rich gas than stars in the solar neighborhood at any given time, and simulations show that stars migrate in the disk over several Gyrs. \citet{baba23} estimated that the sun was born at about 5 kpc from the galactic center. They also noted that when other elements critical for habitable planet formation, such as C, O, Mg, Si, are also considered that planets formed in the inner galaxy have higher condensed mass fraction, higher Fe-to-Si mass fraction, and lower water mass fraction.\footnote{For instance, given the observed range of the C/O ratio amongst nearby stars, the water content in the protoplanetary disk is expected to vary over a large range, with only modestly larger ratios than the solar value producing water-poor systems \citep{john16}.} This hypothesis could account for the sun's anomalous C, O, and Fe abundances relative to nearby stars, but only in part. It would still be necessary to explain why sun-like solar-age nearby stars don't display the same abundance patterns (presumably also from migrating from the inner galactic disk).

The fractional abundances of the heavy $r$-process elements is linked to habitability via the radioactive isotopes $^{235,238}$U and $^{232}$Th, which provide significant heating in Earth's interior for long-lived geophysical processes \citep{gonz01,unter22}. What is surprising in this context is that the sun's $r$-process element abundances are anomalously low. A possible mechanism that limits the $^{235}$U abundance in the biosphere was provided by \citet{coog09}. They argued that natural fission reactors formed on the early earth in localized regions where oxygenic photoautotrophs were growing. The radiation and toxic byproducts from the fission reactors, in turn, would have acted as a negative feedback on these oxygen producers, delaying the rise of oxygen in the atmosphere until $^{235}$U was sufficiently low in abundance to prevent natural fission reactors. A higher initial abundance of $^{235}$U would have slowed the rise of oxygen and thus delayed the arrival of complex metazoan life. Although this mechanism only relates life to $^{235}$U, it would also place restrictions on other heavy $r$-process elements since they are produced together in the same source. In this way, selective pressure on just one $r$-process isotope places pressure on the whole set.

Another possible habitability constraint on the heavy $r$-process elements concerns to toxicity of lead to life \citep{eisler88}. Although, $^{204}$Pb is a pure $s$-process isotope, it only makes up 1.4 percent of the total lead abundance in the Solar System. The other three isotopes, $^{206}$Pb, $^{207}$Pb, and $^{208}$Pb, are the end products of the decay chains of uranium and thorium.

\citet{mel12} suggested that the sun's anomalously low abundances of the refractory elements were an outcome of the formation of the terrestrial planets. Specifically, refractory elements that otherwise would have gone into the sun's envelope during the early accretion phase, instead, were locked away in the inner Solar System planets. What still needs explaining is why other stars with planets don't show this pattern more often, given the high fraction of stars hosting small planets in the solar neighborhood.

The sun's anomalously low lithium abundance is possibly also related to the process of planet formation. \citet{bouv08} proposed a model to explain the lower lithium abundances of stars hosting planets \citep{fig14,gonz15}. In brief, a long-lasting disk produces a slowing torque on a star during the pre-main sequence stage, which causes extra mixing at the base of its convective envelope and, in turn, more rapid destruction of its lithium. However, this doesn't explain the anomalously high solar beryllium abundance. This model also fits with the evidence for the sun being a slow rotator early on. 

While only some measures of the photometric variability and chromospheric activity of the sun are anomalous, the overall pattern shows that the sun is ``quieter" than solar twins. Such variability could be linked to known sun-climate correlations \citep{gray10,le19}. In addition, chromospheric activity is related to the rate of flares, which can detrimental to habitability \citep{berg24}. However, it is difficult to link any one aspect of solar variability to climate, since the mechanism(s) has(have) not been fully elucidated. Proposed mechanisms include UV, solar wind and cosmic rays.

The sun’s relatively moderate activity presents a complex trade-off when evaluating planetary habitability. While frequent, energetic flares can be detrimental to established biospheres, their NUV emission may be a crucial energy source for driving prebiotic chemical networks, suggesting that more active stars could be more conducive to the origin of life \citep{rimm18,xu18,spin23b}. The sun's current quiescence, therefore, may be more favorable for the survival and long-term stability of a complex biosphere than its more volatile youth. This tension extends to broader evolutionary dynamics, where catastrophic events, while disastrous for existing species, have been controversially proposed as long-term drivers of biodiversity by clearing ecological niches, which could ultimately increase the chances of intelligent life emerging \citep{sepk85,raup94}. However, this "creative destruction" hypothesis remains debated, with counterarguments emphasizing the profoundly negative and contingent nature of recovery from such events, questioning whether they are a reliable mechanism for promoting evolutionary innovation. This highlights the intricate and sometimes contradictory environmental conditions required for life's emergence, survival, and long-term evolution.

\subsubsection{Properties of the Solar System}

\citet{livio19} argued that the most likely anthropic explanation for the absence of super-earths as well as planets in tight orbits in the Solar System was the formation of super-earths inside the orbit of Jupiter and their subsequent inward migration and engulfment. He proposed that super-earths formed early and then migrated into the sun, leaving behind just enough solid debris to form the relatively low mass terrestrial planets. This requires fine-tuning of the surface density in the turbulent layer of the disk. An alternative mechanism for causing super-earths to migrate into the sun is the ``Grand-Tack'' scenario \citep{walsh11,bat15}, wherein Jupiter migrates inward and stops at about 1.5 AU and then reverses direction when it captures Saturn in an orbital resonance. This also requires fine-tuning. This second mechanism seems the preferable explanation for the lack of super-earths, since it also links the presence of an outer Jupiter to the lack of super-earths; this is one of the most anomalous aspects of the Solar System's planets. The Grand-Tack scenario also relates to habitability, since it provides a way for volatiles to be delivered to the otherwise dry terrestrial planets \citep{mat16}.

\citet{small18} investigated via N body simulations the effect of a super-earth on the asteroid impact rate. They found that the presence of a super-earth interior to earth's orbit increases the asteroid impact rate on the earth, but the opposite is the case for a super-earth exterior to earth's orbit.

\subsubsection{Galactic context}

In \citet{gonz01} we introduced the Galactic Habitable Zone (GHZ) concept to explore habitability within the galactic context. In the Milky Way the GHZ has an inner boundary mostly set by threats to complex life, which tend to increase towards the galactic center. These include supernovae \citep{thomas18}, nuclear outbursts \citep{balbi17}, passages through giant molecular clouds \citep{kok19}, perturbations of the Oort cloud comets \citep{masi09}, and gamma ray bursts \citep{spin23}.\footnote{In the case of the more luminous long duration gamma ray bursts, there is a metallicity cutoff of about one-third solar metallcity, above which they are far less frequent. For this reason, the outskirts of the Milky Way gradually has become relatively more dangerous relative to the inner region.} The outer boundary is set by the metallicity required to form a habitable planetary system, affecting such properties as the host star variability, the overall system architecture, gas giant properties, and binarity. These factors must be treated together in the evolving Milky Way galaxy \citep{spit17}. Although many studies of the GHZ have been published over the past two decades, the relative importance of each of these factors and the locations of the boundaries are not yet well known.

The fact that most of the stellar mass resides inside the solar circle in galaxy suggests that one or more radial-dependent galactic habitability factors is/are important. In particular, the even lower probability of the Solar System's location relative to the H$_{\rm 2}$ mass in the disk suggests molecular clouds are a habitability factor. At the same time, the Solar System occupies an even more improbable location relative to the full range of its orbit in the galaxy; it is near the highest mean stellar density along its galactic orbit.

The solution to this paradox might be found in another improbable aspect of the Solar System's present location, the z distance from the plane defined by the inner disk. A possible benefit arising from this state is the shielding from UV to soft X-ray radiation provided by intervening dust from the galactic center. AGN outbursts from the Milky Way galaxy's supermassive black hole produce high energy electromagnetic and particle radiation. \citet{haw19} estimated that the most recent large nuclear outburst occurred $3.5 \pm 1$ Myr ago. At that time the Solar System slightly closer to the mid-plane, but on the opposite side. Thus, the Solar System would have received slightly more protection from the galactic center ionizing radiation as it does now.

The Solar System's small V$^{\rm tot}_{\rm LSR}$ value is difficult to understand from an anthropic perspective. A slowly moving star passing by a star forming region spends more time near it, and it also stays closer to the mid-plane. In addition, astrosphere descreening is more likely at lower speeds relative to the ISM \citep{smith09}. However, the relatively cold galactic orbit of the Solar System keeps it from wandering very far into the inner galaxy.

If \citet{lep17} are correct that the Solar System is near the corotation resonance and that it is trapped in a ``corotation zone,'' then at least a couple of the Solar System's galactic properties could be explained by anthropic selection. They argue that our presence within one of the four corotation zones in the disk keeps us from crossing one of the major spiral arms. Spiral arm crossings pose threats to complex life \citep{fili13,kok19}. The combination of the Solar System's proximity to corotation and its low V$^{\rm tot}_{\rm LSR}$ value minimizes the likelihood that its passes through the major spiral arms. However, contrary to anthropic expectations, the small W$_{\odot}$ value increases the time inside or near giant molecular clouds \citep{kok19}.

\section{Conclusions}
\label{concl}
We have compared numerous quantifiable properties of the sun, its planets, and our location in the galaxy to other stars and to exoplanets using the most relevant recent data. We found multiple properties that are anomalous in the statistical sense. The more significant ones include: the sun's photometric variability on short timescales, rotation period, superflare frequency, light element abundances, heavy neutron-capture element abundances, mean eccentricity of detectable two-planet Solar System, size scale of detectable two-planet Solar System, lack of super-earths conditioned on presence of a cold Jupiter, presence of RV-detectable cold Jupiter, z-distance from the galactic mid-plane, low temperature galactic orbit, and the present star density near the sun compared to any other point in the galactic year.

Properties that are typical include: the sun's age, lack of a stellar companion, chromospheric activity, [$\alpha$/Fe], [P/Fe], [Y/Mg] abundance ratios, [Fe/H] value given presence of cold circular Jupiter, and the mean eccentricity of full complement of Solar System planets.

Many of the datasets we employed to prepare comparison samples are of recent vintage. Several of our analyses make use of the {\it Gaia} DR3 dataset, which has been especially useful in deriving stellar ages and galactic motions. Continued exoplanet discoveries from missions such as {\it TESS} and the anticipated bounty of planets from the final {\it Gaia} release will help to refine the probabilities of the Solar System anomalies. Detailed high quality spectroscopic analyses of additional solar twins would be especially helpful, given that several of the sun's anomalous properties involve spectroscopically-determined quantities.

We have also offered plausible anthropic explanations for many of the anomalies, which can guide us in further researching their possible astrobiological significance. One possible application would be to assign to nearby solar twins an index measuring how much they differ from the Solar System in some of the anomalous properties that may be related to planets. This index can then be used to target the stars most similar to the sun for Earth-like planets. Success of such a search would lend support to the putative link between the anomalous properties and the presence of Earth-like planets. They could also be targeted by SETI programs. The galactic anomalous properties, in turn, can be employed to guide research on the Galactic Habitable Zone concept \citep[e.g.,][]{gonz01,line01}.

On the other hand, those properties of the Solar System that we have concluded are typical do not necessarily imply they are not astrobiologically relevant. The fact that they are typical does not, by itself, provide constraints on habitability. In these cases the burden of proof is shifted to demonstrating that habitability declines away from the typical range.

\section*{Data Availability}

This work includes data collected by the {\it Kepler} mission. Funding for {\it Kepler} is provided by the NASA Science Mission directorate. This work has also made use of data from the European Space Agency (ESA) mission Gaia (\url{https://www.cosmos.esa.int/gaia}), processed by the Gaia Data Processing and Analysis Consortium (DPAC, https://www.cosmos.esa.int/web/gaia/dpac/consortium). Funding for DPAC has been provided by national institutions, in particular the institutions participating in the Gaia Multilateral Agreement.

\section*{Acknowledgements}
We thank the anonymous reviewers for suggestions that improved the quality of this work.

\bsp

\label{lastpage}

\end{document}